\documentclass[aps,amssymb,showkeys]{revtex4}

\usepackage{array}
\usepackage{multirow}
\usepackage{amsmath}
\usepackage{graphicx} 
\usepackage{amssymb}
\usepackage[pdftex, bookmarks, colorlinks=true, plainpages = false, citecolor = red, urlcolor = blue, filecolor = blue]{hyperref}
\usepackage{color}

\newcommand{\be}{\begin{equation}}
\newcommand{\ee}{\end{equation}}
\newcommand{\bea}{\begin{eqnarray}}
\newcommand{\eea}{\end{eqnarray}}
\newcommand{\bean}{\begin{eqnarray*}}
\newcommand{\eean}{\end{eqnarray*}}

\allowdisplaybreaks

\begin{document}
\title{Cluster pinch-point densities in polygons}

\date{\today}

\author{Steven M. Flores}
\email{smflores@umich.edu} 
\affiliation{Department of Mathematics, University of Michigan, Ann Arbor, Michigan, 48109-2136, USA}

\author{Peter Kleban}
\email{kleban@maine.edu} 
\affiliation{LASST and Department of Physics \& Astronomy, University of Maine, Orono, Maine, 04469-5708, USA}

\author{Robert M. Ziff}
\email{rziff@umich.edu} 
\affiliation{Michigan Center for Theoretical Physics and Department of Chemical Engineering, University of Michigan, Ann Arbor, Michigan, 48109-2136, USA}

\begin{abstract}  
In a statistical cluster or loop model such as percolation, or more generally the Potts models or O$(n)$ models, a pinch point is a single bulk point where several distinct clusters or loops touch.  In a polygon $\mathcal{P}$ harboring such a model in its interior and with $2N$ sides exhibiting free/fixed side-alternating boundary conditions, boundary clusters anchor to the fixed sides of $\mathcal{P}$.  At the critical point and in the continuum limit, the density (i.e., frequency of occurrence) of pinch-points between $s$ distinct boundary clusters at a bulk point $w\in\mathcal{P}$ is proportional to
\[\langle\psi_1^c(w_1)\psi_1^c(w_2)\ldots\psi_1^c(w_{2N-1})\psi_1^c(w_{2N})\Psi_s(w,\bar{w})\rangle_\mathcal{P}.\]
The $w_i$ are the vertices of $\mathcal{P}$, $\psi_1^c$ is a conformal field theory (CFT) corner one-leg operator, and $\Psi_s$ is a CFT bulk $2s$-leg operator.  In this article, we use the Coulomb gas formalism to construct explicit contour integral formulas for these correlation functions and thereby calculate the density of various pinch-point configurations at arbitrary points in the rectangle, in the hexagon, and for the case $s=N$, in the $2N$-sided polygon at the system's critical point.  Explicit formulas for these results are given in terms of algebraic functions or integrals of algebraic functions, particularly Lauricella functions.  In critical percolation, the result for $s=N=2$ gives the density of red bonds between boundary clusters (in the continuum limit) inside a rectangle.  We compare our results with high-precision simulations of critical percolation and Ising FK clusters in a rectangle of aspect ratio two and in a regular hexagon and find very good agreement.
\end{abstract}

\maketitle

\section{Introduction}\label{intro}

We consider critical bond percolation on a very fine square lattice inside a rectangle $\mathcal{R}$ with wired left and right sides.  Of intrinsic interest to the system are bonds whose activation or deactivation will respectively join or disconnect the percolation boundary cluster anchored to the left side of $\mathcal{R}$ from that anchored to the right side.  Such a bond that connects them is an example of a \emph{red bond} \cite{pikestan}.  Red bonds inherit their name from the following scenario.  If we suppose that only activated bonds conduct electricity and that the wired left and right sides of $\mathcal{R}$ are attached to opposite leads of a battery, then an activated red bond carries the total current and is hottest, and its deactivation stops the flow of current.  Red bonds carry similar significance in other physical scenarios modeled by percolation.  Many of their properties have been studied before, first in context with cluster ramification \cite{stan}.  The average number of red bonds weighted by cluster size was measured in \cite{pikestan}, and the fractal dimension of the set of red bonds is predicted in \cite{saldup,con}, and measured in \cite{apr}.   Further fragmentation properties of percolation clusters are considered in \cite{edgy}.  In this article, we calculate the density (i.e., frequency of occurrence) of red bonds at a given bulk (i.e., interior) point $w\in\mathcal{R}$ and some generalizations which we now explore.

\begin{figure}[t]
\centering
\includegraphics[scale=1.2]{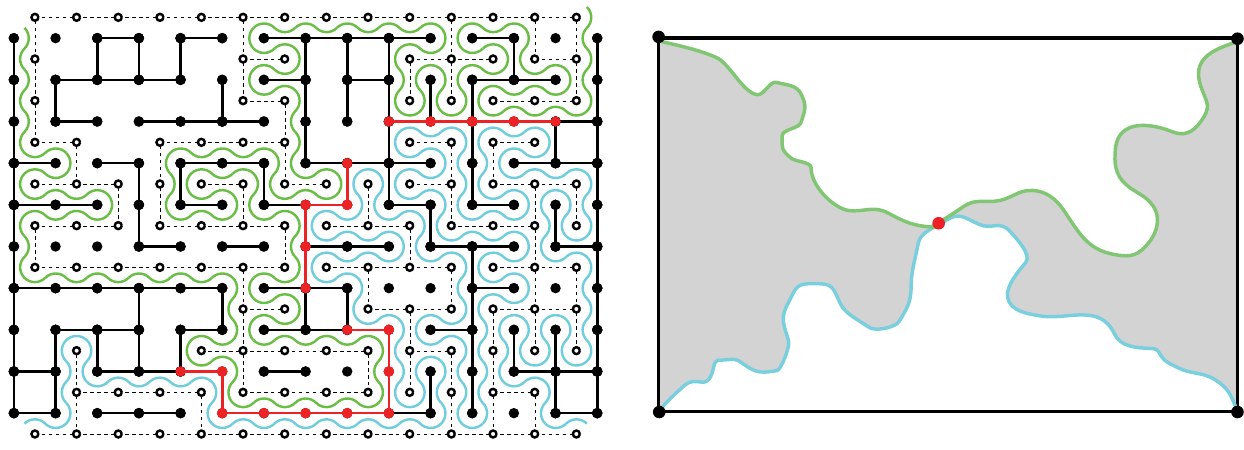}
\caption{A percolation configuration with the red bonds, or two-pinch points (colored red) and the boundary arcs connecting the vertices of the rectangle pairwise (colored green and blue).  The left illustration is a sample in the discrete setting while the right figure shows only the (filled) boundary clusters of a sample in the continuum limit.}
\label{RedBonds}
\end{figure}

In percolation, red bonds are marked by pinch points, or bulk points where distinct percolation clusters touch.  We consider the two boundary arcs (i.e., perimeters of the boundary clusters) of the percolating system in $\mathcal{R}$.  At the center $w$ of a red bond, the two boundary arcs pass very close to each other, separated there by only the red bond (figure \ref{RedBonds}).  In the continuum limit, four distinct boundary arcs appear to emanate from $w$, each ending at a different vertex of $\mathcal{R}$.  In reality, these four curves are not distinct but join pairwise at (or very close to) $w$ to form two boundary arcs.  Each boundary cluster is pinched into a narrow channel between an adjacent pair of boundary arcs, and they touch each other at (or pass very close to) $w$ where the tips of these channels meet (or almost meet).  Thus, we call $w$ a \emph{pinch point} \cite{dup1} (figure \ref{RedBonds}).  The detail of whether or not the red bond at $w$ is activated is lost in the continuum limit where, formally speaking, bonds do not exist but their clusters do.  But the location of the red bond remains.  It is marked by the pinch point at $w$.  Thus, the continuum limit of the density of red bonds in $\mathcal{R}$ equals the density of pinch points between the left and right boundary clusters in $\mathcal{R}$.

The connection between red bonds and pinch points generalizes the problem of computing the red bond density in $\mathcal{R}$ to computing the pinch-point density involving $s$ boundary clusters in a $2N$-sided polygon $\mathcal{P}$.  In particular, we suppose that $\mathcal{P}$ harbors critical percolation in its interior and exhibits a free/fixed side-alternating boundary condition (ffbc).  That is, the boundary condition (bc) of the sides of $\mathcal{P}$ alternate from fixed, or ``wired," (i.e., all bonds activated) to free (i.e., no conditioning imposed on the activation of the bonds).  In this article, we label an ffbc event with the symbol $\varsigma$.  In such an event, the ffbc conditions a boundary cluster to anchor to each wired side.  Now, we define an \emph{$s$-pinch point} to be a bulk point $w\in\mathcal{P}$ where $s$ distinct boundary clusters touch (or pass very close).  In the continuum limit, $2s$ boundary arcs emanate from $w$, each ending at a different vertex of $\mathcal{P}$ (figure \ref{RedBonds}).  Clearly, we must have $1\leq s\leq N$ since at most $N$ distinct boundary clusters can anchor to the fixed sides of $\mathcal{P}$.  When $s=1$ we define a one-pinch point to be a bulk point touched (or approached) by just one of the boundary arcs.  As the continuum limit is approached, the density of pinch-point events decays as a power law of the shrinking lattice spacing (section \ref{boundaryconditions}).  This power, with other scaling exponents, is determined in \cite{saldup,dup1}.  (For $2N$-sided polygons with $N>2$, the density of red bonds is still dominated by pinch points involving two clusters in the large system limit since $s$-pinch points with $s>2$ occur much less often, as discussed in section \ref{simresults}.)

\begin{figure}[t]
\centering
\includegraphics[scale=.6]{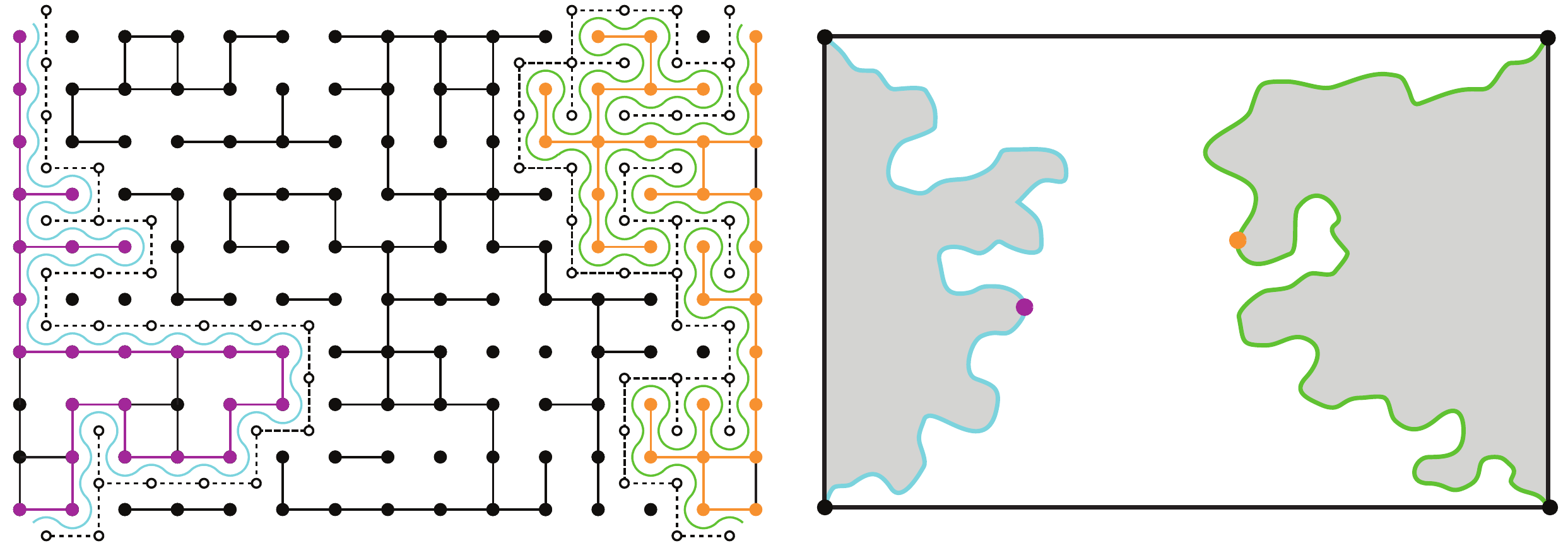}
\caption{An illustration of one-pinch-point events on the perimeters of the boundary clusters (orange and purple) in the discrete (left) and continuum (right) settings.}
\label{1ppfig}
\end{figure}

We obtain another generalization by considering the statistics of the boundary arcs, which fluctuate in $\mathcal{P}$ with the law of multiple-SLE$_\kappa$ \cite{bbk}.  The case of percolation entails $\kappa=6$, but we may consider other $\kappa\in(0,8)$ as well.  In these terms, an $s$-pinch point is a bulk point $w\in\mathcal{P}$ where $s$ distinct multiple-SLE$_\kappa$ curves touch (or pass very near each other).  In particular, a one-pinch point is a bulk point on (or very near) one of these curves (figure \ref{1ppfig}), and the problem of calculating its density generalizes the same problem for when there is one SLE$_\kappa$ curve.  The latter was originally solved in \cite{beffara}.  In our situation with multiple boundary arcs, the regions that a boundary arc can explore in $\mathcal{P}$ are influenced by the presence of the other boundary arcs, so a one-pinch point can be interpreted as measuring the repulsion between the various boundary arcs.  In the case of percolation, this ``repulsion" is not felt until the boundary arcs actually collide due to the locality property of SLE$_\kappa$ with $\kappa=6$ \cite{lawschwer, gruz}.

The range $\kappa\in(0,8)$ describes boundary arcs in many interesting critical lattice models, including those of the $Q$-state Potts model for $Q\leq4$.  As in percolation, an $s$-pinch point is still a bulk point where $s$ distinct boundary clusters touch in the $Q$-state Potts model inside $\mathcal{P}$ with an ffbc.  However, now there are two different types of clusters to consider: FK clusters and spin clusters.  Boundary arcs of the former type are multiple-SLE$_\kappa$ curves with speed $\kappa$ in the dense phase (i.e., $\kappa\in(4,8)$) and related to $Q$ through \cite{sd}
\be\label{kappaQ}Q=4\cos^2(4\pi/\kappa),\quad\kappa\in(4,8),\ee
and boundary arcs of the latter type are multiple-SLE$_\kappa$ curves with ``dual" speed $\hat{\kappa}=16/\kappa$ in the dilute phase (i.e., $\hat{\kappa}\in(0,4]$) and $\kappa$ related to $Q$ through (\ref{kappaQ}).  Scaling exponents and fractal dimensions associated with pinch points are found in \cite{con}.  The generalization of red bonds from percolation to other models is also considered in \cite{apr}.

In this article, we calculate various continuum limit pinch-point densities in the rectangle $\mathcal{R}$ and in the hexagon $\mathcal{H}$ (and for $s=N$ in any $2N$-sided polygon) conditioned on a specified ffbc event $\varsigma$ and for arbitrary $\kappa\in(0,8)$, but before we begin, we refine our definition of a pinch point.  We suppose that a multiple-SLE$_\kappa$ process evolves $2N$ distinct boundary arcs anchored to the vertices of $\mathcal{P}$ until they join pairwise in the long-time limit to form $N$ distinct boundary arcs in one of $C_N$ possible connectivities.  Here, $C_N$ is the $N$-th Catalan number, given by
\be\label{catalan}C_N=\frac{(2N)!}{N!(N+1)!}.\ee
We let $\Lambda$ label a pinch-point event, that is, an event containing all boundary arc configuration samples in which $s$ distinct boundary arcs, each with both endpoints among $2s$ specified vertices of $\mathcal{P}$, pass within a small distance $\delta$ from a specified bulk point $w\in\mathcal{P}$ and the other boundary arcs join the remaining vertices of $\mathcal{P}$ in some particular connectivity.  Then for a specified ffbc event $\varsigma$, the \emph{type-$\Lambda$ $s$-pinch-point density} $\rho_{(\Lambda|\varsigma)}^\mathcal{P}(w)$ is the probability of the pinch-point event $\Lambda$ conditioned on the ffbc event $\varsigma$, and it equals the ratio of the (continuum limit) partition function $Z_{(\Lambda|\varsigma)}^\mathcal{P}$ summing exclusively over samples in $\Lambda\cap\varsigma$ divided by the (continuum limit) partition function $Z_\varsigma^\mathcal{P}$ summing exclusively over samples in $\varsigma$.  

The purpose of this article is to study the asymptotic behavior of the type-$\Lambda$ pinch-point density as $\delta\rightarrow0$.  The asymptotic behavior of the partition functions $Z_{(\Lambda|\varsigma)}^\mathcal{P}$ and $Z_\varsigma^\mathcal{P}$ are supposed to be
\begin{align}\label{ZnulamH}Z_{(\Lambda|\varsigma)}^\mathcal{P}/Z_f&\underset{\delta,\delta_i\rightarrow0}{\sim}c_1^{2N}C_s^2\delta_1^{\theta_1}\ldots\delta_{2N}^{\theta_1}\delta^{2\Theta_s}\Upsilon_{(\Lambda|\varsigma)}^\mathcal{P},& Z_\varsigma^\mathcal{P}/Z_f&\underset{\delta_i\rightarrow0}{\sim}c_1^{2N}\delta_1^{\theta_1}\ldots\delta_{2N}^{\theta_1}\Upsilon_{\varsigma}^\mathcal{P},\end{align}
where $Z_f$ is the free partition function (summing over all samples in the system configuration space), where the functions $\Upsilon_{(\Lambda|\varsigma)}^\mathcal{P}$ and $\Upsilon_\varsigma^\mathcal{P}$ are \emph{universal partition functions}, where $\theta_1$ is the \emph{boundary one-leg weight} associated with the free/fixed boundary condition change (bcc) at each vertex of $\mathcal{P}$, and where $\Theta_s$ is the \emph{bulk $2s$-leg weight} associated with the bulk $s$-pinch-point event.  Also, $c_1$ is a non-universal scaling coefficient associated with each bcc, $C_s$ is a non-universal  scaling coefficient associated with the $s$-pinch-point event (and is \emph{not} the $s$-th Catalan number), and the $i$-th bcc occurs within distance $\delta_i$ from the $i$-th vertex $w_i$ of $\mathcal{P}$.  Then the density $\rho_{(\Lambda|\varsigma)}^\mathcal{P}$ behaves as
\be\label{asyratio}\rho_{(\Lambda|\varsigma)}^\mathcal{P}=Z^{\mathcal{P}}_{(\Lambda|\varsigma)}/Z_\varsigma^{\mathcal{P}}\underset{\delta\rightarrow0}{\sim}C_s^2\delta^{2\Theta_s}\Upsilon_{(\Lambda|\varsigma)}^\mathcal{P}/\Upsilon_\varsigma^\mathcal{P}.\ee
Thus, determining the behavior of $\rho_{(\Lambda|\varsigma)}^\mathcal{P}$ to within a constant amounts to determining the universal partition functions $\Upsilon_{(\Lambda|\varsigma)}^\mathcal{P}$ and $\Upsilon_\varsigma^\mathcal{P}$.

The organization of this article is as follows.  In section \ref{cftdescription}, we identify the universal partition function $\Upsilon_{(\Lambda|\varsigma)}^\mathcal{P}$ with a bulk-boundary CFT correlation function of certain primary operators, and we find an explicit formula for it using the Coulomb gas formalism.  Also in this section, we calculate the $N$-pinch-point weight (defined below) of a $2N$-sided polygon, and we find that it is completely algebraic.  In section \ref{ppdensities}, we compute various $s$-pinch-point densities in the rectangle $(N=2)$ and in the hexagon $(N=3)$.  We find that the formulas for the $(N-1)$-pinch-point densities are given by algebraic factors times Lauricella functions of cross-ratios of (the half-plane conformal images of) the bulk point $w$ and the vertices $w_i$ of $\mathcal{P}$.  In section \ref{simresults}, we compare some of our predictions with high-precision simulations of percolation and Ising FK clusters inside a rectangle and a regular hexagon and find very good agreement.

\section{Conformal field theory description}\label{cftdescription}

\begin{figure}[b]
\centering
\includegraphics[scale=0.7]{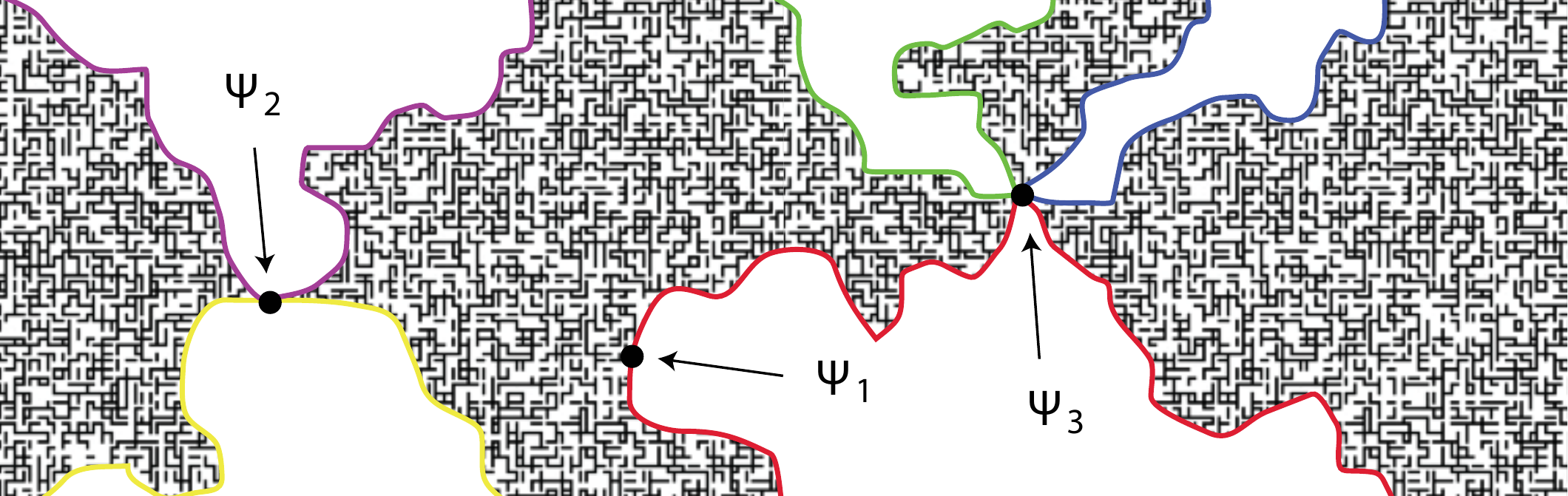}
\caption{An $s$-pinch-point event is induced by the insertion of a bulk $2s$-leg operator.}
\label{PinchPointsFig}
\end{figure}

In the continuum limit, $\Upsilon_{(\Lambda|\varsigma)}^\mathcal{P}$ equals a correlation function of appropriate primary operators belonging to a conformal field theory (CFT) \cite{bpz} of central charge \cite{bauber}
\be\label{centralcharge}c=c(\kappa):=(6-\kappa)(3\kappa-8)/2\kappa.\ee
These primary operators are chosen as follows.  To alternate the bc on the sides of $\mathcal{P}$ from free to fixed to free, etc., we insert a corner one-leg operator $\psi_1^c(w_i)$ at each vertex $w_i$ of $\mathcal{P}$ into the correlation function.  Corner operators are defined in \cite{c1, skfz}, and they are used in section \ref{transform}.  The collection of $2N$ corner one-leg operators introduces $N$ non-crossing boundary arcs that connect the vertices $w_1,\ldots,w_{2N}$ pairwise in one of $C_N$ possible connectivities.  Now, to generate an $s$-pinch point at $w\in\mathcal{P}$, we require $s$ of these arcs to touch at (or come very close to) this point.  One may view this as the event in which $2s$ distinct boundary arcs emanate from $w$, which is conditioned by the insertion of a spinless bulk $2s$-leg operator $\Psi_s(w,\bar{w})$ into the correlation function \cite{rbgw, dup1} (figure \ref{PinchPointsFig}).  Hence, $\Upsilon_{(\Lambda|\varsigma)}^\mathcal{P}$ is given by the $(2N+2)$-point function
\be\label{2N+2point}\Upsilon_{(\Lambda|\varsigma)}^\mathcal{P}=\langle\psi_1^c(w_1)\psi_1^c(w_2)\ldots\psi_1^c(w_{2N-1})\psi_1^c(w_{2N})\Psi_s(w,\bar{w})\rangle_\mathcal{P}.\ee
The standard approach to studying this correlation function is to conformally map the interior of $\mathcal{P}$ onto the upper half-plane $\mathbb{H}$. The half-plane version of this correlation function is
\bea\label{halfplane}\Upsilon_{(\Lambda|\varsigma)}&=&\langle\psi_1(x_1)\psi_1(x_2)\ldots\psi_1(x_{2N-1})\psi_1(x_{2N})\Psi_s(z,\bar{z})\rangle_\mathbb{H}\\
\label{images}&=&\langle\psi_1(x_1)\psi_1(x_2)\ldots\psi_1(x_{2N-1})\psi_1(x_{2N})\Psi_s(z)\Psi_s(\bar{z})\rangle_\mathcal{\mathbb{C}},\eea
where we have used Cardy's method of images \cite{c1} to rewrite the half-plane correlation function on the right side of (\ref{halfplane}) as the whole-plane correlation function (\ref{images}).  Here, $\Psi_s(z)$ has holomorphic weight $\Theta_s$ and antiholomorphic weight zero.
First, we will focus on calculating $\Upsilon_{(\Lambda|\varsigma)}$, and later in section \ref{transform}, we will transform $\Upsilon_{(\Lambda|\varsigma)}$ to $\Upsilon_{(\Lambda|\varsigma)}^\mathcal{P}$.

In the multiple-SLE$_\kappa$ picture, the bulk $2s$-leg operator conditions a specified collection of $2s$ of the $2N$ available multiple-SLE$_\kappa$ curves to grow from their respective origin points at the vertices of $\mathcal{P}$ towards the common bulk point $w\in\mathcal{P}$ until they join pairwise very near $w$ in any one of $C_s$ possible connectivities.  Here, $C_s$ is the $s$-th Catalan number (\ref{catalan}).  The $s$-pinch-point event $\Lambda$, defined above, contains all samples that exhibit any one of these $C_s$ connectivities near the pinch-point.  Each connectivity is equally likely to occur.

\begin{figure}[t]
\centering
\includegraphics[scale=0.3]{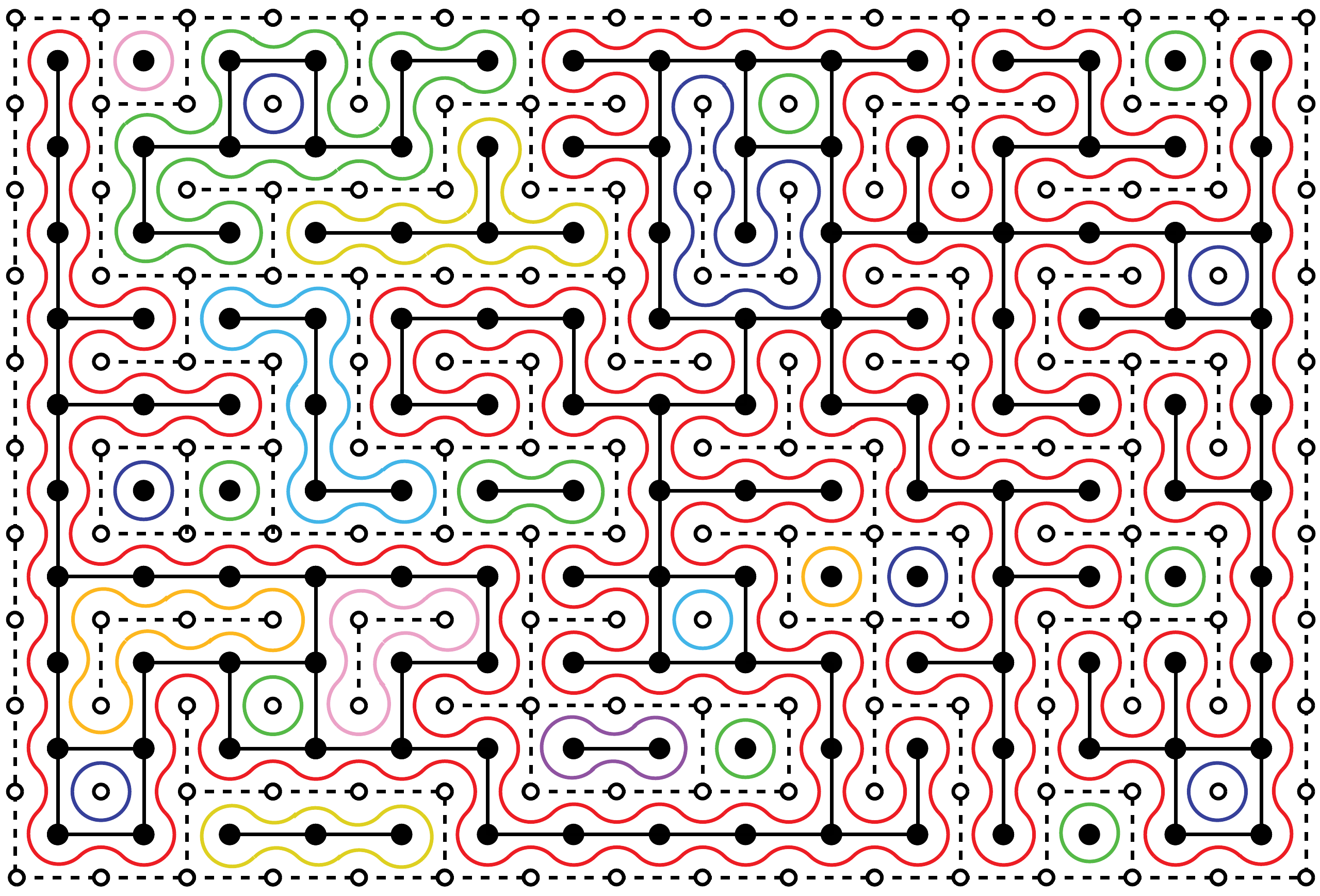}
\caption{The unique loop configuration corresponding to a particular bond configuration in $\mathcal{R}$ with the left/right sides wired.  Boundary loops (surrounding boundary clusters) are red while bulk loops (surrounding bulk clusters) are any color but red.}
\label{LoopRepColor}
\end{figure}

Moreover, we may view each pinch-point sample in $\Lambda$ as a collection of loops that surround the perimeters of the bulk and boundary clusters in $\mathcal{P}$.  These clusters are, for example, FK clusters or spin clusters in a Potts model, and the loops for the former case are shown in figure \ref{LoopRepColor}.  This picture is consistent with the continuum limit of the O$(n)$ model if we set the loop fugacity $n$ equal to \cite{gruz, smir}
\be\label{fugacity}n=n(\kappa):=-2\cos(4\pi/\kappa).\ee
In this interpretation, the endpoints of the boundary arcs are joined pairwise via $N$ exterior arcs to form between one and $N$ boundary loops (red loops in figure \ref{LoopRepColor}) that dodge in an out of $\mathcal{P}$ \cite{skfz, fkz}.  These exterior arcs live outside $\mathcal{P}$ and connect its vertices pairwise, and their connectivity is determined by the choice of ffbc.  All samples in the same ffbc event $\varsigma$ whose boundary arcs join in the $i$-th connectivity inside $\mathcal{P}$ have the same number $p_i$ of boundary loops.  Because the $s$-pinch-point event $\Lambda$ sums over all $C_s$ possible connectivities of the boundary arcs that are conditioned to approach the $s$-pinch-point, we may factor out the fugacity factors associated with the boundary loops to write $\Upsilon_{(\Lambda|\varsigma)}$ in the form
\be\label{factoroutn}\Upsilon_{(\Lambda|\varsigma)}=(n^{p_1}+\ldots+n^{p_{C_s}})\Pi_\Lambda.\ee
The factor $\Pi_\Lambda$ is called the \emph{(half-plane) type-$\Lambda$ pinch-point weight}, and it bears the same partition function interpretation as $\Upsilon_{(\Lambda|\varsigma)}$, but with the boundary loops having fugacity one.  We will elaborate on the relation between an ffbc event and the number of boundary loops in each of its samples further in section \ref{boundaryconditions}.

The bulk $2s$-leg and boundary $s$-leg operators $\Psi_s$ and $\psi_s$ respectively are primary operators of a boundary CFT in the upper half-plane.  The highest-weight vector of their Verma modules respectively belongs to the $(0,s)$ and $(1,s+1)$ (resp.\,$(s,0)$ and $(s+1,1)$) positions of the Kac table in the dense phase (resp.\,dilute phase) of SLE$_\kappa$ \cite{rbgw}.  The Kac weights associated with these positions are given by \cite{bpz}
\be\label{Kacweight}h_{r,s}(\kappa)=\frac{1-c(\kappa)}{96}\left[\left(r+s+(r-s)\sqrt{\frac{25-c(\kappa)}{1-c(\kappa)}}\,\right)^2-4\right]=\frac{1}{16\kappa}\begin{cases}(\kappa r-4s)^2-(\kappa-4)^2&\kappa>4\\(\kappa s-4r)^2-(\kappa-4)^2&\kappa\leq4\end{cases}.\ee
Thus, the conformal weights $\Theta_s$ and $\theta_s$ of the boundary one-leg operator and the bulk $2s$-leg operator in either phase are respectively
\be\label{weights}\Theta_s=\frac{16s^2-(\kappa-4)^2}{16\kappa},\quad\theta_s=\frac{s(2s+4-\kappa)}{2\kappa}.\ee
CFT translates the reducibility of the Verma module associated with each boundary one-leg operator into the following semi-elliptic system of $2N$ PDEs that govern the correlation function (\ref{halfplane}), or equivalently, the pinch-point weight $\Pi_\Lambda$:
\be\label{nullstatebulkbdy}\left[\frac{\kappa}{4}\partial_i^2+\sum_{j\neq i}^{2N}\left(\frac{\partial_j}{x_j-x_i}-\frac{\theta_1}{(x_j-x_i)^2}\right)+\frac{\partial_z}{z-x_i}-\frac{\Theta_s}{(z-x_i)^2}+\frac{\partial_{\bar{z}}}{\bar{z}-x_i}-\frac{\Theta_s}{(\bar{z}-x_i)^2}\right]\Pi_\Lambda=0,\quad i\in\{1,\ldots,2N\}.\ee
The domain of $\Pi_\Lambda$ is such that $x_i<x_j$ when $i<j$ and $z$ and $\bar{z}$ are in the upper and lower half-planes $\mathbb{H}$ and $\mathbb{H}^*$ respectively.  We treat $z$ and $\bar{z}$ as independent holomorphic and antiholomorphic variables until the very end of our calculations, where we set $\bar{z}=z^*$.  (Throughout this article, ``$z^*$" denotes the complex conjugate of $z$.)  In addition, the three conformal Ward identities ensure that $\Pi_\Lambda$ is conformally covariant such that each boundary point $x_i$ has scaling weight $\theta_1$ and the bulk point $z$ and its image point $\bar{z}$ have holomorphic weight $\Theta_s$: 
\bea\label{transl}&&\left[\partial_z+\partial_{\bar{z}}+\sum_{i=1}^{2N}\partial_{x_i}\right]\Pi_\Lambda=0,\\
\label{rotdial}&&\left[z\partial_z+\bar{z}\partial_{\bar{z}}+2\Theta_s+\sum_{i=1}^{2N}(x_i\partial_{x_i}+\theta_1)\right]\Pi_\Lambda=0,\\
\label{sct}&&\left[z^2\partial_z+\bar{z}^2\partial_{\bar{z}}+2\Theta_s(z+\bar{z})+\sum_{i=1}^{2N}(x_i^2\partial_{x_i}+2\theta_1x_i)\right]\Pi_\Lambda=0.\eea
The Ward identities restrict $\Pi_\Lambda$ to a conformally covariant ansatz which may be chosen to be
\be\label{covform}\Pi_\Lambda(x_1,\ldots,x_{2N};z,\bar{z})=|z-\bar{z}|^{-2\Theta_s}\prod_{i=1}^N(x_{2i}-x_{2i-1})^{-2\theta_1}G(\eta_2,\ldots,\eta_{2N-2};\mu,\nu),\ee
where $\{\eta_2,\ldots,\eta_{2N-2},\mu,\nu\}$ is a maximal set of independent cross-ratios that can be formed from the points $x_1,\ldots,x_{2N},z,$ and $\bar{z}$, and where $G$ is an unspecified function differentiable in each independent variable.  We choose
\be\label{crossratios}\eta_i:=\frac{(x_i-x_1)(x_{2N}-x_{2N-1})}{(x_{2N-1}-x_1)(x_{2N}-x_i)},\quad\mu:=\frac{(z-x_1)(x_{2N}-x_{2N-1})}{(x_{2N-1}-x_1)(x_{2N}-z)},\quad\nu:=\frac{(\bar{z}-x_1)(x_{2N}-x_{2N-1})}{(x_{2N-1}-x_1)(x_{2N}-\bar{z})}.\ee
This ansatz reduces the number of variables in the problem from $2N+2$ to $2N-1$.  A standard approach that takes advantage of this reduction is to transform (\ref{nullstatebulkbdy}) into a system of PDEs governing $x_{2N}^{2\theta_1}\Pi_\Lambda$ and take the limit
\be\label{limit}\{x_1,x_2,\ldots x_{2N-2},x_{2N-1},x_{2N},z,\bar{z}\}\rightarrow\{0,\eta_2,\ldots,\eta_{2N-2},1,\infty,\mu,\nu\}.\ee
This gives a system of $2N$ PDEs governing the unknown function $G$ from which we can glean information, ideally exact solutions.  Because we mainly consider the cases $N=1,2$ and 3 in this article, we use the following notation throughout:
\be\label{cross}\eta:=\eta_2,\quad\tau:=\eta_3,\quad\sigma:=\eta_4.\ee

\begin{figure}[b]
\centering
\includegraphics[scale=0.27]{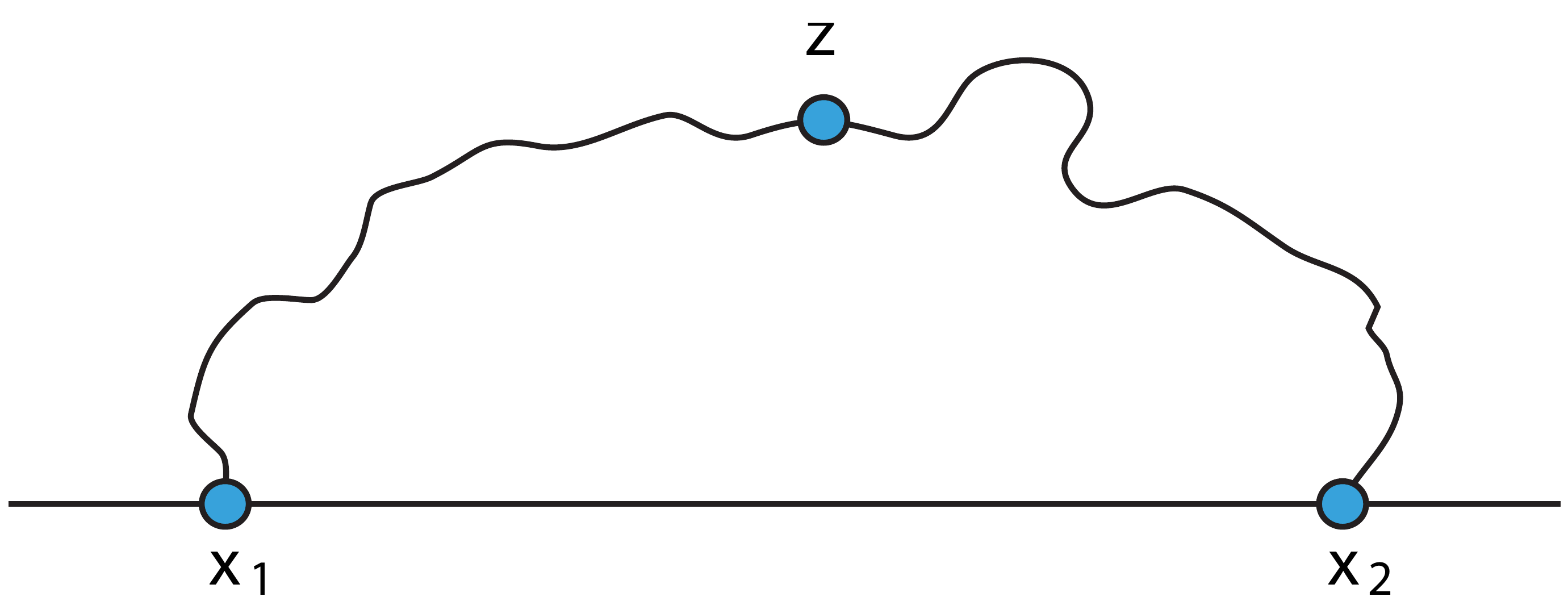}
\caption{The one-pinch-point configuration in the upper half-plane.  We note that the limits $x_2\rightarrow x_1$ and $z\rightarrow x\in\mathbb{R}\setminus\{x_1,x_2\}$ each generate a boundary two-leg operator to leading order.}
\label{2ptpic}
\end{figure}

We can explicitly solve (\ref{nullstatebulkbdy}-\ref{sct}) in the case $s=N=1$ (i.e., the two-gon).  In this one-pinch-point event, a boundary arc $\gamma$ connecting $x_1$ with $x_2$ passes some very small distance $\epsilon$ from the specified bulk point $z\in\mathbb{H}$.  We denote the corresponding pinch-point weight by $\Pi_{12}$.  Substituting the ansatz 
\be \Pi_{12}(x_1,x_2;z,\bar{z})=|z-\bar{z}|^{-2\Theta_1}(x_2-x_1)^{-2\theta_1}G(\upsilon),\hspace{.5cm}\upsilon:=\frac{(x_1-z)(x_2-\bar{z})}{(x_1-\bar{z})(x_2-z)}\ee
(slightly modified from (\ref{covform})) into (\ref{nullstatebulkbdy}) yields a second-order, linear, homogeneous differential equation in $G$.  The general solution is
\be F(x_1,x_2;z,\bar{z})=\frac{(x_2-x_1)^{-2\theta_1+\theta_2}|z-\bar{z}|^{-2\Theta_1+\theta_2}}{|x_1-z|^{8/\kappa-1}|x_2-z|^{8/\kappa-1}}\left[c_1+c_2\,\beta\left(\frac{4}{\kappa},1-\frac{8}{\kappa}\,\,\bigg|\,\,\frac{(x_1-z)(x_2-\bar{z})}{(x_1-\bar{z})(x_2-z)}\right)\right],\ee
where $\beta$ is the incomplete beta function, $c_1$ and $c_2$ are arbitrary real constants, and the weights $\theta_1,\theta_2,$ and $\Theta_1$ are given in (\ref{weights}).  We argue that $c_2=0$ in our application by sending $z\rightarrow x\in\mathbb{R}\setminus\{x_1,x_2\}$.  Because the boundary arc $\gamma$ is conditioned to touch $z$, $\gamma$ will touch the real axis at $x$ in this limit, and two boundary arcs will emanate from $x$.  Thus, the bulk operator $\Psi_1(z)$ must fuse with its image $\Psi_1(\bar{z})$ to create a boundary two-leg operator $\psi_2(x)$ to leading order.  Or instead we send $x_2\rightarrow x_1$.  Then in this limit, the two endpoints of $\gamma$ touch at $x_1$, and the boundary operators $\psi_1(x_1)$ and $\psi_1(x_2)$ fuse to create $\psi_2(x_1)$ to leading order as well (figure \ref{2ptpic}).  In both cases, $\upsilon\rightarrow1$.  Because
\be\beta(a,b\,|\,\upsilon)\underset{\upsilon\rightarrow1}{\sim}-b^{-1}(1-\upsilon)^b\hspace{.5cm}\text{if $b<0$,}\ee
and because $b=-\theta_2<0$ for $\kappa<8$ (\ref{weights}), we see that 
\be F(x_1,x_2;z,\bar{z})\underset{\upsilon\rightarrow1}{\sim}\frac{(x_2-x_1)^{-2\theta_1+\theta_2}|z-\bar{z}|^{-2\Theta_1+\theta_2}}{|x_1-z|^{8/\kappa-1}|x_2-z|^{8/\kappa-1}}\left[c_1+\frac{c_2}{\theta_2}\left(\frac{(x_2-x_1)(z-\bar{z})}{(x_1-\bar{z})(x_2-z)}\right)^{-\theta_2}\right].\ee
To ensure that the bulk-image or boundary-boundary fusion has the two-leg channel at leading order, the second term in the brackets must be absent.  Thus, $c_2=0$, and we find the one-pinch-point weight for an SLE$_\kappa$ connecting the real points $x_1$ and $x_2$:
\be\label{2soln}\Pi_{12}(x_1,x_2;z,\bar{z})=\frac{(x_2-x_1)^{2/\kappa}|z-\bar{z}|^{(8-\kappa)^2/8\kappa}}{|x_1-z|^{8/\kappa-1}|x_2-z|^{8/\kappa-1}}.\ee
If we put $x_1=0$ and $x_2=\infty$ as in the usual setup for SLE$_\kappa$, then we have
\bea\label{1ppPi}\lim_{x_2\rightarrow\infty}x_2^{2\theta_1}\Pi_{12}(0,x_2;z,\bar{z})&=&|z-\bar{z}|^{\kappa/8-1}\sin\arg(z)^{8/\kappa-1}\\
\label{ProbClose}&\asymp&\epsilon^{-2\Theta_1}\mathbb{P}\{\mathcal{B}(\epsilon,z)\cap\gamma\neq\emptyset\},\quad\kappa\in(0,8),\eea
where $\mathbb{P}\{\mathcal{B}(\epsilon,z)\cap\gamma\neq\emptyset\}$ is the probability that $\gamma$ intersects a ball $\mathcal{B}(\epsilon,z)$ centered at $z\in\mathbb{H}$ and of small radius $\epsilon$.  Equation (\ref{ProbClose}) is rigorously proven in \cite{beffara}.  This rigorous result is supposed by physicists to be stronger.  Namely, it is expected to be
\be\mathbb{P}\{\mathcal{B}(\epsilon,z)\cap\gamma\neq\emptyset\}\underset{\epsilon\rightarrow0}{\sim}C\epsilon^{2\Theta_1}|z-\bar{z}|^{\kappa/8-1}\sin\arg(z)^{8/\kappa-1},\quad\kappa\in(0,8)\ee
for some constant $C$.  This is equivalent to the prediction (\ref{asyratio}) when $N=1$.  We note that $\Theta_1>0$ for $\kappa<8$, so this probability goes to zero as $\epsilon\rightarrow0$ as it must.  Below, we will compute this pinch-point weight with another method.

As we observed in this example, it appears to be generally true that the set of pinch-point densities span a proper subspace of the solution space of the system (\ref{nullstatebulkbdy}-\ref{sct}).  This follows from the result (\ref{M}) in the appendix \ref{appendix}.

The system (\ref{nullstatebulkbdy}-\ref{sct}) is very difficult to solve directly when $N>1$, but fortunately the Coulomb gas formalism \cite{df} provides a tractable approach to constructing explicit solutions.  To this end, we write a chiral operator representation for (\ref{images}).  That is, we represent a primary field of holomorphic weight $h$ and antiholomorphic weight zero by a \emph{chiral operator} $V_{\alpha}(z)$ of charge $\alpha$.  A primary field, this chiral operator is defined to be the normal ordering of a $\exp[i\sqrt{2}\alpha\varphi(z)]$ with $\varphi(z)$ the holomorphic part of a massless free boson, and its holomorphic weight is $h=\alpha(\alpha-2\alpha_0)$.  Here, $2\alpha_0$ is the background charge, and it equals $\alpha_++\alpha_-$ with the \emph{screening charges} $\alpha_\pm$ given in (\ref{kaccharge}) below.  Only chiral operators of charge $\alpha$ or $2\alpha_0-\alpha$ have equal holomorphic weights, so we call these two charges \emph{conjugates}.  The two conjugate charges $\alpha_{r,s}^\pm$ associated with the Kac weight $h_{r,s}$ are
\be\label{kaccharge}\alpha_{r,s}^\pm=\frac{1\pm r}{2}\alpha_++\frac{1\pm s}{2}\alpha_-,\hspace{.5cm}\alpha_\pm=\pm\begin{cases}(\sqrt{\kappa}/2)^\pm&\kappa>4\\ (\sqrt{\kappa}/2)^\mp&\kappa\leq4\end{cases},\ee
and we let $V_{r,s}^\pm(z)$ be a chiral operator of charge $\alpha_{r,s}^\pm$.  Adopting dense phase (i.e., $\kappa>4$) notation conventions, we represent $\psi_1(x_i)$ by the chiral operator $V_{1,2}^-(x_i)$, and we represent $\Psi_s(z,\bar{z})$ by the \emph{vertex operator} $V_{0,s}^+(z)\bar{V}_{0,s}^+(\bar{z})$.  The correlation function (\ref{images}) now has total charge
\be\label{notbalanced}2N\alpha_{1,2}^-+2\alpha_{0,s}^+=2\alpha_0+(s-N)\alpha_-.\ee
We wish for this total charge to equal $2\alpha_0$ in order to satisfy the \emph{neutrality condition}.  This is necessary in order for the correlation function to satisfy the conformal Ward identities (\ref{transl}-\ref{sct}).  We see that the neutrality condition is presently satisfied only when $N=s$.

\begin{figure}[t]
\centering
\includegraphics[scale=0.28]{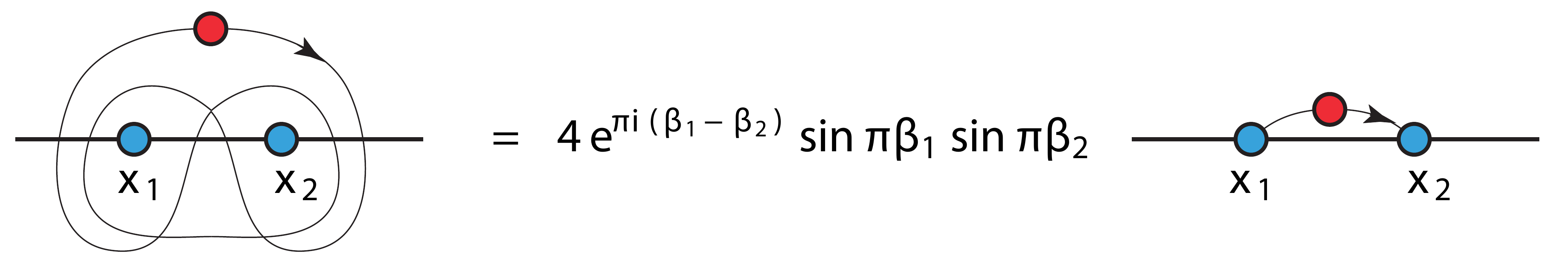}
\caption{The Pochhammer contour entwining the points $x_1$ and $x_2$.  If the numbers $\beta_1$ and $\beta_2$, where $e^{2\pi i\beta_1}$ and $e^{2\pi i\beta_2}$ are the monodromy factors associated with $x_1$ and $x_2$ respectively, are greater than negative one, then a Pochhammer contour may be replaced with the simple contour shown on the right.}
\label{PochhammerContour}
\end{figure}

We momentarily restrict our attention to the case $s=N$, where the correlation function is neutral.  Here, we find an explicit, algebraic formula for the upper half-plane $N$-pinch-point weight in a $2N$-sided polygon:
\bea\label{s=N}\Pi_{\text{$N$-pinch point}}(x_1,\ldots,x_{2N};z,\bar{z})&=&\langle V_{1,2}^-(x_1)\ldots V_{1,2}^-(x_{2N})V_{0,s}^+(z)V_{0,s}^+(\bar{z})V_-(u_1)\ldots V_-(u_{N-s})\rangle\nonumber\\
\label{s=N}&=&|z-\bar{z}|^{(4N+4-\kappa)^2/8\kappa}\prod_{i<j}^{2N}(x_j-x_i)^{2/\kappa}\prod_{i=1}^{2N}|z-x_i|^{1-4(N+1)/\kappa}.\eea
The bulk point $z$ is connected to all boundary points $x_i$ hosting the bccs via the $N$ boundary arcs that touch at $z$.  We note that (\ref{s=N}) is identical to (\ref{2soln}) when $N=1$, as it must be.  This result was also computed in \cite{c2, bencardy} by using other methods.

Next, we consider the cases with $s<N$.  In order for the total charge (\ref{notbalanced}) of the correlation function to equal $2\alpha_0$, we must insert $N-s$ $Q_-$ screening operators, leading to the following modified $(2N+2)$-point function:  
\be\label{screened}\sideset{}{_{\Gamma_1}}\oint\ldots\sideset{}{_{\Gamma_{N-s}}}\oint\langle V_{1,2}^-(x_1)\ldots V_{1,2}^-(x_{2N})V_{0,s}^+(z)V_{0,s}^+(\bar{z})V_-(u_1)\ldots V_-(u_{N-s})\rangle\,du_1\ldots\,du_{N-s}.\ee
After including a useful prefactor discussed below, we find that (\ref{screened}) is given by 
\begin{multline}\label{chiralcorr}\begin{aligned}&\left(\prod_{m=1}^{N-s}\frac{n(\kappa)\Gamma(2-8/\kappa)}{4\exp\pi i(\beta_{1m}-\beta_{2m})\sin\pi\beta_{1m}\sin\pi\beta_{2m}\Gamma(1-4/\kappa)^2}\right)\left(\prod_{i<j}^{2N}(x_j-x_i)^{2/\kappa}\right)\left(\prod_{i=1}^{2N}|z-x_i|^{(\kappa-4s-4)/\kappa}\right)\\
&\hspace{.2cm}\times|z-\bar{z}|^{(\kappa-4s-4)^2/8\kappa}\sideset{}{_{\Gamma_1}}\oint\ldots\sideset{}{_{\Gamma_{N-s}}}\oint du_1\ldots du_{N-s}\left(\prod_{k=1}^{2N}\prod_{l=1}^{N-s}(x_k-u_l)^{-4/\kappa}\right)\left(\prod_{p<q}^{N-s}(u_p-u_q)^{8/\kappa}\right)\end{aligned}\\
\times\left(\prod_{m=1}^{N-s}(z-u_m)^{(4s+4-\kappa)/\kappa}(\bar{z}-u_m)^{(4s+4-\kappa)/\kappa}\right).\end{multline}
Here, $e^{2\pi i\beta_{1m}}$ and $e^{2\pi i\beta_{2m}}$ are the monodromy factors (relative to $u_m$) of the two branch points entwined by the $m$-th contour, $n(\kappa)$ is the O$(n)$ model loop fugacity (\ref{fugacity}), and we choose the branch of the logarithm with $\arg z\in[-\pi,\pi)$ for all complex $z$ so that each branch cut parallels the real axis.   Every pinch-point weight $\Pi_\Lambda$ will be some linear combination of functions of the form (\ref{chiralcorr}), with each term using a different set of contours $\{\Gamma_m\}$.  A proof that (\ref{chiralcorr}) solves the system (\ref{nullstatebulkbdy}-\ref{sct}) is given in section \ref{appendix}.

What remains is to determine a collection of closed, non-intersecting integration contours $\{\Gamma_m\}_{m=1}^{N-s}$ appropriate for a particular type-$\Lambda$ pinch-point event.  The simplest closed contour along which an integration is nonzero is a closed Pochhammer contour entwining only a pair among the branch points $x_1,\ldots,x_{2N},z$, and $\bar{z}$ of the integrand, as shown figure \ref{PochhammerContour}.  Throughout this article, we take each $\Gamma_m$ to be such a contour.  

Now we explain convenience of the prefactor in (\ref{chiralcorr}).  A first reason involves the limit $x_{i+1}\rightarrow x_i$ that sends the $2N$-sided polygon to a $(2N-2)$-sided polygon.  After multiplying the half-plane weight $\Pi_\Lambda$ by $(x_{i+1}-x_i)^{2\theta_1}$ and taking this limit, this product goes to either zero or an $s$-pinch-point weight for the half-plane conformal image of a $(2N-2)$-sided polygon with vertices sent to $x_1,\ldots,x_{i-1},x_{i+2},\ldots,x_{2N}$.  Meanwhile, one can show that (\ref{chiralcorr}) goes to either zero or the same expression except with all factors containing $x_i,x_{i+1},$ and $u_{N-s}$ omitted, the $u_{N-s}$ integration omitted, a factor of $\beta(-4/\kappa,-4/\kappa)^{-1}=\Gamma(2-8/\kappa)/\Gamma(1-4/\kappa)^2$ (with $\beta(a,b)$ the Euler beta function) and $[4\sin^2(4\pi/\kappa)]^{-1}$ omitted, and possibly a fugacity factor $n(\kappa)$ omitted.  So to within a factor of $n(\kappa)$, we retain the same normalization as that for the case of the $2N$-sided polygon with $N\mapsto N-1$.  

A second reason involves the cases $\kappa=4/a$ for some positive integer $a$.  At these special values, all of the powers in the integrand of (\ref{chiralcorr}) are integers.  As a result, the $m$-th Pochhammer contour entwining two branch points, now with their respective monodromy factors $e^{2\pi i\beta_{1m}}$ and $e^{2\pi i\beta_{2m}}$ equaling one, disintegrates into two pairs of oppositely-oriented loops.  One pair surrounds the first point, the other pair surrounds the second, and each integration thus gives zero.  More precisely, one can show that the $(N-s)$-fold integral in (\ref{chiralcorr}) is $O((\kappa-4/a)^{N-s})$ as $\kappa\rightarrow 4/a$ with $a\in\mathbb{Z}^+$.  Meanwhile, the complete prefactor (with each $\beta_{1m}$ and $\beta_{2m}$ equaling $-4/\kappa$ or $4(s+1)/\kappa-1$) is
\be\label{overallfactor}\prod_{m=1}^{N-s}\left(-\frac{\cos(4\pi/\kappa)\Gamma(2-8/\kappa)}{2\sin\pi\beta_{1m}\sin\pi\beta_{2m}\,\Gamma(1-4/\kappa)^2}\right)=O((\kappa-4/a)^{-(N-s)})\quad\text{as $\kappa\rightarrow4/a$, $a\in\mathbb{Z}^+$.}\ee
Therefore, the product of the prefactor (\ref{overallfactor}) with the integral in (\ref{chiralcorr}) is finite and nonzero in the limit.  We note that $n(\kappa)\Gamma(2-8/\kappa)$ is also finite when $\kappa\rightarrow8/a$ with $a$ an odd integer greater than one.  These two cases cover all of the singularities of the gamma functions appearing in (\ref{overallfactor}) for $\kappa\in(0,8)$.

A third reason involves simplifying the contours.  In some cases, a Pochhammer contour entwining two branch points may be replaced by a simple curve that starts and ends at the those points (figure \ref{PochhammerContour}) \cite{mf}.  Each branch point has some monodromy $e^{2\pi i\beta}$ relative to each integration variable: for each $x_i$, $\beta=-4/\kappa$, and for $z$ and $\bar{z}$, $\beta=4(s+1)/\kappa-1$.  The former power is greater than negative one only when $\kappa>4$, so only then can we replace a Pochhammer contour entwining an $x_i$ by a simple curve.  Otherwise, such a replacement yields a divergent integral.  The latter power is greater than negative one for all $\kappa\in(0,8)$, so any Pochhammer contour entwining $z$ with $\bar{z}$ can be replaced by a simple curve.  Replacement of the $m$-th contour by a simple curve cancels the factor of $4\exp\pi i(\beta_{1m}-\beta_{2m})\sin\pi\beta_{1m}\sin\pi\beta_{2m}$ in the denominator of (\ref{chiralcorr}).  Throughout this article, we will explicitly use simple curves in place of Pochhammer contours and omit these factors.  If one of these replacements creates a divergent integral, then we implicitly revert back to using a Pochhammer contour for that integral, and we include the omitted factor.

\section{Calculation of Pinch-point densities}\label{ppdensities}

In this section, we calculate the explicit formula for the type-$\Lambda$ $s$-pinch-point density $\rho_{(\Lambda|\varsigma)}^\mathcal{P}$ conditioned on the ffbc event $\varsigma$ for various $s$-pinch-point events $\Lambda$ for either a rectangle $\mathcal{R}$ $(N=2)$ or for a hexagon $\mathcal{H}$ $(N=3)$.  We proceed in four steps, as enumerated below.
\begin{enumerate}
\item We compute the half-plane pinch-point weight $\Pi_\Lambda$ for the pinch-point event $\Lambda$ of interest.  
\item We construct from $\Pi_\Lambda$ the universal partition function $\Upsilon_{(\Lambda|\varsigma)}$ that sums exclusively over the event $\Lambda\cap\varsigma$.  
\item We transform $\Upsilon_{(\Lambda|\varsigma)}$ into the universal partition function $\Upsilon_{(\Lambda|\varsigma)}^\mathcal{P}$ with the appropriate polygon $\mathcal{P}$ for its domain.  
\item With $\mathcal{P}=\mathcal{R}$ or $\mathcal{H}$, we divide $\Upsilon^\mathcal{P}_{(\Lambda|\varsigma)}$ by the universal partition function $\Upsilon_\varsigma^{\mathcal{P}}$ that sums exclusively over the ffbc event $\varsigma$ to obtain formulas for pinch-point densities in $\mathcal{P}$.
\end{enumerate}

\subsection{Half-plane pinch-point weights}\label{uhpppweights}

\subsubsection{The case $N=2,s=2$ and $N=3,s=3$}

First, we consider the two-pinch-point weight $\Pi_{1234}$ for $N=2$ and the three-pinch-point weight $\Pi_{123456}$ for $N=3$.  The subscripts on each weight indicate the indices of the points $x_i$ that are connected to the pinch point by a boundary arc, and in this case, these are all of the available vertices.  The weights are given by (\ref{s=N}) with $N=2$ and $N=3$ respectively, and both may be expressed in the covariant form (\ref{covform}) which we will find convenient later.  We find that the half-plane $(s=N=2)$-pinch-point weight is
\begin{multline}
\label{2ppcovform}\Pi_{1234}(x_1,\ldots,x_4;z,\bar{z})=[(x_2-x_1)(x_4-x_3)]^{1-6/\kappa}|z-\bar{z}|^{\kappa/8-6/\kappa-1}\\
\times \eta^{8/\kappa-1}(1-\eta)^{2/\kappa}|\mu-\nu|^{24/\kappa-2}[\mu\nu(\eta-\mu)(\eta-\nu)(1-\mu)(1-\nu)]^{1/2-6/\kappa},\end{multline}
and the half-plane $(s=N=3)$-pinch-point weight is
\begin{multline}
\label{3ppcovform}\Pi_{123456}(x_1,\ldots,x_6;z,\bar{z})=[(x_2-x_1)(x_4-x_3)(x_6-x_5)]^{1-6/\kappa}|z-\bar{z}|^{\kappa/8-16/\kappa-1}\\
\begin{aligned}
&\times[\eta(\sigma-\tau)]^{8/\kappa-1}[\tau\sigma(\tau-\eta)(\sigma-\eta)(1-\eta)(1-\tau)(1-\sigma)]^{2/\kappa}|\mu-\nu|^{48/\kappa-3}\\
&\times[\mu\nu(\eta-\mu)(\eta-\nu)(\tau-\mu)(\tau-\nu)(\sigma-\mu)(\sigma-\nu)(1-\mu)(1-\nu)]^{1/2-8/\kappa},\end{aligned}\end{multline}
where the cross-ratios $\eta,\tau,\sigma,\mu,\nu$ are defined in (\ref{crossratios}, \ref{cross}).  The correct normalization of these pinch-point weights depends on bulk-boundary fusion coefficients, but because it is not needed for our purposes, we ignore it in this article.

Before we calculate $s$-pinch-point weights for $s<N$, we comment on the normalizations of these weights too.  When $s<N$, virtually all samples in the $s$-pinch-point event $\Lambda$ will have at least one interval $(x_i,x_{i+1})$ with its endpoints mutually connected by a boundary arc that does not pass near the pinch-point.  The fugacity of this boundary arc is one since we are working with pinch-point weights, so when we send $x_{i+1}\rightarrow x_i$ (after multiplying by $(x_{i+1}-x_i)^{2\theta_1}$ first), we must recover an $s$-pinch-point weight (independent of $x_i$) in a system with bccs at the remaining $2(N-1)$ points on the real axis.  Continuing this process until no such intervals remain, we eventually reach an $s$-pinch-point weight in a system with bccs at the remaining $2s$ points on the real axis.  The weight of this event is given by (\ref{s=N}) with $N=s$.  Therefore, the $s$-pinch-point weights with $s<N$ are normalized so that they equal the $s$-pinch-point weight (\ref{s=N}) with $N\mapsto s$ after this sequence of $N-s$ limits is taken.

\subsubsection{The case $N=2,s=1$}

Next, we consider one-pinch-point events with $N=2$ boundary arcs.  Here, one boundary arc $\gamma_1$ connects the points $x_i$ and $x_j$ with a bulk point $z\in\mathbb{H}$ while the other boundary arc $\gamma_2$ connects the remaining points $x_k$ and $x_l$.  The weight $\Pi_{ij:kl}$ of this event is given by (\ref{chiralcorr}) with $N=2$ and $s=1$.  

\begin{figure}[t]
\centering
\includegraphics[scale=0.27]{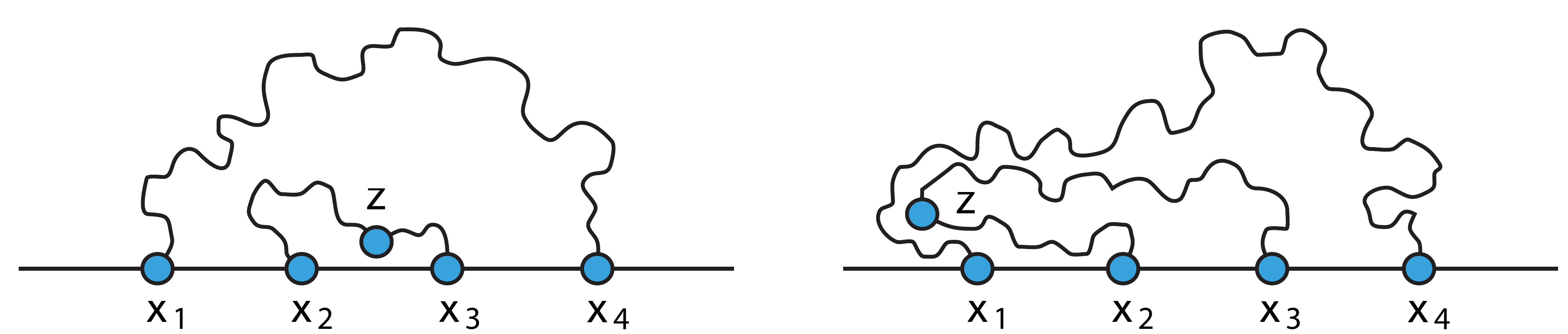}
\caption{The pinch-point configuration for $\Pi_{23:41}$.  The left (resp.\,right) figure shows that a boundary two-leg (resp.\,four-leg) operator is generated to leading order when $z$ approaches the intervals $(x_1,x_2),(x_2,x_3)$, and $(x_3,x_4)$ (resp.\,the interval $(x_4,x_1)$).}
\label{2leg4leg}
\end{figure}

The formula for $\Pi_{ij:kl}$ has a single contour integral $\Gamma_{ij:kl}$, and the contour is chosen so that the chiral operators exhibit specific fusion rules that depend on which vertices are connected to $z$ through $\gamma_1$.   For example, we consider $\Pi_{23:41}$ (figure \ref{2leg4leg}).  If we let the bulk point $z$ approach a boundary point $x$ in the segment $(x_1,x_2)$, then $\gamma_1$ must touch $(x_1,x_2)$ at $x$ in this limit, which is equivalent to placing a boundary two-leg operator $\psi_2(x)$ there.  Therefore the bulk operator $\Psi_1(z)$ must fuse with its image $\Psi_1(\bar{z})$ to give $\psi_2(x)$ to leading order.  Now, when the two chiral operators $V_{0,1}^+(z)$ and $V_{0,1}^+(\bar{z})$ fuse, their product is a boundary chiral operator with charge $2\alpha_{0,1}^+=\alpha_{1,3}^+$ which carries the weight $\theta_2$ of a boundary two-leg operator as required (\ref{weights}).  The same is true of the intervals $(x_2,x_3)$ and $(x_3,x_4)$.

Next, we let the bulk point $z$ approach a boundary point $x$ in the interval $(x_4,x_1)$.  Because $\gamma_2$ joins $x_1$ with $x_4$, topological considerations show that both $\gamma_1$ and $\gamma_2$ must touch $(x_4,x_1)$ at $x$ in this limit (figure \ref{2leg4leg}).  Therefore the leading operator of the ensuing bulk-image fusion must be a boundary four-leg operator.  Above, we saw that the total charge of the bulk-image pair equals that of a boundary two-leg operator.  But if we add the screening charge $\alpha_-$, then this total charge becomes $2\alpha_{0,1}^++\alpha_-=\alpha_{1,5}^+$, which is that of a chiral operator with the desired boundary four-leg weight $\theta_4$ (\ref{weights}).  The screening charge is pulled in with the bulk-image fusion only if $\Gamma_{23:41}$ contracts to a point in the process.  Thus, $\Gamma_{23:41}$ must be a simple curve starting at $\bar{z}$ and ending at $z$.

In order for $\Pi_{23:41}$ to be a continuous function of $z$ and $\bar{z}$, each point of $\Gamma_{23:41}$ must reside on the same Riemann sheet of the integrand, so $\Gamma_{23:41}$ can only cross the real axis through a specific segment $(x_i,x_{i+1})$.  (Here, $x_5:=x_1$.)  This segment must be $(x_4,x_1)$ in order to ensure that $\Gamma_{23:41}$ contracts to a point when we let $z$ and $\bar{z}$ approach a point in $(x_4,x_1)$.  This choice creates another desired effect.  In the event of a bulk-image fusion over $(x_1,x_4)\setminus\{x_2,x_3\}$, $\Gamma_{23:41}$ does not contract to a point, the screening charge is not drawn in, and an undesired boundary four-leg operator in $(x_1,x_4)$, which  would contradict the assertions of the previous paragraph, is not produced.

By cyclically permuting the indices, we find four one-pinch-point weights:
\be\label{1pinchpointweights}\{\Pi_{12:34},\Pi_{23:41},\Pi_{34:12},\Pi_{41:23}\}.\ee
Each weight is given by (\ref{chiralcorr}) with $N=2,s=1$, and $\Gamma_{ij:kl}$ a simple curve connecting $z$ with $\bar{z}$ and crossing the real axis only through $(x_k,x_l)$.  In the formula for each weight, we order the differences in the factors of the integrand so that the branch cuts do not intersect $\Gamma_{ij:kl}$ and the integrand restricted to $\Gamma_{ij:kl}$ is therefore a continuous function of $x_1,\ldots,x_4,z,\bar{z},$ and $u:=u_1$.

\begin{figure}[t]
\centering
\includegraphics[scale=0.27]{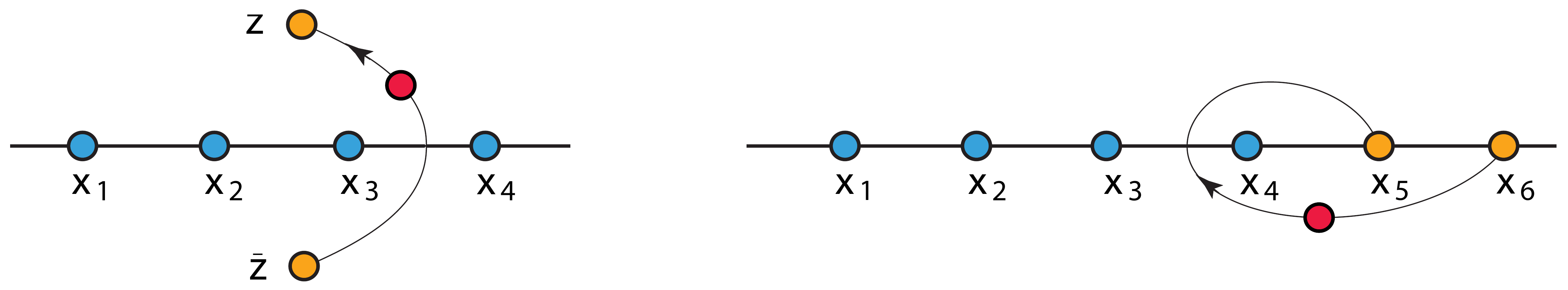}
\caption{The contour used for the one-pinch-point weight $\Pi_{12:34}$.  To facilitate calculation, we at times place $z$ and $\bar{z}$ at adjacent locations $x_5,x_6$ respectively on the real axis as in the right figure.  (In each figure of this article, a blue (resp.\,orange, resp.\,red) circle marks a point of charge $\alpha_{1,2}^-$ (resp.\,$\alpha_{0,s}^+$, resp.\,$\alpha_-$) in the dense phase.)}
\label{RectPaths}
\end{figure}

It is useful to decompose these one-pinch-point weights (\ref{1pinchpointweights}) into a linear combination of the integrals (with $x_5:=z$ and $x_6:=\bar{z}$)
\be\label{Ii}I_i:=\beta(-4/\kappa,-4/\kappa)^{-1}\sideset{}{_{x_{i-1}}^{x_i}}\int du\,\,\mathcal{N}\left[\prod_{j=1}^4(x_j-u)^{-4/\kappa}(x_5-u)^{8/\kappa-1}(x_6-u)^{8/\kappa-1}\right],\quad i\in\{1,\ldots,6\},\ee
in order to explicitly show that these weights are real (or at least share a common phase, as they must be for physical reasons) and to express them in terms of Lauricella functions.  The operator ``$\mathcal{N}$" orders the differences in the factors of the integrand so that $I_i$ is real.  ($I_1$ is a sum of integrations from $x_0:=x_6$ to $\infty$ and from $-\infty$ to $x_1$.)  Because $\arg(z)\in[-\pi,\pi)$ for $z\in\mathbb{C}$, the integrand has a branch cut that starts at each $x_j$ with $j<i$ (resp.\,$j\geq i$) and points leftward (resp.\,rightward) along the real axis.  For simplicity, we momentarily suppose that $x_5$ and $x_6$ are real as we decompose each one-pinch-point weight into a linear combination of the various $I_i$ times algebraic factors.  For example, we can use figure \ref{RectPaths} to find the decomposition
\be\label{Pi1234A}\Pi_{12:34}=A\bigg[2i\sin\bigg(\frac{4\pi}{\kappa}\bigg)I_5+e^{4\pi i/\kappa}I_6\bigg](x_6-x_5)^{\kappa/8+8/\kappa-2}\prod_{i<j}^4(x_j-x_i)^{2/\kappa}\prod_{i=1}^4(x_5-x_i)^{1/2-4/\kappa}(x_6-x_i)^{1/2-4/\kappa}.\ee
The proportionality constant $A$ will be determined below.  Now, to show that the one-pinch-point weights have a constant phase that may be adjusted to unity, we seek a basis of integrals for the span of $\{I_1,\ldots,I_6\}$ that are real when $x_5=z$ and $x_6=\bar{z}$.  Integrating the screening charge along a contour parallel to and immediately above (resp.\,below) the real axis gives the $-$ (resp.\,$+$) branch of the linear relation
\be\label{abovebelow}I_1+e^{\pm4\pi i/\kappa}I_2+e^{\pm8\pi i/\kappa}I_3+e^{\pm12\pi i/\kappa}I_4+e^{\pm16\pi i/\kappa}I_5-e^{\pm8\pi i/\kappa}I_6=0,\ee
which allows us to write $I_5$ and $I_6$ in terms of $I_1,\ldots,I_4$.  Then because the integral 
\be\label{I41} \mathcal{I}_1:=\beta(-4/\kappa,-4/\kappa)^{-1}\sideset{}{_{x_4}^{x_1}}\int du\,\,\mathcal{N}\left[(x_5-u)^{8/\kappa-1}(x_6-u)^{8/\kappa-1}\prod_{j=1}^4(u-x_i)^{-4/\kappa}\right]=I_5-e^{8\pi i/\kappa}I_6+I_1\ee
is real when $x_5=z$ and $x_6=\bar{z}$, we find a real basis $\{\mathcal{I}_1,\mathcal{I}_2:=I_2,\mathcal{I}_3:=I_3,\mathcal{I}_4:=I_4\}$ for the span of $\{I_i\}_{i=1}^6$.  We anticipate that the coefficients found from decomposing the $\Pi_{ij:kl}$ over this basis will share a common phase.

\begin{figure}[t]
\centering
\includegraphics[scale=0.27]{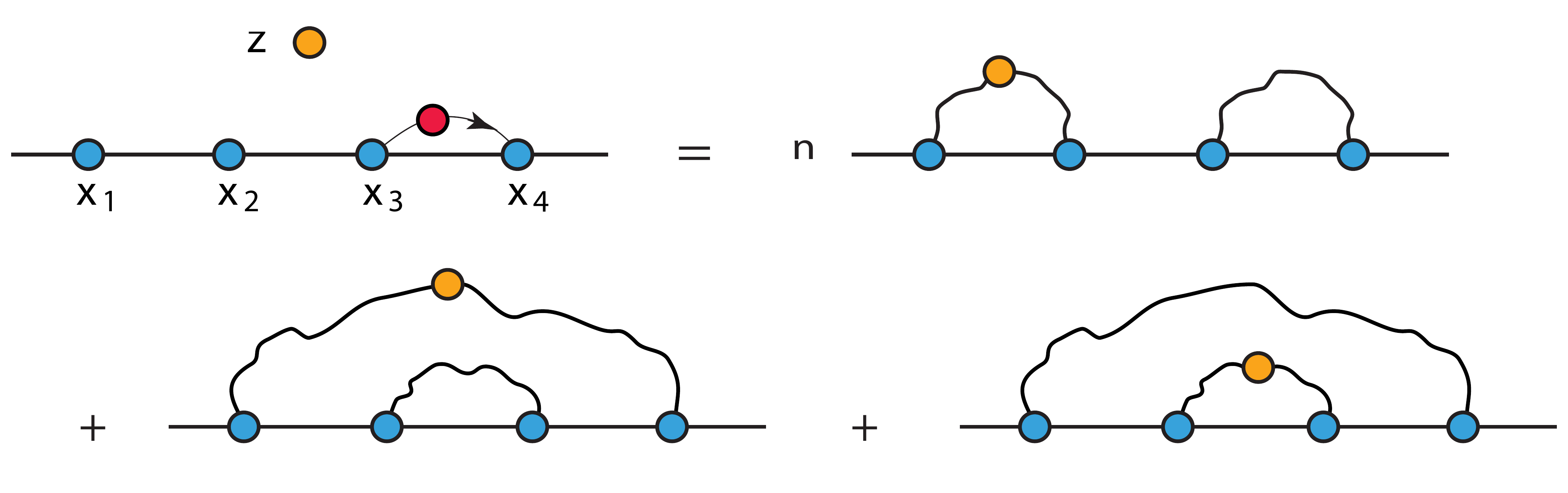}
\caption{The decomposition of (\ref{pre34}) into a linear combination of the weights $\Pi_{12:34},\Pi_{41:23}$, and $\Pi_{23:41}$ as given in (\ref{first34}).}
\label{RectNeutralPair}
\end{figure}

However,  it is more useful for our purposes (of calculating one-pinch-point weights for the hexagon later) to compute this decomposition via a different approach in which the four integrals $\mathcal{I}_i$ arise naturally as conformal blocks.   We consider the one-pinch-point weight $\Pi_{12}(x_1,x_2;z,\bar{z})$, given by $\langle\psi_1(x_1)_{[2]}\psi_1(x_2)\Psi_1(z)_{[2]}\Psi_1(\bar{z})\rangle$.  Here, the bracketed subscript between a pair of primary fields indicates the unique fusion channel propagating between that pair, so ``$[s]$" indicates the $s$-leg channel when $s>0$ and the identity channel when $s=0$.  To increase $N$ from one to two, we insert the charge-neutral collection $\int_{x_3}^{x_4} du\,V_{1,2}^-(x_3)V_{1,2}^-(x_4)V_-(u)$ with $x_2<x_3<x_4$ into the chiral operator representation of this four-point function.  We find
\be\label{pre34}\langle\psi_1(x_1)_{[2]}\psi_1(x_2)\psi_1(x_3)_{[0]}\psi_1(x_4)\Psi_1(z)_{[2]}\Psi_1(\bar{z})\rangle=nJ(x_1,\ldots x_4;z,\bar{z})\,\mathcal{I}_4(x_1,\ldots x_4;z,\bar{z}),\ee
where the function $J$ is given by (\ref{chiralcorr}) with $N=2$ and $s=1$:
\be J(x_1,\ldots,x_4;z,\bar{z}):=|z-\bar{z}|^{\kappa/8+8/\kappa-2}\prod_{i<j}^4(x_j-x_i)^{2/\kappa}\prod_{i=1}^4|z-x_i|^{1-8/\kappa}.\ee
The new pair of boundary one-leg operators at $x_3$ and $x_4$ fuse through only the identity channel because the screening charge is integrated along a simple curve connecting $x_3$ with $x_4$.  The original boundary one-leg operators at $x_1$ and $x_2$ still fuse through only the two-leg channel.  Three of the four one-pinch-point events ($ij$:$kl$) are consistent with these fusion rules, (41:23), (12:34), and (23:41), so (\ref{pre34}) must be a linear combination of the pinch-point weights $\Pi_{12:34},\Pi_{41:23},$ and $\Pi_{23:41}$.  Indeed, this linear combination is (figure \ref{RectNeutralPair})
\be\label{first34}\Pi_{41:23}+n\Pi_{12:34}+\Pi_{23:41}=nJ(x_1,\ldots x_4;z,\bar{z})\,\mathcal{I}_4(x_1,\ldots x_4;z,\bar{z}).\ee
As usual, $n$ is the loop fugacity (\ref{fugacity}) of the O$(n)$ model.

The coefficients of the linear combination on the left side of (\ref{first34}) are found in the following way.  First, to find the coefficient of $\Pi_{12:34}$, we send $x_4\rightarrow x_3$ on both sides of (\ref{first34}).  (We always implicitly multiply by $(x_{i+1}-x_i)^{2\theta_1}$ before sending $x_{i+1}\rightarrow x_i$ so that the limit exists.)  Then $\Pi_{41:23},\Pi_{23:41}\rightarrow0$ while $\Pi_{12:34}\rightarrow\Pi_{12}$ (\ref{2soln}), and $nJ\times\mathcal{I}_4\rightarrow n\Pi_{12}$.  This justifies the coefficient of $n$ that dresses $\Pi_{12:34}$ in (\ref{first34}).  Next, to find the coefficient of $\Pi_{41:23}$, we send $x_3\rightarrow x_2$.  On the left side of (\ref{first34}), $\Pi_{12:34},\Pi_{23:41}\rightarrow0$ while $\Pi_{41:23}\rightarrow\Pi_{14}$, or really $\Pi_{12}$ with $x_2\mapsto x_4$.  On the right side, we use (\ref{abovebelow}) to write $nJ\times\mathcal{I}_4$ as a linear combination of $nJ\times I_1,nJ\times I_3,nJ\times I_5$, and $nJ\times I_6$.  (These are the four $nJ\times I_i$ that have either both or neither bounds of integration among $\{x_2,x_3\}$.  Again, $x_5=z$ and $x_6=\bar{z}$.)  All of the integrals in this combination except $nJ\times I_3$ vanish in this limit, while $nJ\times I_3$ goes to $n\Pi_{12}$.  Because $nJ\times I_3$ carries a coefficient of $n^{-1}$ in this linear combination, the right side of (\ref{first34}) becomes $\Pi_{12}$ with $x_2\mapsto x_4$.  This justifies the coefficient of one that dresses $\Pi_{41:23}$ in (\ref{first34}).  The same reasoning gives the coefficient of one for $\Pi_{23:41}$ in (\ref{first34}).   

Cyclically permuting the indices in (\ref{first34}) gives three more equations relating the four one-pinch-point weights (\ref{1pinchpointweights}) to the four integrals $\mathcal{I}_i$.  Upon inverting these equations to isolate the weights, we find
\be\label{1rectpp}\Pi_{ij:kl}=J\left[\frac{2\,\mathcal{I}_j+(n^2-2)\,\mathcal{I}_l-n\,\mathcal{I}_i-n\,\mathcal{I}_k}{n^2-4}\right].\ee
For each index $i$, we can multiply $J\times \mathcal{I}_i$ by $(x_2-x_1)^{6/\kappa-1}(x_4-x_3)^{6/\kappa-1}|z-\bar{z}|^{1-\kappa/8}$ to arrive with a function $G_i$ that depends only on cross-ratios $\eta,\mu,$ and $\nu$, according to (\ref{covform}).  After making the replacement $(x_1,x_2,x_3,x_4,z,\bar{z})\mapsto(0,\eta,1,\infty,\mu,\nu)$, we find
\bean\label{firstG}G_1(\eta,\mu,\nu)&=&\frac{(\eta|\mu-\nu|)^{8/\kappa-1}(1-\eta)^{2/\kappa}}{(\mu\nu(\eta-\mu)(\eta-\nu)(1-\mu)(1-\nu))^{4/\kappa-1/2}}F_D\bigg(1-\frac{4}{\kappa};\frac{4}{\kappa},1-\frac{8}{\kappa},1-\frac{8}{\kappa};2-\frac{8}{\kappa}\,\bigg|\,1-\eta,1-\mu,1-\nu\bigg),\\
G_2(\eta,\mu,\nu)&=&\frac{(\mu\nu|\mu-\nu|^2)^{4/\kappa-1/2}(1-\eta)^{2/\kappa}}{((\eta-\mu)(\eta-\nu)(1-\mu)(1-\nu))^{4/\kappa-1/2}}F_D\bigg(1-\frac{4}{\kappa};\frac{4}{\kappa},1-\frac{8}{\kappa},1-\frac{8}{\kappa};2-\frac{8}{\kappa}\,\bigg|\,\eta,\frac{\eta}{\mu},\frac{\eta}{\nu}\bigg),\\
G_3(\eta,\mu,\nu)&=&\frac{(\eta^2(1-\mu)(1-\nu)|\mu-\nu|^2)^{4/\kappa-1/2}}{(\mu\nu(\eta-\mu)(\eta-\nu))^{4/\kappa-1/2}(1-\eta)^{6/\kappa-1}}F_D\bigg(1-\frac{4}{\kappa};\frac{4}{\kappa},1-\frac{8}{\kappa},1-\frac{8}{\kappa};2-\frac{8}{\kappa}\,\bigg|\,1-\eta,\frac{1-\eta}{1-\mu},\frac{1-\eta}{1-\nu}\bigg),\\
\label{lastG}G_4(\eta,\mu,\nu)&=&\frac{(\eta|\mu-\nu|)^{8/\kappa-1}(1-\eta)^{2/\kappa}}{(\mu\nu(\eta-\mu)(\eta-\nu)(1-\mu)(1-\nu))^{4/\kappa-1/2}}F_D\bigg(1-\frac{4}{\kappa};\frac{4}{\kappa},1-\frac{8}{\kappa},1-\frac{8}{\kappa};2-\frac{8}{\kappa}\,\bigg|\,\eta,\mu,\nu\bigg).\eean
We have expressed each $G_i$ in terms of the Lauricella function $F_D$, defined as \cite{ex}
\be\label{Laur}F_D(a,b_1,\ldots, b_m,c\,|\,x_1,\ldots, x_m):=\frac{\Gamma(a)}{\Gamma(c)\Gamma(c-a)}\sideset{}{_0^1}\int t^{a-1}(1-t)^{c-a-1}(1-x_1t)^{-b_1}\ldots(1-x_mt)^{-b_m}\,dt,\ee
by writing the integration variable $u$ of $\mathcal{I}_i$ as the following M\"{o}bius transformation of the integration variable $t$ in (\ref{Laur}):
\begin{align} &i=1:\quad u=\frac{t-1}{t},& &i=2:\quad u=\eta\,t,\\
&i=3:\quad u=1-(1-\eta)t,& & i=4:\quad u=\frac{1}{t}.\end{align}
These transformations are chosen so that each $F_D$ has $m=3$ arguments with the first argument between zero and one and the last two arguments being complex conjugates of each other.  These choices ensure that each $F_D$ is real.  Thus, the half-plane pinch-point weights $\Pi_{ij:kl}$, expressed in the covariant form (\ref{covform}), are given by
\be\label{1HRectweight}\Pi_{ij:kl}(x_1,\ldots,x_4;z,\bar{z})=[(x_2-x_1)(x_4-x_3)]^{1-6/\kappa}|z-\bar{z}|^{\kappa/8-1}\left[\frac{2G_j+(n^2-2)G_l-nG_i-nG_k}{n^2-4}\right](\eta,\mu,\nu),\ee
with $\eta,\mu,\nu,$ and $n$ defined in (\ref{crossratios}, \ref{cross}, \ref{fugacity}) respectively.  We note that our normalization in (\ref{1HRectweight}) ensures that $(x_l-x_k)^{2\theta_1}\Pi_{ij:kl}\rightarrow\Pi_{ij}$ (\ref{2soln}) as $x_l\rightarrow x_k$.  Now, if we let $\kappa$ approach $4/a$ with $a\in\mathbb{Z}^+$ so that $n\rightarrow\pm2$, then the limit $\Pi_{ij}$ remains finite.  Thus (\ref{1HRectweight}) must be finite when $n=\pm2$, although showing this explicitly appears to be difficult.  Comparing (\ref{1HRectweight}) with (\ref{Pi1234A}), we find that $A=-in/\sqrt{4-n^2}$.

\subsubsection{The case $N=3,s=2$}

Next, we consider two-pinch-point events with $N=3$ boundary arcs.  Here, two boundary arcs $\gamma_1$ and $\gamma_2$, with endpoints respectively at $x_i,x_j\in\mathbb{R}$ and $x_k,x_l\in\mathbb{R}$, touch at a bulk point $z$, and the remaining boundary arc $\gamma_3$ has endpoints at $x_m$ and $x_n$.  We note that this setup restricts the allowed boundary arc connectivities to those in which $\gamma_3$ does not separate $\gamma_1$ from $\gamma_2$, so $x_m$ and $x_n$ must be either adjacent or among $\{x_6,x_1\}$.  The weight $\Pi_{ijkl:mn}$ of this event is given by (\ref{chiralcorr}) with $N=3$ and $s=2$.  By cyclically permuting the indices, we find six such two-pinch-point configurations.  

\begin{figure}[b]
\centering
\includegraphics[scale=0.27]{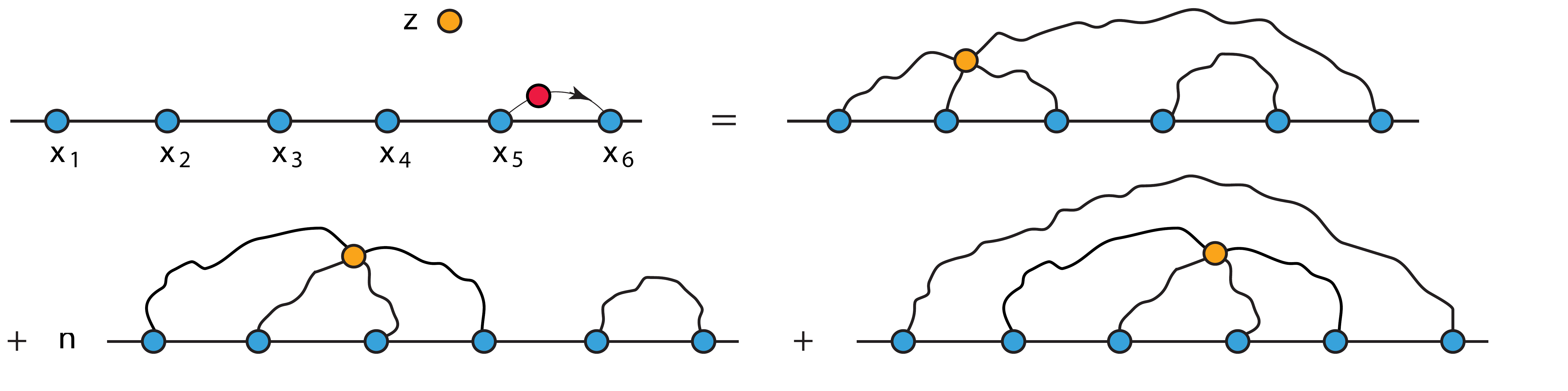}
\caption{The decomposition of (\ref{pre56}) into a linear combination of the weights $\Pi_{6123:45},\Pi_{1234:56}$, and $\Pi_{2345:61}$, as given in (\ref{first56}).}
\label{HexNeutralPair2pp}
\end{figure}

The formula for $\Pi_{ijkl:mn}$ contains a single contour integral $\Gamma_{ijkl:mn}$ that is determined via the same reasoning that was used for the case $N=2$ and $s=1$ above.  We summarize the argument.  The two-pinch-point event is conditioned by the insertion of a bulk four-leg operator $\Psi_2(z)$.  Topological considerations (as can be understood upon examining figure \ref{HexNeutralPair2pp}) show that fusing this operator with its image across any interval $(x_a,x_b)$ with $(a,b)\neq(m,n)$ (resp.\,$(a,b)=(m,n)$) must, to leading order, give rise to a boundary four-leg (resp.\,six-leg) operator $\psi_4$ (resp.\,$\psi_6$).  Because $2\alpha_{0,2}^+=\alpha_{1,5}^+$ is the charge of a chiral operator with the boundary four-leg weight $\theta_4$, this requirement is already satisfied when $(a,b)\neq(m,n)$.  If $\Gamma_{ijkl:mn}$ is a simple curve with endpoints at $z$ and $\bar{z}$ and crossing the real axis only through $(x_m,x_n)$, then the screening charge is drawn into a bulk-image fusion across this interval, shifting the product to a chiral operator with net charge $2\alpha_{0,2}^++\alpha_-=\alpha_{1,7}^+$.  This operator has the desired boundary six-leg weight $\theta_6$ (\ref{weights}).

To express the two-pinch-point weights in terms of Lauricella functions, we write them as linear combinations of the six real integrals
\be\mathcal{K}_i:=\beta(-4/\kappa,-4/\kappa)^{-1}\sideset{}{_{x_{i-1}}^{x_i}}\int\,du\,\mathcal{N}\left[(z-u)^{12/\kappa-1}(\bar{z}-u)^{12/\kappa-1}\prod_{j=1}^6(u-x_j)^{-4/\kappa}\right],\quad i=1,\ldots,6.\ee
(As before, the operator ``$\mathcal{N}$" orders the differences in the factors of the integrand so that the integrand is real, and $\mathcal{K}_1$ is integrated from $x_0:=x_6$ to $\infty$ and then from $-\infty$ to $x_1$.)  To proceed, we consider the two-pinch-point weight $\Pi_{1234}$.  Inserting the charge-neutral collection $\int_{x_5}^{x_6} du\,V_{1,2}^-(x_5)V_{1,2}^-(x_6)V_-(u)$ into the chiral representation of its six-point function $\langle\psi_1(x_1)_{[2]}\psi_1(x_2)\psi_1(x_3)_{[2]}\psi_1(x_4)\Psi_1(z)_{[4]}\Psi_1(\bar{z})\rangle$ with $x_4<x_5<x_6$, we get the conformal block
\be\label{pre56}\langle\psi_1(x_1)_{[2]}\psi_1(x_2)\psi_1(x_3)_{[2]}\psi_1(x_4)\psi_1(x_5)_{[0]}\psi_1(x_6)\Psi_1(z)_{[4]}\Psi_1(\bar{z})\rangle=nL(x_1,\ldots, x_6;z,\bar{z})\,\mathcal{K}_6(x_1,\ldots, x_6;z,\bar{z}),\ee
where the pre-factor $L$ is given by (\ref{chiralcorr}) with $N=3$ and $s=2$:
\be\label{prefactorch5}L(x_1,\ldots,x_6;z,\bar{z}):=|z-\bar{z}|^{\kappa/8+18/\kappa-3}\prod_{i<j}^6(x_j-x_i)^{2/\kappa}\prod_{i=1}^6|z-x_i|^{1-12/\kappa}.\ee
After following the reasoning that led to (\ref{first34}), we find (figure \ref{HexNeutralPair2pp})
\be\label{first56}\Pi_{6123:45}+n\Pi_{1234:56}+\Pi_{2345:61}=nL(x_1,\ldots,x_6;z,\bar{z})\,\mathcal{K}_6(x_1,\ldots,x_6;z,\bar{z}).\ee
Another five equations relating the six weights with the six integrals $\mathcal{K}_i$ is found by cyclically permuting the indices in (\ref{first56}).  These equations may be simultaneously solved to give
\be\label{hex2pp}\Pi_{ijkl:mn}=nL\left[\frac{(2-n^2)(\mathcal{K}_i+\mathcal{K}_m)+n(\mathcal{K}_j+\mathcal{K}_l)-2\,\mathcal{K}_k+(n^3-3n)\mathcal{K}_n}{(n^2-4)(n^2-1)}\right].\ee
(We note the present double-use of $n$ as an index and as the loop fugacity of the O$(n)$ model.)  To finish, we seek a form for the weights that expresses the $\mathcal{K}_i$ in terms of Lauricella functions and exhibits the conformally covariant ansatz of (\ref{covform}).  To this end, we define the function
\bea\label{G1}H_i(x_1,\ldots,x_6;z,\bar{z})&:=&[(x_2-x_1)(x_4-x_3)(x_6-x_5)]^{2\theta_1}|z-\bar{z}|^{2\Theta_2}L\times\mathcal{K}_i\quad\quad\\
\label{G2}&=&[(x_2-x_1)(x_4-x_3)]^{6/\kappa-1}|z-\bar{z}|^{1+6/\kappa-\kappa/8}L'\times\mathcal{K}'_i,\eea
with $L$ and $\mathcal{K}_i$ adjusted to respective quantities $L'$ and $\mathcal{K}_i'$ that are finite in the limit $x_6\rightarrow\infty$:
\bea L'(x_1,\ldots,x_6;z,\bar{z})&:=&(x_6-x_5)^{2/\kappa-1}L,\\
\mathcal{K}'_i(x_1,\ldots,x_6;z,\bar{z})&:=&(x_6-x_5)^{4/\kappa}\mathcal{K}_i.\eea
According to (\ref{covform}), $H_i$ is strictly a function of the cross-ratios $\eta,\tau,\sigma,\mu$, and $\nu$.  After making the replacement $(x_1,x_2,x_3,x_4,x_5,x_6,z,\bar{z})\mapsto(0,\eta,\rho,\sigma,1,\infty,\mu,\nu)$,
we find
\begin{multline}\label{Gi}H_i(\eta,\tau,\sigma,\mu,\nu)=\mathcal{K}'_i(\eta,\tau,\sigma,\mu,\nu)[\eta(\sigma-\tau)]^{8/\kappa-1}[\tau\sigma(\tau-\eta)(\sigma-\eta)(1-\eta)(1-\tau)(1-\sigma)]^{2/\kappa}\\
\times|\mu-\nu|^{24/\kappa-2}[\mu\nu(\mu-\eta)(\nu-\eta)(\mu-\tau)(\nu-\tau)(\mu-\sigma)(\nu-\sigma)(\mu-1)(\nu-1)]^{1/2-6/\kappa},\end{multline}
with each $\mathcal{K}_i'(\eta,\tau,\sigma,\mu,\nu):=\mathcal{K}_i'(0,\eta,\tau,\sigma,1,\infty;\mu,\nu)$ finite and equaling a Lauricella function $F_D$ times algebraic prefactors:
\bea\label{Ifirst}\mathcal{K}_1'(\eta,\tau,\sigma,\mu,\nu)&\propto& F_D\bigg(\{\chi_j\}\,\bigg|\,1-\eta,1-\tau,1-\sigma,1-\mu,1-\nu\bigg),\\
\mathcal{K}_2'(\eta,\tau,\sigma,\mu,\nu)&\propto& \eta^{1-8/\kappa}\tau^{-4/\kappa}\sigma^{-4/\kappa}\mu^{12/\kappa-1}\nu^{12/\kappa-1}F_D\bigg(\{\chi_j\}\,\bigg|\,\eta,\frac{\eta}{\tau},\frac{\eta}{\sigma},\frac{\eta}{\mu},\frac{\eta}{\nu}\bigg),\\
\mathcal{K}_3'(\eta,\tau,\sigma,\mu,\nu)&\propto& \eta^{1-8/\kappa}\tau^{4/\kappa-1}(\tau-\eta)^{1-8/\kappa}(\sigma-\eta)^{-4/\kappa}(1-\eta)^{-4/\kappa}(\mu-\eta)^{12/\kappa-1}(\nu-\eta)^{12/\kappa-1}\nonumber\\
&\times& F_D\bigg(\{\chi_j\}\,\bigg|\,1-\frac{\eta}{\tau},\frac{\sigma(\tau-\eta)}{\tau(\sigma-\eta)},\frac{\tau-\eta}{\tau(1-\eta)},\frac{\mu(\tau-\eta)}{\tau(\mu-\eta)},\frac{\nu(\tau-\eta)}{\tau(\nu-\eta)}\bigg),\\
\mathcal{K}_4'(\eta,\tau,\sigma,\mu,\nu)&\propto&\tau^{-4/\kappa}(\tau-\eta)^{1-8/\kappa}(\sigma-\eta)^{4/\kappa-1}(\sigma-\tau)^{1-8/\kappa}(1-\tau)^{-4/\kappa}(\mu-\tau)^{12/\kappa-1}(\nu-\tau)^{12/\kappa-1}\nonumber\\
&\times& F_D\bigg(\{\chi_j\}\,\bigg|\,\frac{\sigma-\tau}{\sigma-\eta},\frac{\eta(\sigma-\tau)}{\tau(\sigma-\eta)},\frac{(1-\eta)(\sigma-\tau)}{(1-\tau)(\sigma-\eta)},\frac{(\mu-\eta)(\sigma-\tau)}{(\mu-\tau)(\sigma-\eta)},\frac{(\nu-\eta)(\sigma-\tau)}{(\nu-\tau)(\sigma-\eta)}\bigg),\\
\mathcal{K}_5'(\eta,\tau,\sigma,\mu,\nu)&\propto&(1-\eta)^{-4/\kappa}(1-\tau)^{-4/\kappa}(1-\sigma)^{1-8/\kappa}(1-\mu)^{12/\kappa-1}(1-\nu)^{12/\kappa-1}\nonumber\\
&\times& F_D\bigg(\{\chi_j\}\,\bigg|\,1-\sigma,\frac{1-\sigma}{1-\eta},\frac{1-\sigma}{1-\tau},\frac{1-\sigma}{1-\mu},\frac{1-\sigma}{1-\nu}\bigg),\\
\label{Ilast}\mathcal{K}_6'(\eta,\tau,\sigma,\mu,\nu)&\propto& F_D\bigg(\{\chi_j\}\,\bigg|\,\eta,\tau,\sigma,\mu,\nu\bigg).\eea
Again, we have expressed each $\mathcal{K}_i'$ in terms of $F_D$ by writing its integration variable $u$ as the following M\"{o}bius transformation of the integration variable $t$ in (\ref{Laur}):
\begin{align} &i=1:\quad u=\frac{t-1}{t}, &&i=2:\quad u=\eta\,t, \\
&i=3:\quad u=\frac{\eta\tau}{\tau-(\tau-\eta)t}, && i=4:\quad u=\frac{\eta(\sigma-\tau)t-\tau(\sigma-\eta)}{(\sigma-\tau)t-(\sigma-\eta)}, \\
&i=5:\quad u=1-(1-\sigma)t, &&i=6:\quad u=\frac{1}{t}. \end{align}
The transformations are chosen so that the first three arguments of each $F_D$ is between zero and one and the last two arguments are complex conjugates.  These choices ensure that each $F_D$ is real.   Each $F_D$ uses the same set of seven parameters:
\be\label{params}\{\chi_j\}_{j=1}^7=\left\{1-\frac{4}{\kappa};\frac{4}{\kappa},\frac{4}{\kappa},\frac{4}{\kappa},1-\frac{12}{\kappa},1-\frac{12}{\kappa};2-\frac{8}{\kappa}\right\}.\ee
Combining (\ref{hex2pp}) and (\ref{G1}), we find that each weight is given by the conformally covariant formula 
\begin{multline}\label{2pphexcov}\Pi_{ijkl:mn}(x_1,\ldots,x_6;z,\bar{z})=[(x_2-x_1)(x_4-x_3)(x_6-x_5)]^{1-6/\kappa}|z-\bar{z}|^{\kappa/8-6/\kappa-1}\\
\times\left[\frac{n(2-n^2)(H_i+H_m)+n^2H_j-2nH_k+n^2H_l+n^2(n^2-3)H_n}{(n^2-4)(n^2-1)}\right](\eta,\tau,\sigma,\mu,\nu),\end{multline}
with each $H_i$ explicitly given among (\ref{Gi}-\ref{Ilast}), $\eta,\tau,\sigma,\mu,$ and $\nu$ given by (\ref{crossratios}, \ref{cross}), and $n$ given by (\ref{fugacity}).   We note that our normalization in (\ref{hex2pp}) ensures that $(x_n-x_m)^{2\theta_1}\Pi_{ijkl:mn}\rightarrow\Pi_{ijkl}$ (\ref{2ppcovform}) as $x_n\rightarrow x_m$.  If we let $\kappa$ approach the zeros $4/a$ and $12/(3a\pm1)$, with $a\in\mathbb{Z}^+$, of the denominator of (\ref{2pphexcov}) in this relation, then the limit $\Pi_{ijkl}$ remains finite.  Therefore, (\ref{2pphexcov}) must be finite when $n=\pm2,\pm1$, although this seems to be very difficult to prove directly.

\subsubsection{The case $N=3,s=1$}

Last, we consider one-pinch-point events with $N=3$ boundary arcs.   Here, a boundary arc $\gamma_1$ connects $x_i,x_j$ and $z$, another $\gamma_2$ connects $x_k$ and $x_l$, and the last $\gamma_3$ connects $x_m$ and $x_n$.  We denote the half-plane weight of this event by $\Pi_{ij:kl:mn}$, and it is given by (\ref{chiralcorr}) with $N=3$ and $s=1$.  This formula contains a double contour integral, and the contours must not intersect in order to guarantee a solution of the system of null-state PDEs (appendix \ref{appendix}).  According to the discussion preceding (\ref{catalan}), there are $C_3=5$ possible boundary arc connectivities, and in each, $z$ may touch any one of the three boundary arcs to give a total of fifteen possible one-pinch-point events.

\begin{figure}[b]
\centering
\includegraphics[scale=0.27]{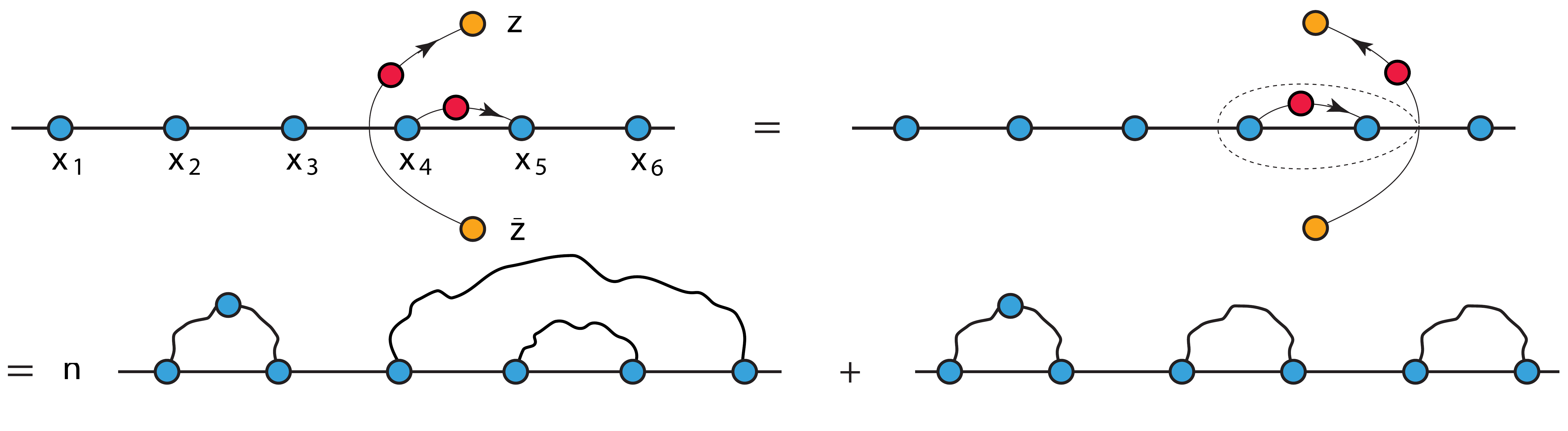}
\caption{The decomposition of the integral in (\ref{2configs}) into a linear combination of $\Pi_{12:34:56}$ and $\Pi_{12:36:45}$.  The contour $\Gamma_{12}$ connecting $z$ with $\bar{z}$ in the top-left illustration can be deformed into the contour that is a vertical reflection of $\Gamma_{12}$ across $[x_5,x_6]$ plus the dashed contour in the top-right illustration.  According to (\ref{loop}), integration along the dashed contour gives zero.}
\label{HexPaths}
\end{figure}

Now we associate certain choices of integration contours with particular linear combinations of these configurations.  In the previously considered cases with one screening charge, we noted that a half-plane pinch-point weight with a simple contour connecting $z$ with $\bar{z}$ by crossing a specified interval $(x_a,x_b)$ corresponds to a specified pinch-point event, and now we investigate to what extent this remains true in our present situation with two screening charges.  We suppose that $\gamma_1$ connects $x_1,x_2$, and $z$.  Then $x_3,\ldots,x_6$ are connected pairwise by the two remaining boundary arcs in one of two possible ways.  In both cases, topological considerations show that fusion of the bulk two-leg operator $\Psi_1(z)$ with its image across the intervals $(x_3,x_4)$ or $(x_5,x_6)$ must give rise to a boundary four-leg operator to leading order.  Hence, we choose the first contour $\Gamma_{12}$ to be a simple curve connecting $z$ and $\bar{z}$ and crossing the real axis only through $(x_3,x_4)$.  A natural choice for the second contour would be the same as for the first but crossing $(x_5,x_6)$ instead, yet this is not allowed because otherwise this contour would intersect $\Gamma_{12}$ at $z$ and $\bar{z}$.  We suppose that the second contour is $[x_4,x_5]$ instead.  Now, it is easy to show that
\be\label{loop}\oint_{\Gamma}\int_{x_4}^{x_5}(u_2-u_1)^{8/\kappa}\prod_{i=1,2}(u_i-x_4)^{-4/\kappa}(x_5-u_i)^{-4/\kappa}\ldots\,du_2\,du_1\,=0,\ee
where $u_1$ is integrated around a simple loop $\Gamma$ surrounding $[x_4,x_5]$ and the ellipsis stands for the rest of the integrand in (\ref{chiralcorr}) with $N=3$ and $s=1$.  Then this identity (\ref{loop}) allows us to deform $\Gamma_{12}$ into a simple curve crossing the real axis only through $(x_5,x_6)$, so a bulk-image fusion across $(x_5,x_6)$ will produce a boundary four-leg operator there as well (figure \ref{HexPaths}).  
The formula that follows from these contour choices sum over both possible connectivities of the two boundary arcs joining $x_3,\ldots,x_6$ pairwise.  Their relative coefficients can be found in the usual way.  Thus, we have from (\ref{chiralcorr}) with $N=3$ and $s=1$ that
\be\label{2configs}\Pi_{12:34:56}+n\Pi_{12:36:45}=\frac{n|z-\bar{z}|^{\kappa/8+8/\kappa-2}}{i\beta(-4/\kappa,-4/\kappa)^2\sqrt{4-n^2}}\prod_{i<j}^6(x_j-x_i)^{2/\kappa}\prod_{i=1}^6|z-x_i|^{1-8/\kappa}\sideset{}{_{\Gamma_{12}}}\int\sideset{}{_{x_4}^{x_5}}\int\,du_1\,du_2\,\mathcal{N}\bigg[\,\,\ldots\,\,\bigg].\ee
The ellipsis stands for the rest of the integrand in (\ref{chiralcorr}), and the normalization follows from requiring that we recover the two-pinch-point weight $\Pi_{12:34}$ with $x_6\mapsto x_4$ upon sending $x_5\rightarrow x_4$.  Although this density is a natural observable, the left side is not a single one-pinch-point configuration.  Cyclic permutation of the indices generates only five more equations involving just twelve of the fifteen possible weights.  One of the missing weights is $\Pi_{14:23:56}$, and the other two missing weights are generated by rotating the hexagon.

\begin{table}[t!]
\centering
\begin{tabular}{c|cccc}
&\begin{tabular}{l}$M\times\mathcal{K}_{12;34}\propto$\\\resizebox{3.3cm}{.8cm}{\includegraphics{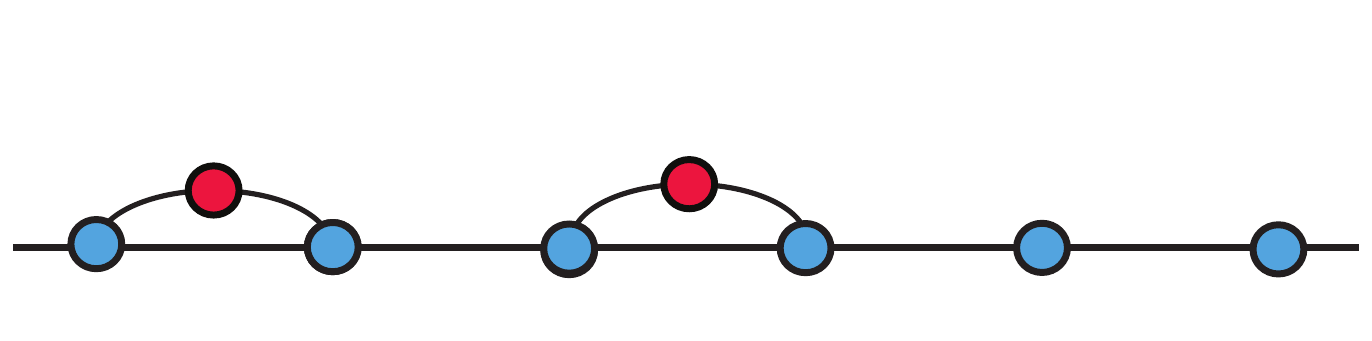}}\end{tabular}
&\begin{tabular}{l}$M\times\mathcal{K}_{12;45}\propto$\\\resizebox{3.3cm}{.8cm}{\includegraphics{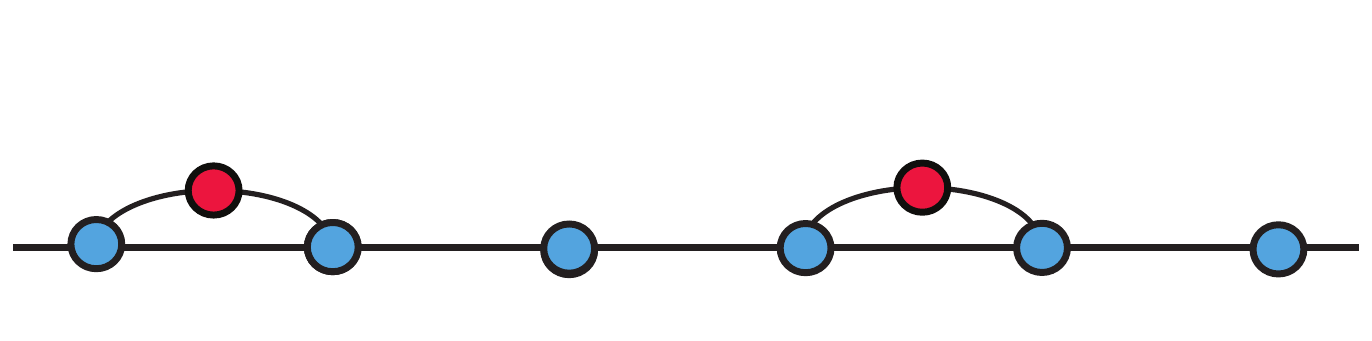}}\end{tabular}
&\begin{tabular}{l}$M\times\mathcal{K}_{12;56}\propto$\\\resizebox{3.3cm}{.8cm}{\includegraphics{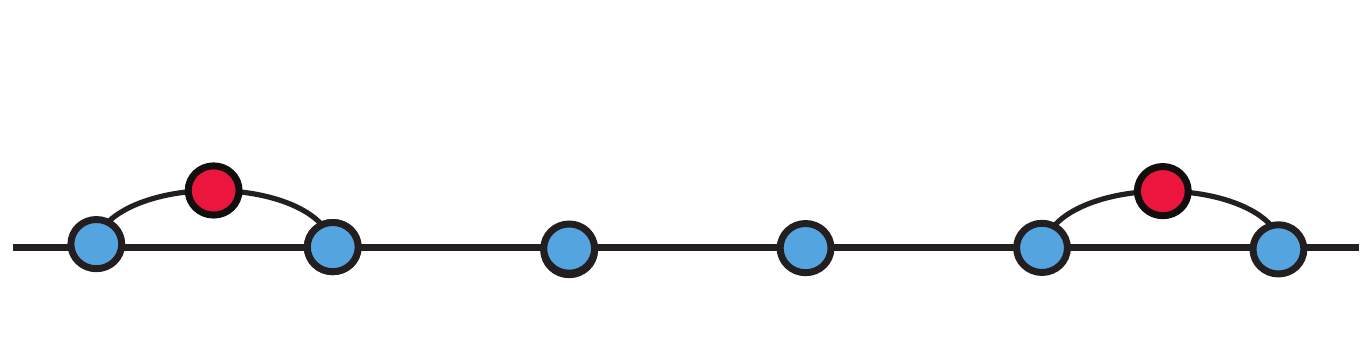}}\end{tabular}
&\begin{tabular}{l}$M\times\mathcal{K}_{12;63}\propto$\\\resizebox{3.3cm}{.8cm}{\includegraphics{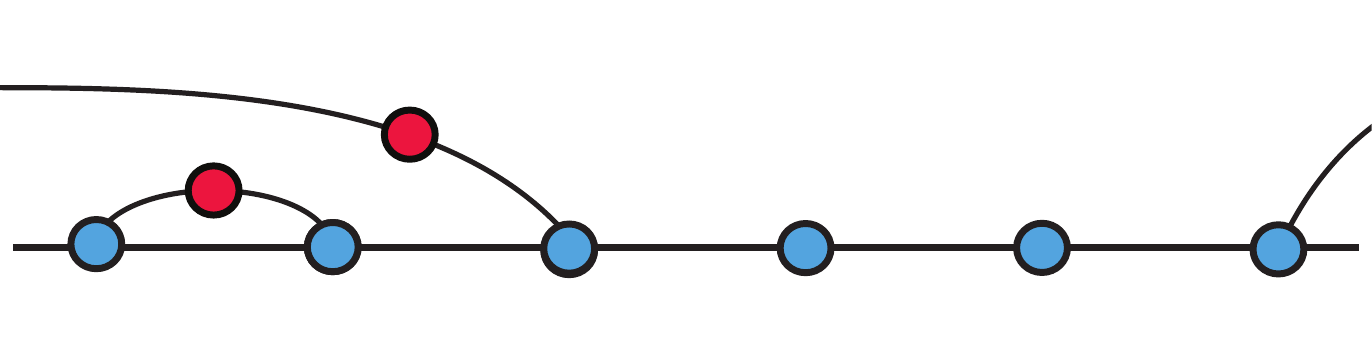}}\end{tabular}\\
\hline
\hline
\resizebox{3.5cm}{.8cm}{\includegraphics{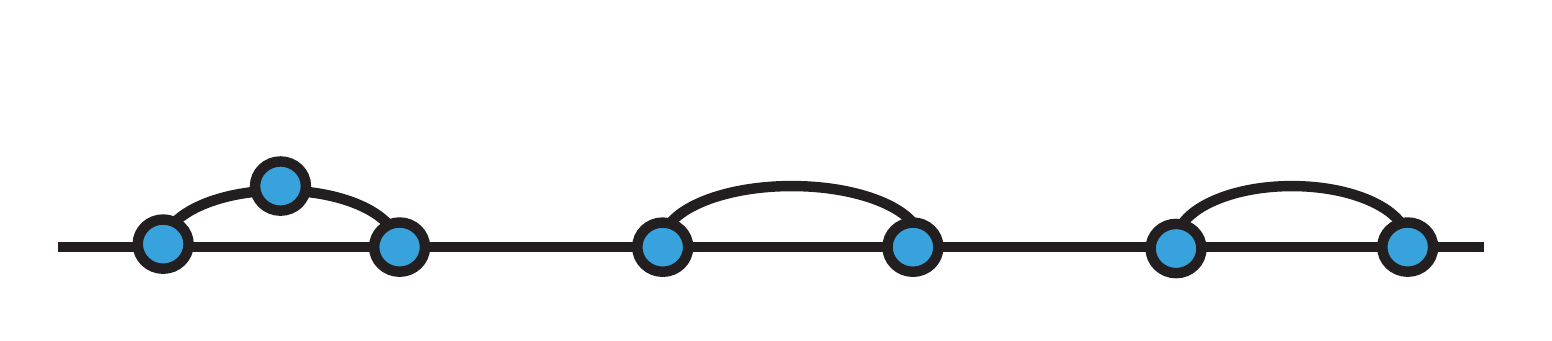}}&0&0&0&0\\
\resizebox{3.5cm}{.8cm}{\includegraphics{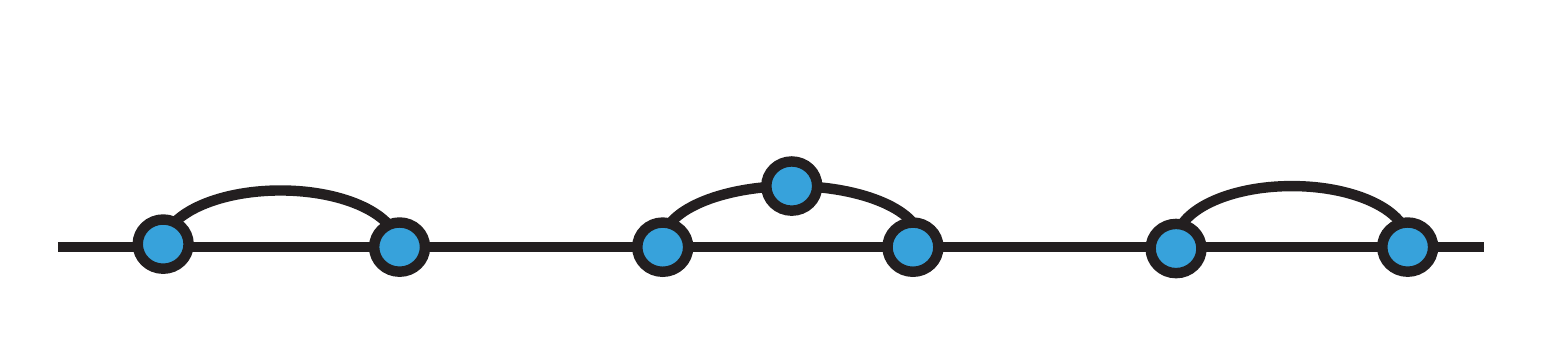}}&0&$n$&$n^2$&$n$\\
\resizebox{3.5cm}{.8cm}{\includegraphics{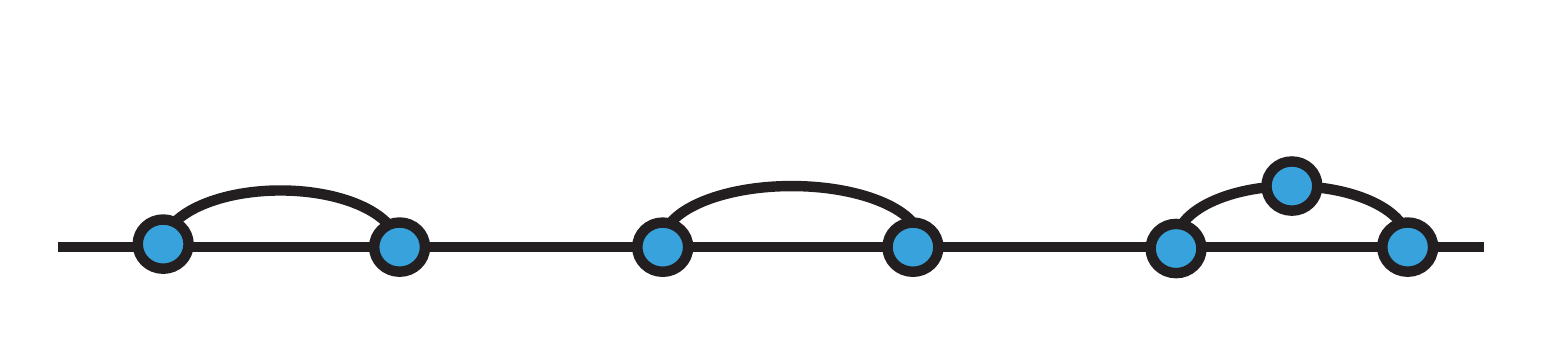}}&$n^2$&$n$&0&$n$\\
\resizebox{3.5cm}{.8cm}{\includegraphics{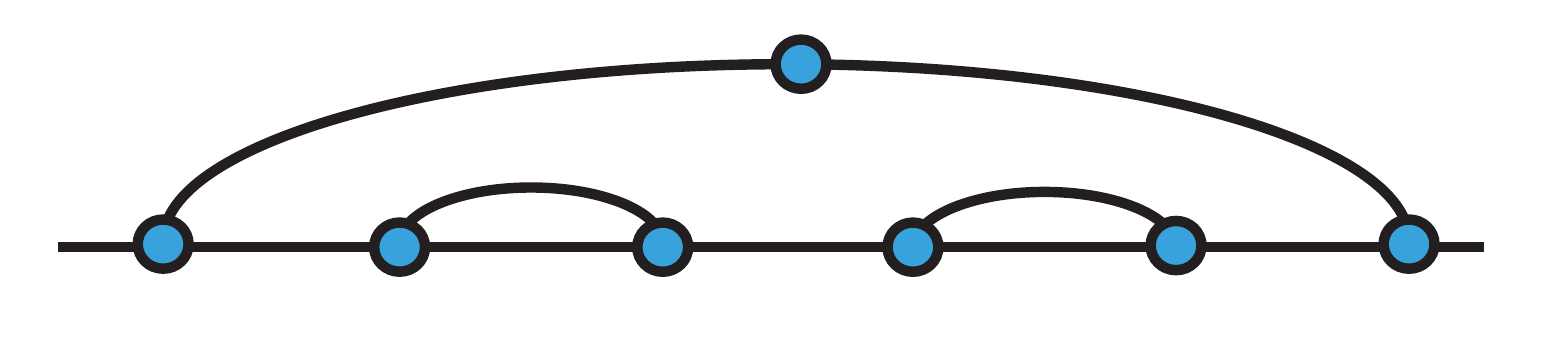}}&0&1&1&1\\
\resizebox{3.5cm}{.8cm}{\includegraphics{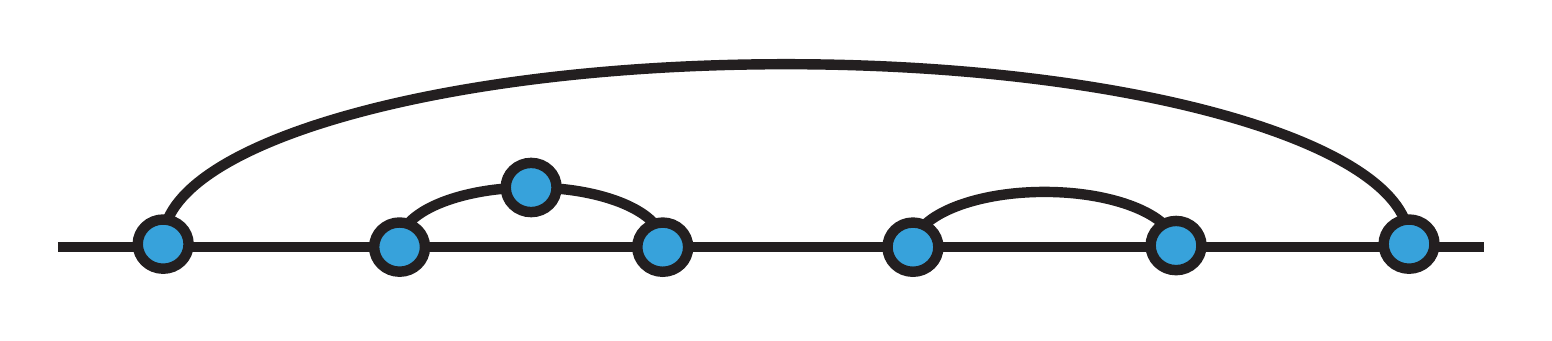}}&1&$n$&1&0\\
\resizebox{3.5cm}{.8cm}{\includegraphics{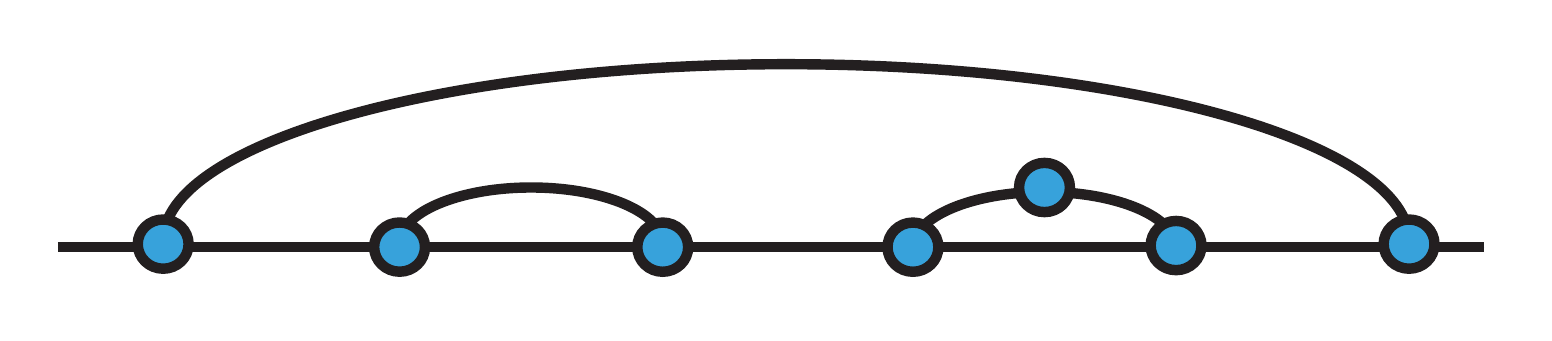}}&1&0&1&0\\
\resizebox{3.5cm}{.8cm}{\includegraphics{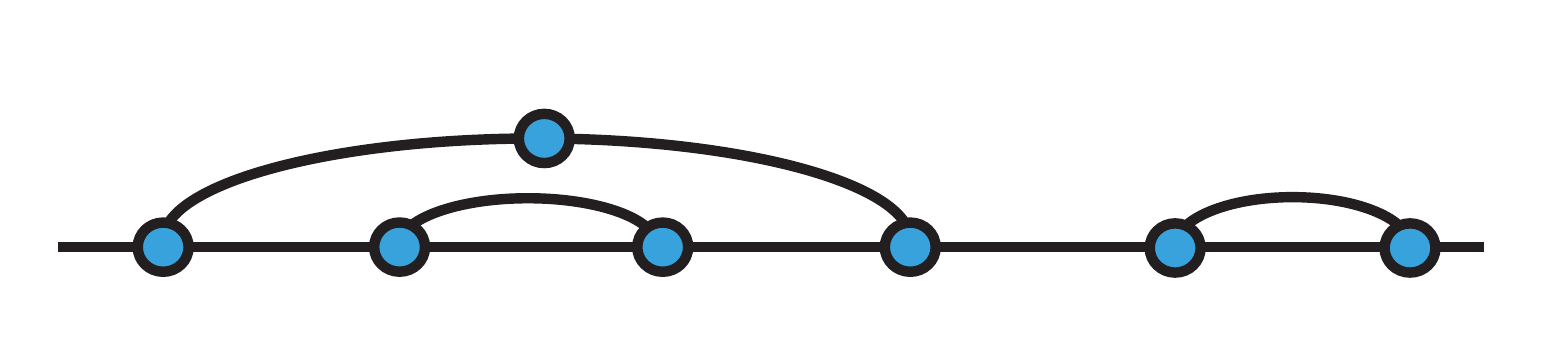}}&0&$n$&$n$&1\\
\resizebox{3.5cm}{.8cm}{\includegraphics{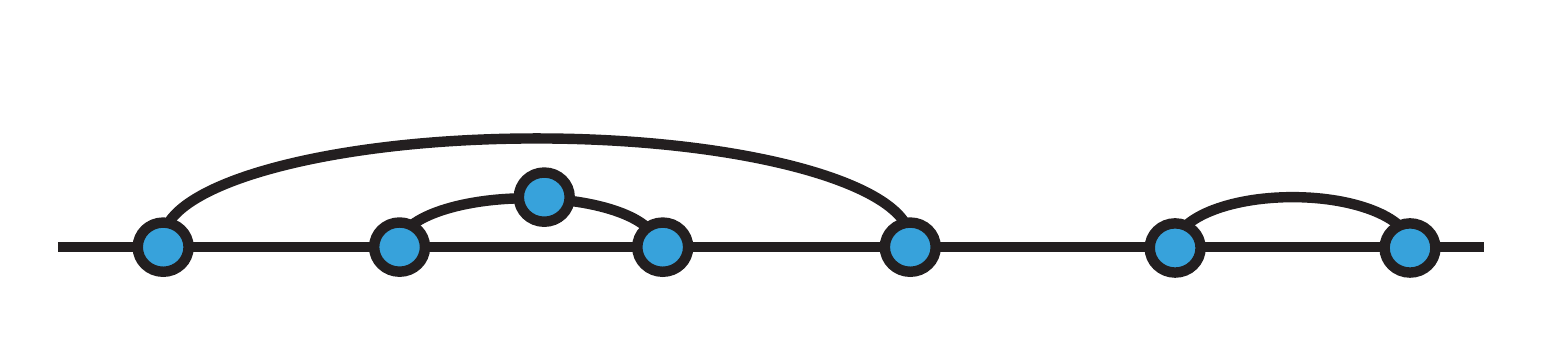}}&0&1&$n$&1\\
\resizebox{3.5cm}{.8cm}{\includegraphics{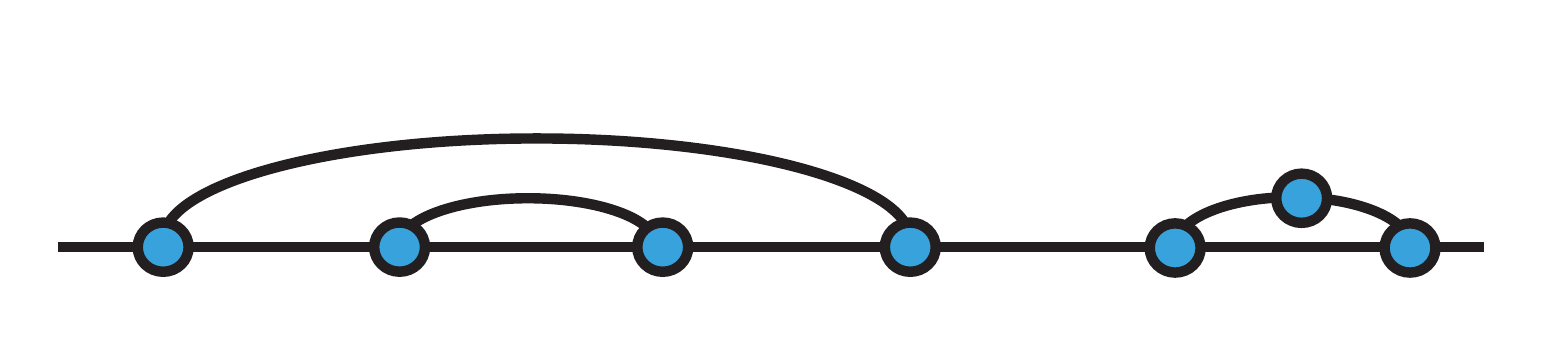}}&$n$&1&0&1\\
\resizebox{3.5cm}{.8cm}{\includegraphics{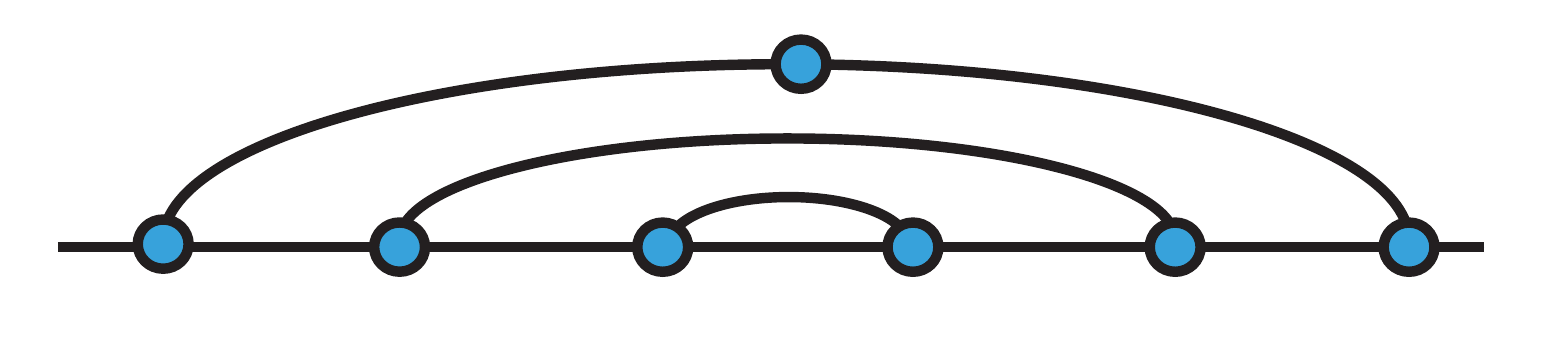}}&$n$&1&0&1\\
\resizebox{3.5cm}{.8cm}{\includegraphics{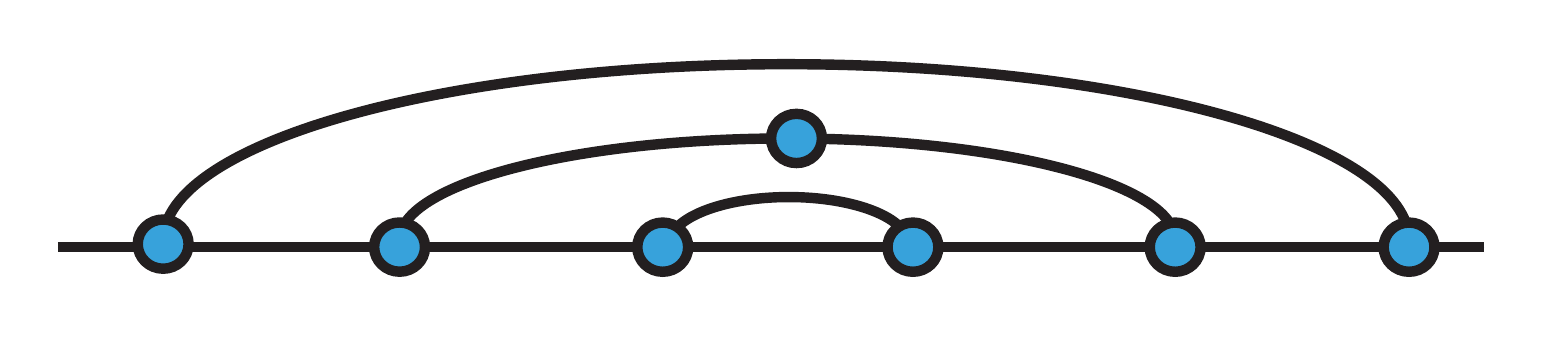}}&$n$&1&0&1\\
\resizebox{3.5cm}{.8cm}{\includegraphics{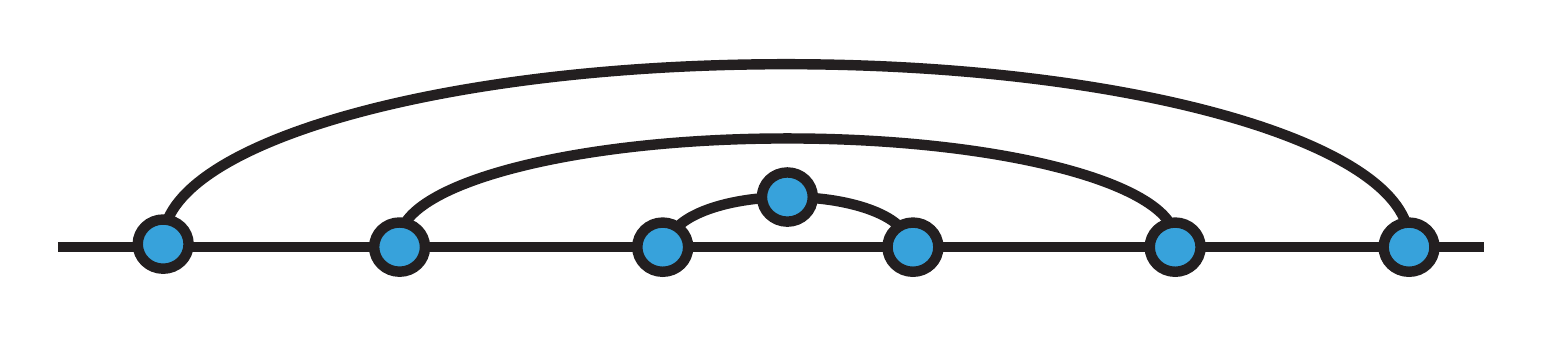}}&0&1&$n$&1\\
\resizebox{3.5cm}{.8cm}{\includegraphics{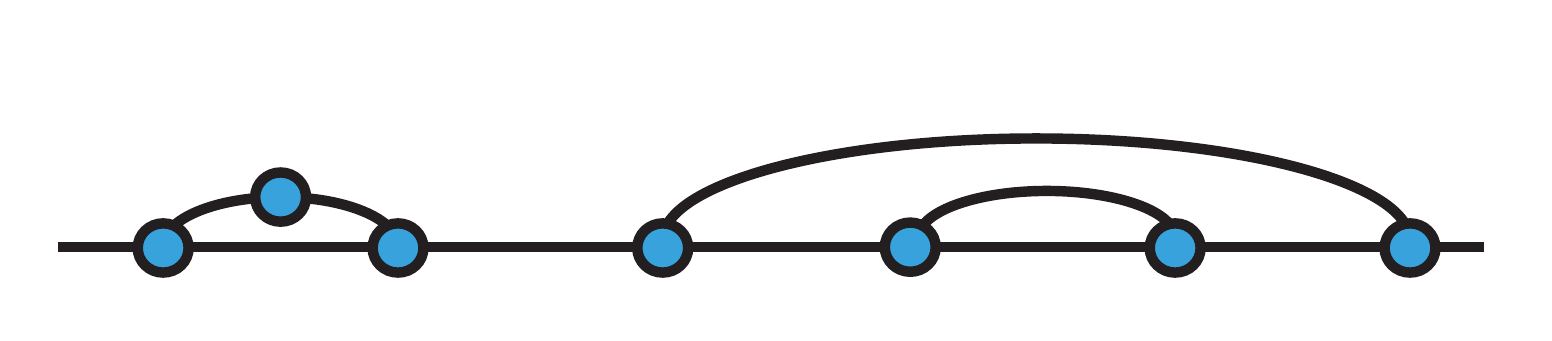}}&0&0&0&0\\
\resizebox{3.5cm}{.8cm}{\includegraphics{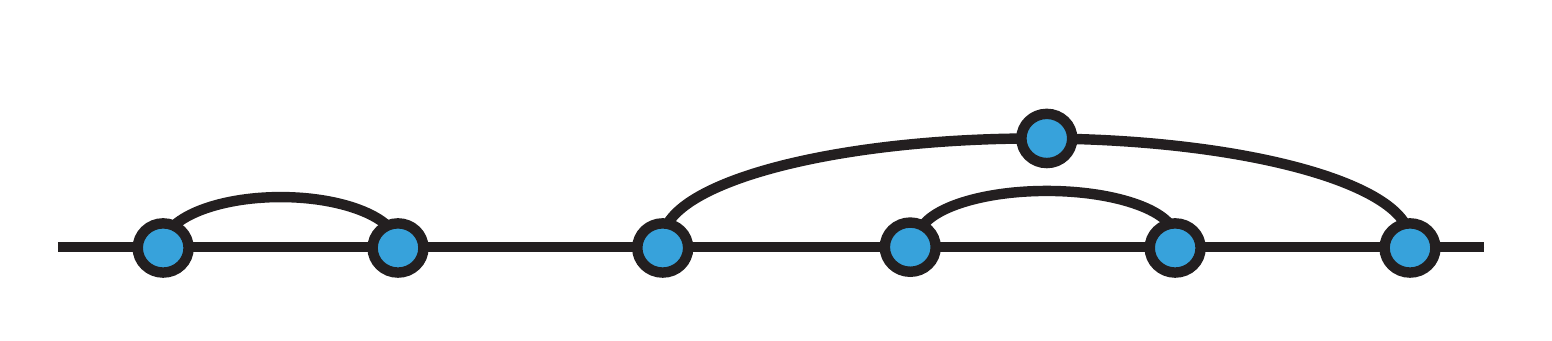}}&$n$&$n^2$&$n$&0\\
\resizebox{3.5cm}{.8cm}{\includegraphics{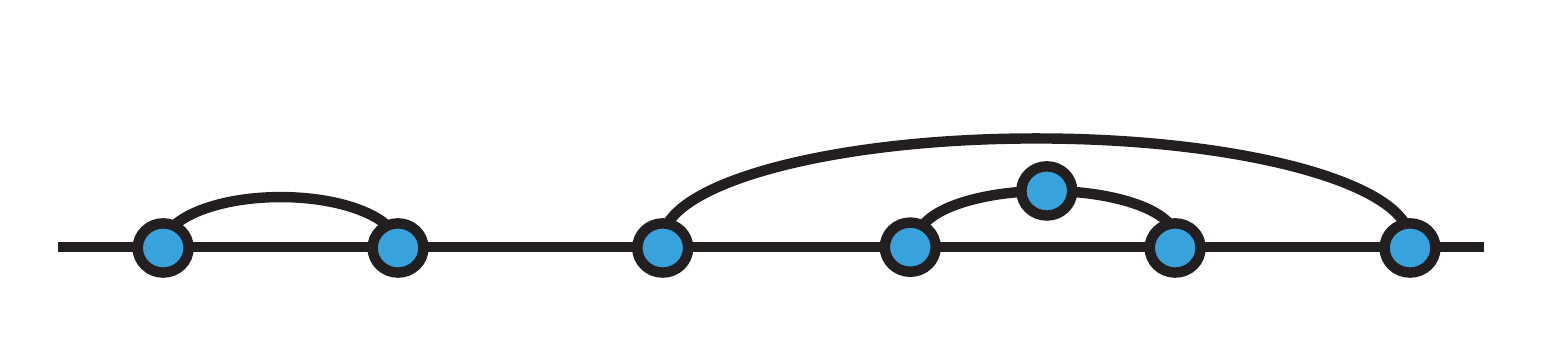}}&$n$&0&$n$&$n^2$\\
\end{tabular}
\caption{The $M\times\mathcal{K}_{kl;mn}$ (columns) may be decomposed into linear combinations of the fifteen one-pinch-point densities (rows).  (Relative) coefficients for four of the fifteen decompositions are shown in this table.}
\label{Coefficients}
\end{table}

In order to isolate all fifteen weights, we pursue our second strategy of splicing charge-neutral pairs into simpler correlation functions.  We begin with the one-pinch-point weight $\Pi_{12}(x_i,x_j;z,\bar{z})$, given by $\langle\psi_1(x_i)_{[0]}\psi_1(x_j)\Psi_1(z)_{[2]}\Psi_1(\bar{z})\rangle$.  We insert into its chiral representation a first charge-neutral collection $\int_{x_k}^{x_l}V_{1,2}^-(x_k)V_{1,2}^-(x_l)$ $V_-(u_1)\,du_1$, chosen so that $x_k$ and $x_l$ are not separated within the real axis by $x_i$ and $x_j$, and then we insert a second charge-neutral collection $\int_{x_m}^{x_n}V_{1,2}^-(x_m)V_{1,2}^-(x_n)V_-(u_2)\,du_2$, chosen so that $x_m$ and $x_n$ are not separated within the (one-point compactified) real axis by $x_i,x_j,x_k$, or $x_l$.  We find fifteen distinct conformal blocks, each with $n=m+1$ and $l=k+1$ (i.e., the inserted boundary arcs are not nested) or $l=k+3$ (i.e., the inserted boundary arcs are nested) (The case $n=7$ or $l=7,8,9$ is identified with $n=1$ or $l=1,2,3$ respectively).  Each block has the form
\be\langle\psi_1(x_i)_{[2]}\psi_1(x_j)\psi_1(x_k)_{[0]}\psi_1(x_l)\psi_1(x_m)_{[0]}\psi_1(x_n)\Psi_1(z)_{[2]}\Psi_1(\bar{z})\rangle=n^2M(x_1,\ldots,x_6;z,\bar{z})\,\mathcal{K}_{kl;mn}(x_1,\ldots,x_6;z,\bar{z}),\ee
where $\mathcal{K}_{kl;mn}$ is the real-valued integral
\begin{multline}\mathcal{K}_{kl;mn}:=\beta(-4/\kappa,-4/\kappa)^{-2}\sideset{}{_{x_k}^{x_l}}\int\sideset{}{_{x_m}^{x_n}}\int \,du_1\,du_2\,\,\mathcal{N}\Bigg[(u_1-z)^{8/\kappa-1}(u_1-\bar{z})^{8/\kappa-1}\\
\left.\times\,\,(u_2-z)^{8/\kappa-1}(u_2-\bar{z})^{8/\kappa-1}(u_2-u_1)^{8/\kappa}\prod_{i=1}^6(u_1-x_i)^{-4/\kappa}(u_2-x_i)^{-4/\kappa}\right],\end{multline}
and where $M$ is given by (\ref{chiralcorr}):
\be M(x_1,\ldots,x_6;z,\bar{z}):=|z-\bar{z}|^{\kappa/8+8/\kappa-2}\prod_{i<j}^6(x_j-x_i)^{2/\kappa}\prod_{i=1}^6|z-x_i|^{1-8/\kappa}.\ee
The $\mathcal{K}_{kl;mn}$ can be decomposed into linear combinations of the fifteen crossing weights in the usual way.  This decomposition is shown for four of the $\mathcal{K}_{kl;mn}$ in the top row of table \ref{Coefficients}, and the other eleven are found by cyclically permuting the indices.  We thus find an invertible system of fifteen equations with the fifteen weights as unknowns.  The formulas that follow from this inversion are complicated, and we leave their further investigation to the interested reader.

\subsection{Half-plane universal partition functions}\label{boundaryconditions}

In this section, we construct half-plane universal partition functions from the weights computed in section \ref{uhpppweights}.  We complete these calculations only for the cases $N=2$ and 3.  However, the method is clearly generalizable to polygons with more sides.  In this section, our figures and some of our language suggest that we have conformally mapped the upper half-plane onto the interior of a $2N$-sided polygon $\mathcal{P}$ with the vertices numbered counterclockwise in ascending order starting with the bottom-left vertex, with the $i$-th vertex $w_i$ the image of $x_i$, with $x_i<x_j$ whenever $i<j$, and with the bottom side of $\mathcal{P}$ sitting flush against the real axis.  This language streamlines the discussion of this section, and the implementation of this transformation is postponed to the next section \ref{transform}.

We begin with a background discussion.  In the present context, the Kac operator $\psi_1(x_i)$ bears two complementary interpretations.  First, it is a \emph{boundary one-leg operator} because it conditions the system so that a single boundary arc anchors to the point $x_i$ on the real axis.  As described in section \ref{cftdescription}, the $2N$ boundary one-leg operators in (\ref{halfplane}) generate $N$ non-crossing boundary arcs that cross the upper half-plane and join the $x_i$ pairwise in one of $C_N$ possible connectivities, with $C_N$ given by (\ref{catalan}).  If $\lambda$ is an event in which the boundary arcs join in a specified connectivity, then we call the (continuum) O$(n)$ model partition function summing exclusively over all samples in $\lambda$ and with boundary arcs having fugacity one the ``weight" $\Pi_\lambda$ of the event $\lambda$.  Because only closed loops should enjoy the full loop fugacity $n$ and the boundary arcs do not close into loops (yet), we have endowed them with fugacity one.  

Because the bulk loops have fugacity $n$ but the boundary arcs only have fugacity one, the weight $\Pi_\lambda$ is almost, but not quite, a physical continuum O$(n)$ partition function.  Indeed, we must account for how we physically condition the boundary arcs to anchor to the specified boundary points, and this matter will endow the boundary arcs with the proper fugacity $n$.  For instance, in the dilute $2\leq Q\leq4$ Potts model, boundary arcs are scaling limits of interfaces between clusters with all spins in, say, state $A$ and clusters with all spins in any state but $A$, so we condition a boundary arc to anchor to $x_i$ by changing the bc there from wired to spin $A$ to freely assuming all spins except $A$.   And in the dense $1\leq Q\leq4$ Potts model, boundary arcs are scaling limits of FK-cluster perimeters, so we condition a boundary arc to anchor to $x_i$ by changing the bc there from wired to spin $A$ to free.  Thus, we find a second interpretation of $\psi_1(x_i)$ as a \emph{bcc operator} because it changes the bc at $x_i$ from fixed to free or vice versa.  (In our application, $\psi_1$ will also sum over the $Q$ possible spin types for the adjacent fixed segment whenever we are working with the $Q$-state Potts model.)

\begin{figure}[t]
\centering
\includegraphics[scale=0.4]{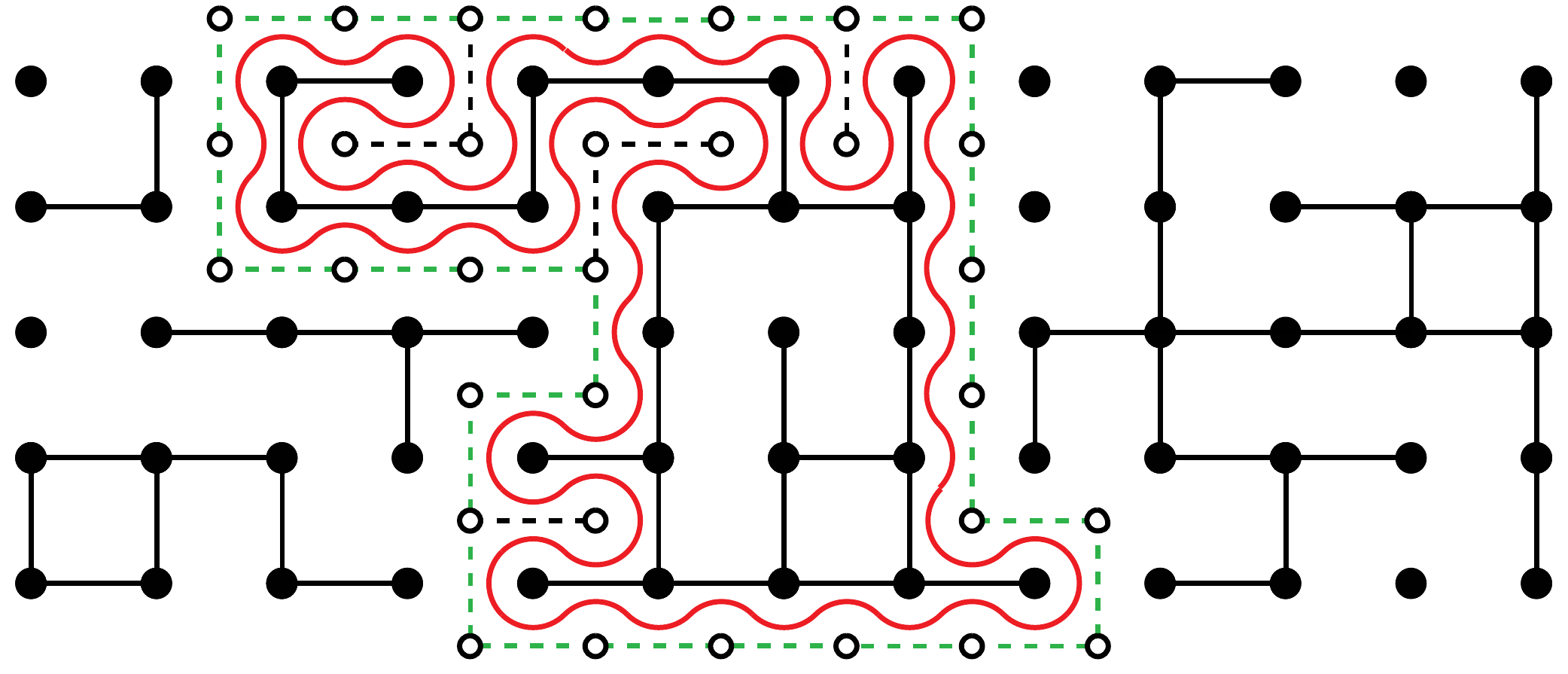}
\caption{A boundary loop (red) surrounding an FK boundary cluster. If all sites inside (resp. outside) of the red curve are (resp. are not) in spin state $A$, then the green loop of activated dual bonds is a boundary loop surrounding a spin boundary cluster.}
\label{FKboundaryloop}
\end{figure}

In our application, each vertex $w_i$ of the polygon (or in the present setting, its half-plane conformal image $x_i$) under consideration will host a free-to-fixed or fixed-to-free bcc.  This sets up an ffbc, of which there are many to consider.  For example, we can condition a pair of fixed sides to be \emph{independently wired,} that is, not constrained to exhibit the same state, or to be \emph{mutually wired,} that is, constrained to exhibit the same state.  By taking different combinations of these options, we generate many different ffbc events.  We denote a specified ffbc event by $\varsigma$.

Because its boundary arcs have fugacity one, the weight $\Pi_\lambda$ is independent of our choice of ffbc event $\varsigma$.  To promote $\Pi_\lambda$ into a true O$(n)$ partition function that sums exclusively over $\lambda\cap\varsigma$ and endows the boundary arcs with fugacity $n$, we close the boundary arcs into boundary loops by connecting the vertices of the polygon pairwise via the non-crossing exterior arcs mentioned in the discussion preceding (\ref{factoroutn}).  (In our present setting, the exterior arcs actually live in the lower half-plane.)  How this is done is described in \cite{skfz}, and we summarize the details for the rectangle and the hexagon.  We connect the endpoints of a fixed segment that is independently wired with a single exterior arc, and we connect each endpoint of every segment in a collection of mutually wired segments to another endpoint of another segment in that collection via an exterior arc.  The exterior arcs close the boundary arcs into $l_{\lambda,\varsigma}\in\{1,\ldots,N\}$ distinct boundary loops with each boundary loop contributing a fugacity factor of $n$ (figure \ref{FKboundaryloop}).  Thus, the universal partition function summing exclusively over the event $\lambda\cap\varsigma$ is $n^{l_{\lambda,\varsigma}}\Pi_\lambda$.  By summing over either all boundary arc connectivity events $\lambda$ or only those $\lambda$ with samples in the $s$-pinch-point event $\Lambda$, we find the respective half-plane universal partition functions
\be\label{sumitup}\Upsilon_\varsigma=\sum_\lambda n^{l_{\lambda,\varsigma}}\Pi_\lambda,\quad\Upsilon_{(\Lambda|\varsigma)}=\sum_{\lambda\,\cap\,\Lambda\neq\emptyset}n^{l_{\lambda,\varsigma}}\Pi_\Lambda.\ee
There are $C_N$ and $C_s$ terms in the left and right sum of (\ref{sumitup}) respectively, with $C_N$ and $C_s$ the $N$-th and $s$-th Catalan numbers respectively (\ref{catalan}).  By factoring the pinch-point weight $\Pi_\Lambda$ out of the right sum, we obtain (\ref{factoroutn}).

\begin{figure}[b]
\centering
\includegraphics[scale=0.27]{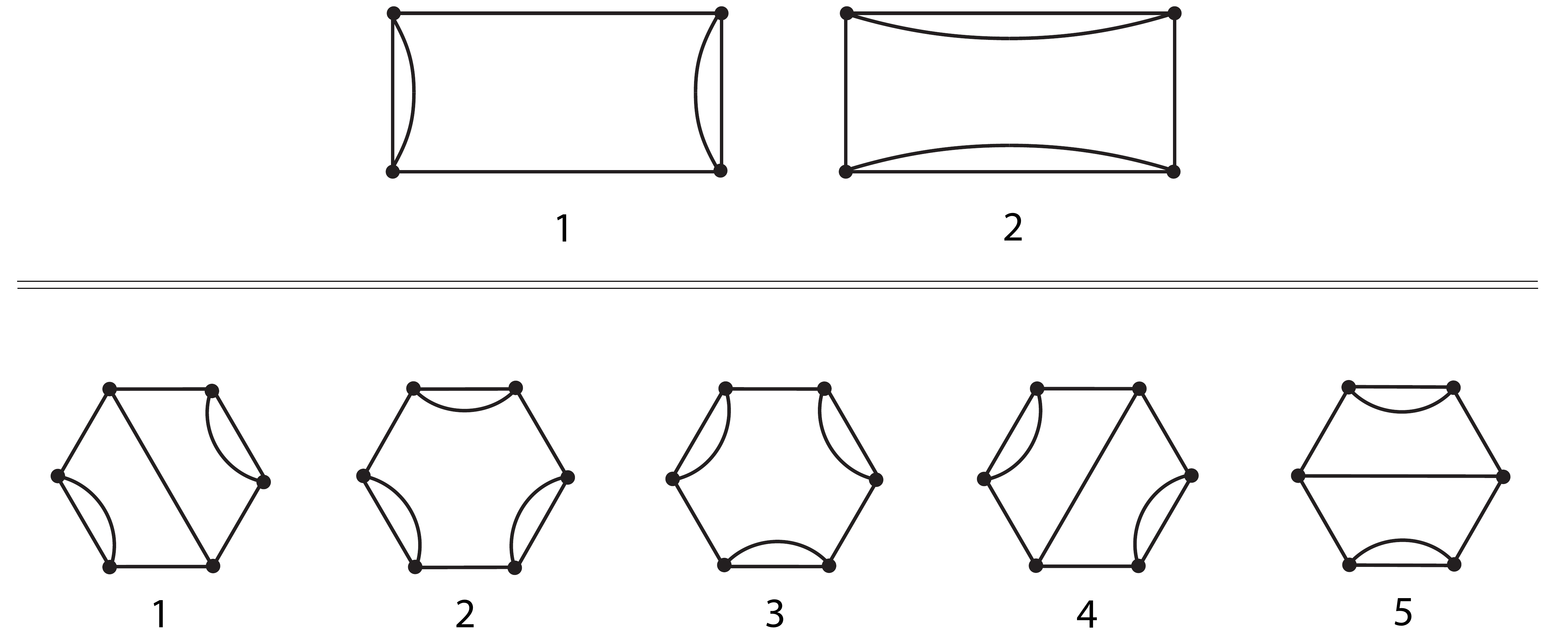}
\caption{An illustration of the labeling that we will use for the boundary arc connectivities of $\mathcal{R}$ and $\mathcal{H}$.}
\label{Xings}
\end{figure}

\begin{figure}[t]
\centering
\includegraphics[scale=0.3]{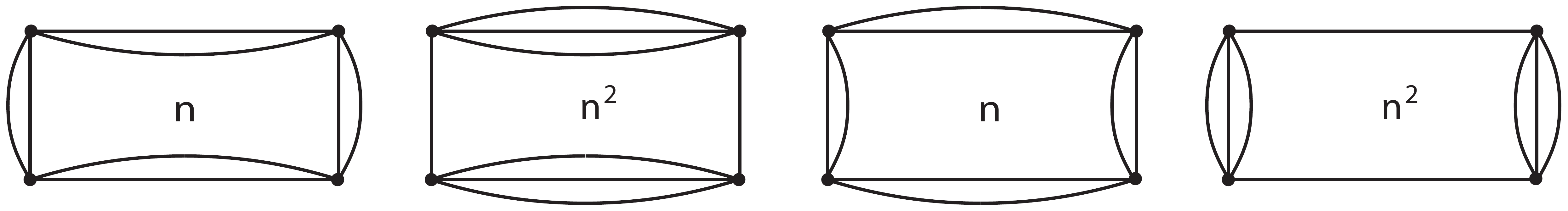}
\caption{The four boundary loop configurations for $\mathcal{R}$.  Each boundary loop contributes a factor of $n$.}
\label{RectBoundaryLoops}
\end{figure}

First, we sum (\ref{sumitup}) for pinch-point events in the rectangle $\mathcal{R}$ with the left/right sides wired.  When $N=2$, there are two possible boundary arc connectivity events, enumerated according to figure \ref{Xings}, and there are two possible ffbc events, also enumerated according to figure \ref{Xings} after reflecting the interior arcs outwards into exterior arcs.  Thus, $\varsigma_1$ is the independent wiring event and $\varsigma_2$ is the mutual wiring event.  Both ffbc events are possible in the dense phase of the Potts model while only the latter is possible in the dilute phase.  If $\Lambda$ is a specified one-pinch-point event, then the boundary arcs connect in exactly one way, so there is only one term in the sum (\ref{sumitup}).  Using figure \ref{RectBoundaryLoops}, we find that in the independent wiring ffbc event $\varsigma_1$,
\begin{align}\label{partitionind}\Upsilon_{(12:34|1)}&=n\Pi_{12:34},&\Upsilon_{(34:12|1)}&=n\Pi_{34:12},\\
\Upsilon_{(41:23|1)}&=n^{2}\Pi_{41:23},&\Upsilon_{(23:41|1)}&=n^{2}\Pi_{23:41},\end{align}
and we find that in the mutual wiring ffbc event $\varsigma_2$,
\begin{align}\label{partitionmut}\Upsilon_{(12:34|2)}&=n^{2}\Pi_{12:34},&\Upsilon_{(34:12|2)}&=n^{2}\Pi_{34:12},\\
\Upsilon_{(41:23|2)}&=n\Pi_{41:23},&\Upsilon_{(23:41|2)}&=n\Pi_{23:41},\end{align}
with the one-pinch-point half-plane weight $\Pi_{ij:kl}$ and the loop fugacity $n$ given in (\ref{1HRectweight}) and (\ref{fugacity}) respectively.  If  $\Lambda$ is the two-pinch-point event, then for the mutual wiring and independent wiring events, we find
\be\label{partition1234}\Upsilon_{(1234|1)}=\Upsilon_{(1234|2)}=(n+n^2)\Pi_{1234},\ee
with the two-pinch-point half-plane weight $\Pi_{1234}$ and the loop fugacity $n$ given in (\ref{2ppcovform}) and (\ref{fugacity}) respectively.

The half-plane universal partition function $\Upsilon_\varsigma$ summing exclusively over the ffbc event $\varsigma$ can be computed from (\ref{sumitup}) by using appropriate generalizations of Cardy's formula for horizontal and vertical crossings in the rectangle \cite{c3}.  The boundary arc connectivity events are labeled in figure \ref{Xings}, so with $\Upsilon_i:=\Upsilon_{\varsigma_i}$ and $\Pi_i:=\Pi_{\lambda_i}$, we find
\begin{align}\label{nobulk}\Upsilon_1&=n\Pi_1+n^2\Pi_2,&\Upsilon_2&=n^2\Pi_1+n\Pi_2,\end{align}
where the weights $\Pi_1$ and $\Pi_2$ are given by
\be\label{Pii}\Pi_i(x_1,\ldots,x_4)=\frac{\Gamma(12/\kappa-1)\Gamma(4/\kappa)}{\Gamma(8/\kappa)\Gamma(8/\kappa-1)}[(x_4-x_2)(x_3-x_1)]^{1-6/\kappa}G_i\left(\frac{(x_2-x_1)(x_4-x_3)}{(x_3-x_1)(x_4-x_2)}\right),\quad i=1,2,\ee
and where $G_1$ and $G_2$ are given by \cite{bbk}
\be\label{Pi1}G_1(\eta)=G_2(1-\eta)=\eta^{2/\kappa}(1-\eta)^{1-6/\kappa}\,_2F_1\,\left(\frac{4}{\kappa},1-\frac{4}{\kappa};\frac{8}{\kappa}\,\bigg|\,\eta\right),\ee
with $_2F_1$ the Gauss hypergeometric function.  

\begin{figure}[t]
\centering
\includegraphics[scale=0.27]{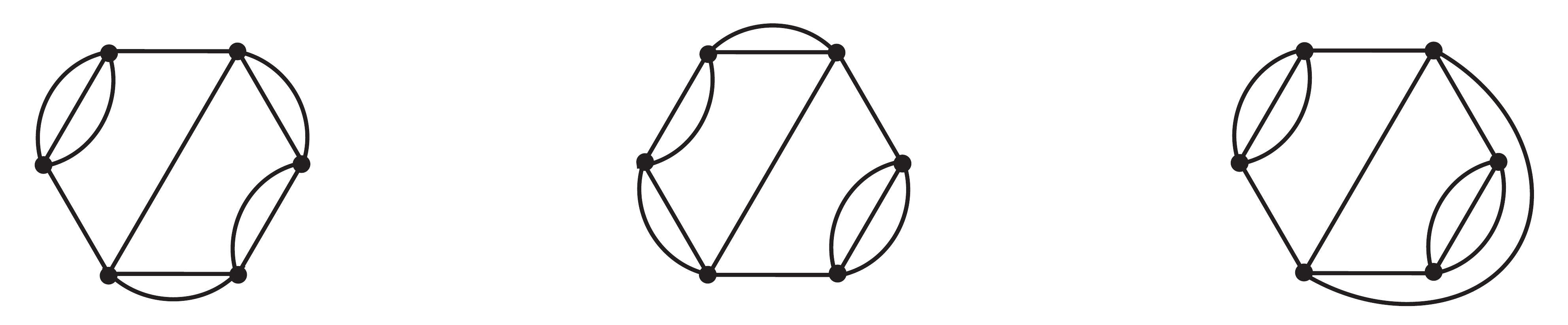}
\caption{Three possible exterior arc connectivities of $\mathcal{H}$ out of five possibilities.  Left to right, there are two, two, and three boundary loops, giving rise to boundary loop fugacity factors of $n^2,n^2$, and $n^3$ respectively.}
\label{HexBoundaryLoops}
\end{figure}

Next, we sum (\ref{sumitup}) for pinch-point events in the hexagon $\mathcal{H}$ with the bottom and top left/right sides wired.  When $N=3$, there are five possible ffbc events.  They are enumerated according to figure \ref{Xings}, so $\varsigma_3$ is the independent wiring event, $\varsigma_2$ is the mutual wiring event, and $\varsigma_1$ (resp.\,$\varsigma_4$, resp.\,$\varsigma_5$) is the event with the bottom and top-left (resp.\,bottom and top-right, resp.\,top-left and top-right) sides mutually wired and the remaining fixed side independently wired.  We call any of the events $\varsigma_1,\varsigma_4$, and $\varsigma_5$ a \emph{mixed ffbc}, and as usual, only the mutual wiring event is possible in the dilute phase.  If $\Lambda$ is a specified one-pinch-point event, then there is only one term in the sum (\ref{sumitup}).  With $p$ the number of boundary loops in each sample of $\Lambda\cap\varsigma$ (figure \ref{HexBoundaryLoops}), we find
\be\label{1Z}\Upsilon_{(ij:kl:mn|\varsigma)}=n^{p}\Pi_{ij:kl:mn},\quad p\in\{1,2,3\},\ee
with the loop fugacity $n$ given by (\ref{fugacity}) and with the half-plane weight $\Pi_{ij:kl:mn}$ given by inverting the system of equations that is partly shown in table \ref{Coefficients}.  
Next, if $\Lambda$ is a specified two-pinch-point event, then there are two terms in the sum (\ref{sumitup}) since the two boundary arcs touching at the two-pinch point can be separated into two nonintersecting boundary arcs in two ways.  The first and second way has $p_1$ and $p_2$ boundary loops respectively, with $\{p_1, p_2\}=\{1, 2\}$ or $\{2, 3\}$.  Therefore,
\be\label{2Z}\Upsilon_{(ijkl:mn|\varsigma)}=(n^{p_1}+n^{p_2})\Pi_{ijkl:mn},\quad\text{$\{p_1, p_2\}=\{1, 2\}$ or $\{2, 3\},$}\ee
with the two-pinch-point half-plane weight $\Pi_{ijkl:mn}$ and the loop fugacity $n$ given by (\ref{2pphexcov}) and (\ref{fugacity}) respectively. 
Finally, if $\Lambda$ is the three-pinch-point event, then there are five terms in the sum (\ref{sumitup}) since the three boundary arcs touching at the three-pinch point can be separated into any of the five possible connectivities shown in figure \ref{Xings}.  If $\varsigma$ is either the independent wiring ffbc $\varsigma_3$ or the mutual wiring ffbc $\varsigma_2$, then we have
\be\label{3Z}\Upsilon_{(123456|2)}=\Upsilon_{(123456|3)}=(n+3n^2+n^3)\Pi_{123456},\quad\text{(independent or mutual wiring ffbc)}\ee
and if $\varsigma$ is any of the mixed ffbcs $\varsigma_1,\varsigma_4$, or $\varsigma_5$, then we have
\be\label{3Zboth}\Upsilon_{(123456|1)}=\Upsilon_{(123456|4)}=\Upsilon_{(123456|5)}=(2n+2n^2+n^3)\Pi_{123456},\quad\text{(mixed ffbc)}\ee
with the three-pinch-point half-plane weight $\Pi_{123456}$ and the loop fugacity $n$ given in (\ref{3ppcovform}) and (\ref{fugacity}) respectively. 

There are many combinations of pinch-point events $\Lambda$ and ffbcs events $\varsigma$ for the hexagon.  We give explicit formulas for some type-$\Lambda$ pinch-point densities with the independent wiring event $\varsigma_3$.  Except for the three-pinch-point density, which is too rare to accurately measure, these results are verified via simulation in section \ref{simresults}.  First, the following combination sums exclusively over all samples with a specified one-pinch point on a boundary arc connecting vertex one with vertex two:
\be\label{1ppindwire}\Upsilon_{(12:34:56|3)}+n^2\Upsilon_{(12:36:45|3)}=n^3[\Pi_{12:34:56}+n\Pi_{12:36:45}],\quad\text{(all fixed sides independently wired).}\ee
The linear combination on the left side is chosen so that the right side equals $n^3$ times (\ref{2configs}).  Next, the universal partition function for a two-pinch point between two boundary arcs that connect vertices $w_6,w_1,w_2,$ and $w_3$ of the hexagon is given by (\ref{2Z}).  We find 
\be\label{2ppindwire}\Upsilon_{(6123:45|3)}=(n+n^2)\Pi_{6123:45},\quad\text{(all fixed sides independently wired)}\ee
with $n$ and $\Pi_{6123:45}$ given in (\ref{fugacity}) and (\ref{2pphexcov}) respectively.  Finally, the three-pinch-point partition function $\Upsilon_{(123456|3)}$ is already given in (\ref{3Z}).

To finish, we give the half-plane universal partition function $\Upsilon_i:=\Upsilon_{\varsigma_i}$ summing exclusively over the ffbc event $\varsigma_i$.  This is done as follows \cite{fkz}.  If we label the points $x_1,\ldots, x_6$ by $a,b,c,d,e,$ and $f$ in any way with $x_6=e$ or $f$, then the ffbc event $\varsigma_i$ with its exterior arcs pairing $x_1,\ldots,x_6$ into $\{a,b\},\{c,d\},\{e,f\}$ is given by
\begin{multline}\label{Z3Hex}\Upsilon_i(x_1,\ldots,x_6)=
n^3\beta(-4/\kappa,-4/\kappa)^{-2}\prod_{j<k}^5(x_k-x_j)^{2/\kappa}\prod_{j=1}^5(x_6-x_j)^{1-6/\kappa}\\
\times\int_a^b\int_c^d\mathcal{N}\left[\prod_{j=1}^5(u_1-x_j)^{-4/\kappa}(u_2-x_j)^{-4/\kappa}(u_1-x_6)^{12/\kappa-2}(u_2-x_6)^{12/\kappa-2}(u_2-u_1)^{8/\kappa}\right]\,du_1\,du_2.\end{multline}
As usual, $\mathcal{N}$ orders the differences in the integrand so that the integral is real.  If $\kappa\leq4$, then we replace the simple integration contours with Pochhammer contours that entwine the endpoints and we divide by an extra factor of $4\sin^2(4\pi/\kappa)$.

\subsection{Transforming the universal partition functions}\label{transform}

In this section, we transform the half-plane universal partition functions to universal partition functions for system in the appropriate $2N$-sided polygon.  

We let $Z_{(\Lambda|\varsigma)}$ (resp.\,$Z_\varsigma$) be the half-plane partition function summing exclusively over the pinch-point and ffbc event $\Lambda\cap\varsigma$ (resp.\,ffbc event $\varsigma$), where the $j$-th bcc occurs within a small distance $\epsilon_j$ from the specified point $x_j$ on the real axis in the ffbc event $\varsigma$.  Now, if we send the upper half-plane onto a simply connected domain $\mathcal{D}$ with a smooth boundary via a conformal bijection $f$, then conformal invariance says that
\begin{align} Z_{(\Lambda|\varsigma)}^\mathcal{D}(w_1,\ldots,w_{2N};w)&=Z_{(\Lambda|\varsigma)}(x_1,\ldots,x_{2N};z),& Z_\varsigma^\mathcal{D}(w_1,\ldots,w_{2N})&=Z_\varsigma(x_1,\ldots,x_{2N}),\end{align}
where $Z_{(\Lambda|\varsigma)}^\mathcal{D}$ (resp.\,$Z_\varsigma^\mathcal{D}$) is the partition function summing exclusively over the event  $\Lambda\cap\varsigma$ (resp.\,$\varsigma$) in the image system, and where $w_j=f(x_j)\in\partial\mathcal{D}$ and $w=f(x)$.  As the half-plane partition functions $Z_{(\Lambda|\varsigma)}$ and $Z_\varsigma$ have the respective asymptotic behaviors
\begin{align} Z_{(\Lambda|\varsigma)}/Z_f&\underset{\epsilon_j,\epsilon\rightarrow0}{\sim}C_s^2c_1^{2N}\epsilon_1^{\theta_1}\ldots\epsilon_{2N}^{\theta_1}\epsilon^{2\Theta_s}\Upsilon_{(\Lambda|\varsigma)},& Z_\varsigma/Z_f&\underset{\epsilon_j\rightarrow0}{\sim}C_s^2c_1^{2N}\epsilon_1^{\theta_1}\ldots\epsilon_{2N}^{\theta_1}\Upsilon_\varsigma,\end{align}
the image partition functions $Z_{(\Lambda|\varsigma)}^\mathcal{D}$ and $Z_\varsigma^\mathcal{D}$ have the respective asymptotic behaviors
\begin{align} Z_{(\Lambda|\varsigma)}^\mathcal{D}/Z_f&\underset{\delta_j,\delta\rightarrow0}{\sim}C_s^2c_1^{2N}\delta_1(w_1)^{\theta_1}\ldots\delta_{2N}(w_{2N})^{\theta_1}\delta(w)^{2\Theta_s}\Upsilon_{(\Lambda|\varsigma)}^\mathcal{D},& Z_\varsigma^\mathcal{D}/Z_f\underset{\delta_j\rightarrow0}{\sim}c_1^{2N}\delta_1(w_1)^{\theta_1}\ldots\delta_{2N}(w_{2N})^{\theta_1}\Upsilon_\varsigma^\mathcal{D},\end{align}
with $\delta_j(w_j)=\epsilon_j|\partial f(x_j)|,$ with $\delta(w)=\epsilon|\partial f(x)|$, and with $\Upsilon_{(\Lambda|\varsigma)}^\mathcal{D}$ and $\Upsilon_\varsigma^\mathcal{D}$ respectively given by the usual conformal covariance transformation laws
\begin{align}\label{covarrule2}\Upsilon_{(\Lambda|\varsigma)}^\mathcal{D}&:=|\partial f(x_1)|^{-\theta_1}\ldots|\partial f(x_{2N})|^{-\theta_1}|\partial f(z)|^{-2\Theta_s}\Upsilon_{(\Lambda|\varsigma)},& \Upsilon_\varsigma^\mathcal{D}&:=|\partial f(x_1)|^{-\theta_1}\ldots|\partial f(x_{2N})|^{-\theta_1}\Upsilon_\varsigma.\end{align}
Although it is conformally invariant, the type-$\Lambda$ pinch-point density given by the ratio $Z_{(\Lambda|\varsigma)}^\mathcal{D}/Z_\varsigma^\mathcal{D}$ is unnatural because the radii of the disks containing either the pinch-point or the bccs varies with their locations in the closure of $\mathcal{D}$.  For this reason, it is natural to replace $\delta(w)$ and each $\delta_j(w_j)$ with small numbers $\delta$ and $\delta_j$ that are independent of $w$ and $w_j$ respectively.  The type-$\Lambda$ pinch-point density that results is conformally covariant instead of conformally invariant.

Now we let $\mathcal{D}$ be an equiangular $2N$-sided polygon $\mathcal{P}$ with vertices at $w_1,\ldots,w_{2N}$.  This is the standard setting for this article.  To replace $\delta(w)$ with $\delta$ as prescribed above is valid as long as $w$ is sufficiently far from the vertices.  But to replace the other $\delta_j(w_j)$ with $\delta_j$ is not valid since the derivative of $f$ blows up at each vertex.  Instead, we replace
\begin{align}\label{redefpartf}Z_{(\Lambda|\varsigma)}^\mathcal{P}&\quad\longrightarrow\quad Z_{(\Lambda|\varsigma)}^\mathcal{P}\underset{\delta_j,\delta\rightarrow0}{\sim}C_s^2c_1^{2N}\delta_1^{\theta_1}\ldots\delta_{2N}^{\theta_1}\delta^{2\Theta_s}\Upsilon_{(\Lambda|\varsigma)}^\mathcal{P},& Z_\varsigma^\mathcal{P}&\quad\longrightarrow\quad Z_\varsigma^\mathcal{P}\underset{\delta_j\rightarrow0}{\sim}c_1^{2N}\delta_1^{\theta_1}\ldots\delta_{2N}^{\theta_1}\delta^{2\Theta_s}\Upsilon_\varsigma^\mathcal{P},\end{align}
where $\Upsilon_{(\Lambda|\varsigma)}^\mathcal{P}$ and $\Upsilon_\varsigma^\mathcal{P}$ are the respective correlation functions 
\begin{align}\Upsilon_{(\Lambda|\varsigma)}^\mathcal{P}&=\langle\psi_1^c(w_1)\ldots\psi_1^c(w_{2N})\Psi_s(w,\bar{w})\rangle_\mathcal{P},&\Upsilon_\varsigma^\mathcal{P}&=\langle\psi_1^c(w_1)\ldots\psi_1^c(w_{2N})\rangle_\mathcal{P},\end{align}
that use corner one-leg operators in place of boundary one-leg operators at the vertices of $\mathcal{P}$.  We define the corner one-leg operator by
\be\label{corner}\psi_1^c(w_j)=\lim_{\varepsilon_j\rightarrow0}\left(\frac{\phi}{\pi}\varepsilon_j^{1-\pi/\phi}\right)^{\theta_1}\psi_1(w_j+\varepsilon_je^{i\alpha}),\ee
where $\phi$ is the interior angle of $\mathcal{P}$ at the vertex $w_j$, and where $\varepsilon_je^{i\alpha}$ is a complex number (with $\varepsilon_j\in\mathbb{R}$) such that $w_j+\varepsilon_je^{i\alpha}$ is on a side of $\mathcal{P}$ and very close to $w_j$.  Because $\mathcal{P}$ is an equiangular $2N$-sided polygon, $\phi=(N-1)\pi/N$.  It is easy to check that if we use corner one-leg operators in place of ordinary boundary one-leg operators, then our redefined partition functions (\ref{redefpartf}) are finite.

\begin{figure}[t]
\centering
\includegraphics[scale=0.27]{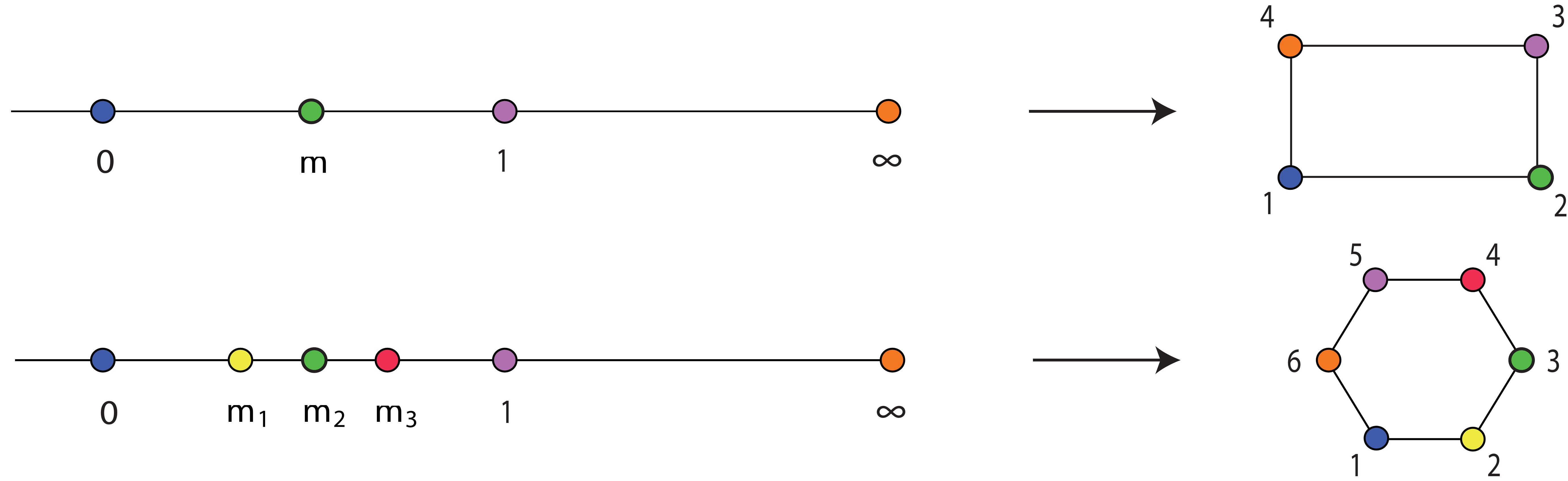}
\caption{The transformation of the upper half-plane to the interior of the rectangle and the hexagon and our enumeration of the vertices of either polygon.}
\label{SwchMapAlt}
\end{figure}

First, we transform the $N=2$ half-plane universal partition functions into universal partition functions for the system in the rectangle $\mathcal{R}$ (figure \ref{SwchMapAlt}).  The rectangle has aspect ratio $R$, vertices $w_1=0,w_2=R,w_3=R+i,$ and $w_4=i$, and interior angles $\phi=\pi/2$.  We therefore have
\begin{multline}\label{6ptR}\Upsilon_{(\Lambda|\varsigma)}^\mathcal{R}=\langle\psi_1^c(0)\psi_1^c(R)\psi_1^c(R+i)\psi_1^c(i)\Psi_s(z,\bar{z})\rangle_{\mathcal{R}}=\lim_{\varepsilon_j\rightarrow0}(16\varepsilon_1\varepsilon_2\varepsilon_3\varepsilon_4)^{-\theta_1}\\
\times\langle\psi_1(\varepsilon_1)\psi_1(R-\varepsilon_2)\psi_1(R+i-\varepsilon_3)\psi_1(i+\varepsilon_4)\Psi_s(z,\bar{z})\rangle_{\mathcal{R}}.\end{multline}
We use the Schwarz-Christoffel map that conformally sends the upper half-plane onto $\mathcal{R}$ and whose continuous extension to the real axis sends the points $\{0,m,1,\infty\}$ counterclockwise to the respective vertices $\{w_1,w_2,w_3,w_4\}$:
\be\label{trans} f(z)=\frac{1}{2K'(m)}\int_0^z\zeta^{-1/2}(m-\zeta)^{-1/2}(1-\zeta)^{-1/2}\,d\zeta,\quad f^{-1}(w)=m\,\text{sn}(w\,K'\,|\,m)^2.\ee
Here, $K(m)$ is the complete elliptic integral of the first kind, $K'(m):=K(1-m)$, and $m\in(0,1)$ is the modular parameter of the transformation, related one-to-one with $R\in(0,\infty)$ through $R=K(m)/K'(m).$  This transformation (and its continuous extension to the real axis) is invertible, so we define $x_j'$ to be the half-plane pre-image of the $j$-th vertex of $\mathcal{R}$ shifted by a small amount $\varepsilon_j$ as shown in (\ref{6ptR}).  These points are given in \cite{sk} to leading order, and they are respectively near zero, $m,$ one, and infinity.  Conformal covariance gives
\be\Upsilon^{\mathcal{R}}_{(\Lambda|\varsigma)}(m;w,\bar{w})=|\partial f(z)|^{-2\Theta_s}\lim_{\varepsilon_j\rightarrow0}(16\varepsilon_1\varepsilon_2\varepsilon_3\varepsilon_4)^{-\theta_1}\prod_{j=1}^4 |\partial f(x_j')|^{-\theta_1}\Upsilon_{(\Lambda|\varsigma)}(w_1',w_2',w_3',w_4';w,\bar{w}),\ee
with $\Upsilon_{(\Lambda|\varsigma)}$ given in section \ref{boundaryconditions}.  The right side may be computed explicitly, and we find (with $z=f^{-1}(w)$) that
\begin{multline}\label{2pprect}\Upsilon^{\mathcal{R}}_{(\Lambda|\varsigma)}(m;w,\bar{w})=|2mK'\text{sn}(wK'\,|\,m)\text{cn}(wK'\,|\,m)\text{dn}(wK'\,|\,m)|^{[16s^2-(\kappa-4)^2]/8\kappa}\\
\hspace{.5cm}\times[m(1-m)]^{6/\kappa-1}K'^{24/\kappa-4}\lim_{x\rightarrow\infty}x^{6/\kappa-1}\Upsilon_{(\Lambda|\varsigma)}(0,m,1,x;z,\bar{z}).\end{multline}
We note that $\eta\rightarrow m$ as $\varepsilon_1,\ldots,\varepsilon_4\rightarrow0$.  The universal partition function $\Upsilon_\varsigma^\mathcal{R}$ for $\mathcal{R}$ is found from (\ref{nobulk}) in the same way after dropping all covariance factors associated with the bulk point $w$.  The result is \cite{skfz}
\be\left\{\begin{array}{l}\Upsilon_1^\mathcal{R}(m) \\ \Upsilon_2^\mathcal{R}(m)\end{array}\right\}=n^{2}K'(m)^{24/\kappa-4}\left\{\begin{array}{l}F(1-m) \\ F(m)\end{array}\right\},\quad F(m):=\,_2F_1\left(2-\frac{12}{\kappa},1-\frac{4}{\kappa};2-\frac{8}{\kappa}\,\Bigg|\,m\right).\ee

Next, we transform the $N=3$ half-plane universal partition functions into universal partition functions for the system in the hexagon $\mathcal{H}$ with vertices $w_1=0,w_2>0$, $w_3,\ldots,w_6$ in the upper half-plane, and interior angles $\phi=2\pi/3$.  We have
\begin{multline}\label{8ptH}\Upsilon^{\mathcal{H}}_{(\Lambda|\varsigma)}=\langle\psi_1^c(w_1)\psi_1^c(w_2)\psi_1^c(w_3)\psi_1^c(w_4)\psi_1^c(w_5)\psi_1^c(w_6)\Psi_s(w,\bar{w})\rangle_{\mathcal{H}}\\
=\lim_{\varepsilon_j\rightarrow0}\left(\frac{729\sqrt{\varepsilon_1\varepsilon_2\varepsilon_3\varepsilon_4\varepsilon_5\varepsilon_6}}{64}\right)^{-\theta_1}\left\langle\prod_{j=1}^6\psi_1(w_j+\varepsilon_je^{(j-1)\pi i/3})\Psi_s(w,\bar{w})\right\rangle_{\mathcal{H}}.\end{multline}
We use the Schwarz-Christoffel map that sends the upper half-plane conformally onto $\mathcal{H}$ and whose continuous extension to the real axis sends the points $\{0,m_1,m_2,m_3,1,\infty\}$ clockwise to the respective vertices $\{w_1=0,w_2,\ldots,w_6\}$ of $\mathcal{H}$ (figure \ref{SwchMapAlt}):
\be\label{hextrans} f(w)=\frac{2}{3}\int_0^w\zeta^{-1/3}(m_1-\zeta)^{-1/3}(m_2-\zeta)^{-1/3}(m_3-\zeta)^{-1/3}(1-\zeta)^{-1/3}\,d\zeta.\ee
This transformation (and its extension to the real axis) is invertible, so we define the point $x_j'$ to be the half-plane pre-image of the $j$-th vertex shifted by $\varepsilon_je^{(j-1)\pi i/3}$.  To leading order in $\varepsilon_j$, these points are
\bea\label{firstuhex}x_1'&=&f^{-1}(w_1+\varepsilon_1e^{0\pi i/3})=\sqrt{m_1m_2m_3}\,\varepsilon_1^{3/2},\\
x_2'&=&f^{-1}(w_2+\varepsilon_2e^{1\pi i/3})=m_1+\sqrt{m_1(m_2-m_1)(m_3-m_1)(1-m_1)}\,\varepsilon_2^{3/2},\\
x_3'&=&f^{-1}(w_3+\varepsilon_3e^{2\pi i/3})=m_2+\sqrt{m_2(m_2-m_1)(m_3-m_2)(1-m_2)}\,\varepsilon_3^{3/2},\\
x_4'&=&f^{-1}(w_4+\varepsilon_4e^{3\pi i/3})=m_3+\sqrt{m_3(m_3-m_1)(m_3-m_2)(1-m_3)}\,\varepsilon_4^{3/2},\\
x_5'&=&f^{-1}(w_5+\varepsilon_3e^{4\pi i/3})=1+\sqrt{(1-m_1)(1-m_2)(1-m_3)}\,\varepsilon_5^{3/2},\\
\label{lastuhex}x_6'&=&f^{-1}(w_6+\varepsilon_3e^{5\pi i/3})=-\varepsilon_6^{-3/2},\eea 
and they are respectively near zero, $m_1,m_2,m_3$, one, and infinity.  Conformal covariance gives 
\begin{multline}\Upsilon^{\mathcal{H}}_{(\Lambda|\varsigma)}(m_1,m_2,m_3;w,\bar{w})=|\partial f(z)|^{-2\Theta_s}\lim_{\varepsilon_j\rightarrow0}\left(\frac{729\sqrt{\varepsilon_1\varepsilon_2\varepsilon_3\varepsilon_4\varepsilon_5\varepsilon_6}}{64}\right)^{-\theta_1}\\
\times\prod_{j=1}^6|\partial f(x_j')|^{-\theta_1}\Upsilon_{(\Lambda|\varsigma)}(w_1',w_2',w_3',w_4',w_5',w_6';w,\bar{w}),\end{multline}
with $\Upsilon_{(\Lambda|\varsigma)}$ given in section \ref{boundaryconditions}.  The right side may be computed explicitly, and we find (with $z=f^{-1}(w)$) that
\begin{multline}\label{2pphex}\Upsilon^{\mathcal{H}}_{(\Lambda|\varsigma)}(m_1,m_2,m_3;w,\bar{w})=|27z(m_1-z)(m_2-z)(m_3-z)(1-z)/8|^{[16s^2-(\kappa-4)^2]/24\kappa}\\
\begin{aligned}\times&[m_1m_2m_3(m_2-m_1)(m_3-m_1)(m_3-m_2)(1-m_1)(1-m_2)(1-m_3)]^{(6-\kappa)/2\kappa}\\
\times&\lim_{x\rightarrow\infty}x^{6/\kappa-1}\Upsilon_{(\Lambda|\varsigma)}(0,m_1,m_2,m_3,1,x;z,\bar{z}).\end{aligned}\end{multline}
We note that $(\eta,\tau,\sigma)\rightarrow(m_1,m_2,m_3)$ as $\varepsilon_1,\ldots,\varepsilon_6\rightarrow0$.
The universal partition function $\Upsilon_\varsigma^\mathcal{H}$ for $\mathcal{H}$ is found from (\ref{Z3Hex}) in the same way after dropping all covariance factors associated with the bulk point $w$.  The result is
\begin{multline}\Upsilon_\varsigma^\mathcal{H}(m_1,m_2,m_3)=[m_1m_2m_3(m_2-m_1)(m_3-m_1)(m_3-m_2)(1-m_1)(1-m_2)(1-m_3)]^{(10-\kappa)/2\kappa}\\
\times n^3\beta(-4/\kappa,-4/\kappa)^{-2}\sideset{}{_a^b}\int\sideset{}{_c^d}\int\mathcal{N}\left[(u_2-u_1)^{8/\kappa}\prod_{i=1}^2u_i^{-4/\kappa}(1-u_i)^{-4/\kappa}\prod_{j=1}^3(m_j-u_i)^{-4/\kappa}\right]\,du_2\,du_1,\end{multline}
where $\{a,b\}$ is the pair of finite endpoints for one of the two exterior arcs and $\{c,d\}$ is that for the other exterior arc.  These endpoints are among $\{0,m_1,m_2,m_3,1\}$.  In the event of the ffbc with all fixed sides independently wired ($\varsigma_3$) (\ref{Z3Hex}), we have $a=0,b=m_1,c=m_2,$ and $d=m_3$. 

\subsection{Pinch-point densities in polygons}\label{ppdens}

In this section, we give explicit expressions for some of these densities for the rectangle $\mathcal{R}$ $(N=2$) and for the hexagon $\mathcal{H}$ $(N=3$).  The behavior of the type-$\Lambda$ pinch-point density $\rho_{(\Lambda|\varsigma)}^\mathcal{P}$ in the polygon $\mathcal{P}$ is given by (\ref{asyratio}) with the bulk $s$-leg exponent $\Theta_s$ given by (\ref{weights}) and with the universal partition functions $\Upsilon_{(\Lambda|\varsigma)}^\mathcal{P}$ and $\Upsilon_\varsigma$ computed in section \ref{transform} for the rectangle and the hexagon.

\begin{figure}[b]
 \centering
\includegraphics[scale=0.35]{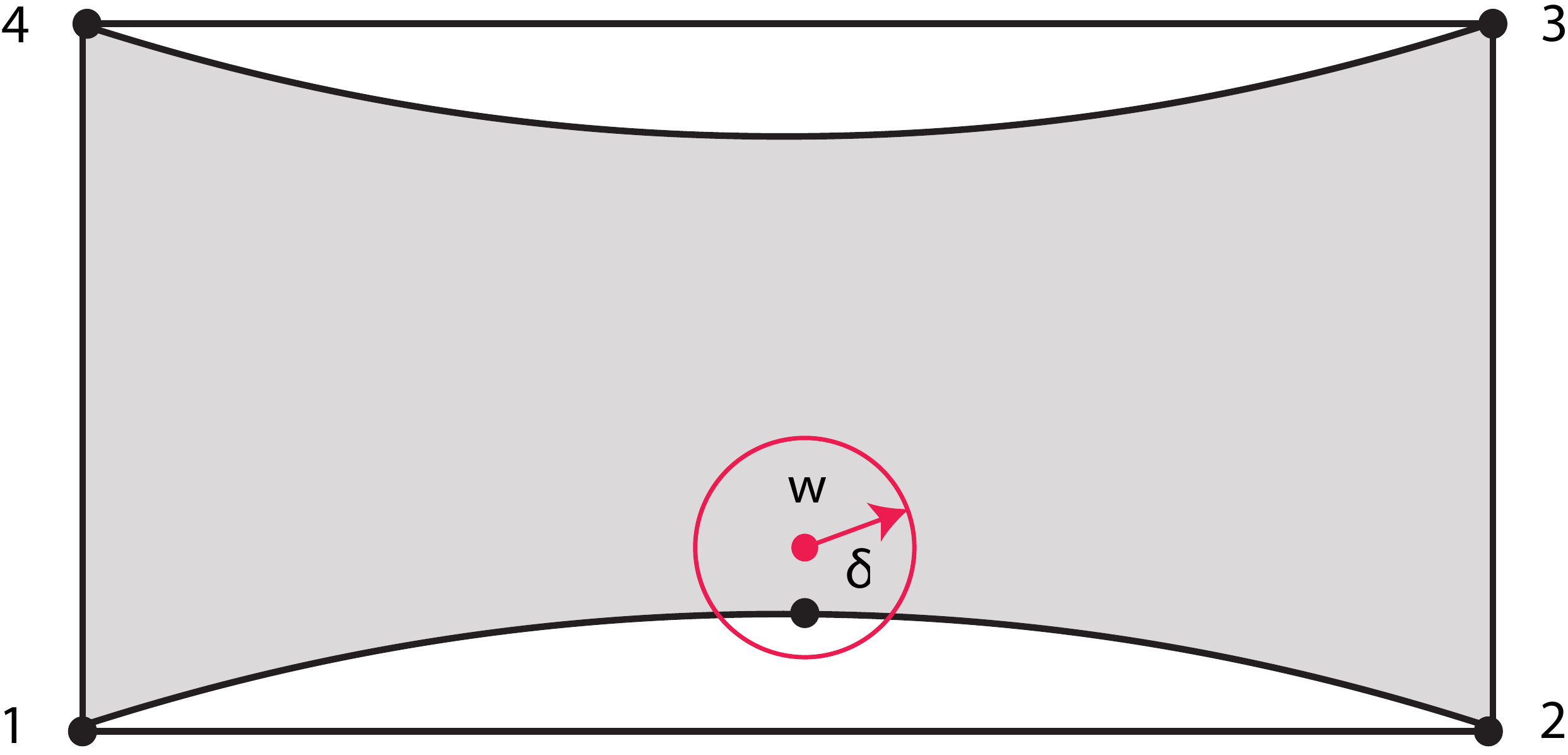}
\caption{Illustration of the type-$(12:34)$ pinch-point configuration in the rectangle.  The boundary cluster is shaded gray.}
\label{Rect1DensityConfig}
\end{figure}
\subsubsection{A one-pinch-point density for the rectangle}
\begin{figure}[h!]
 \centering
\includegraphics[scale=0.35]{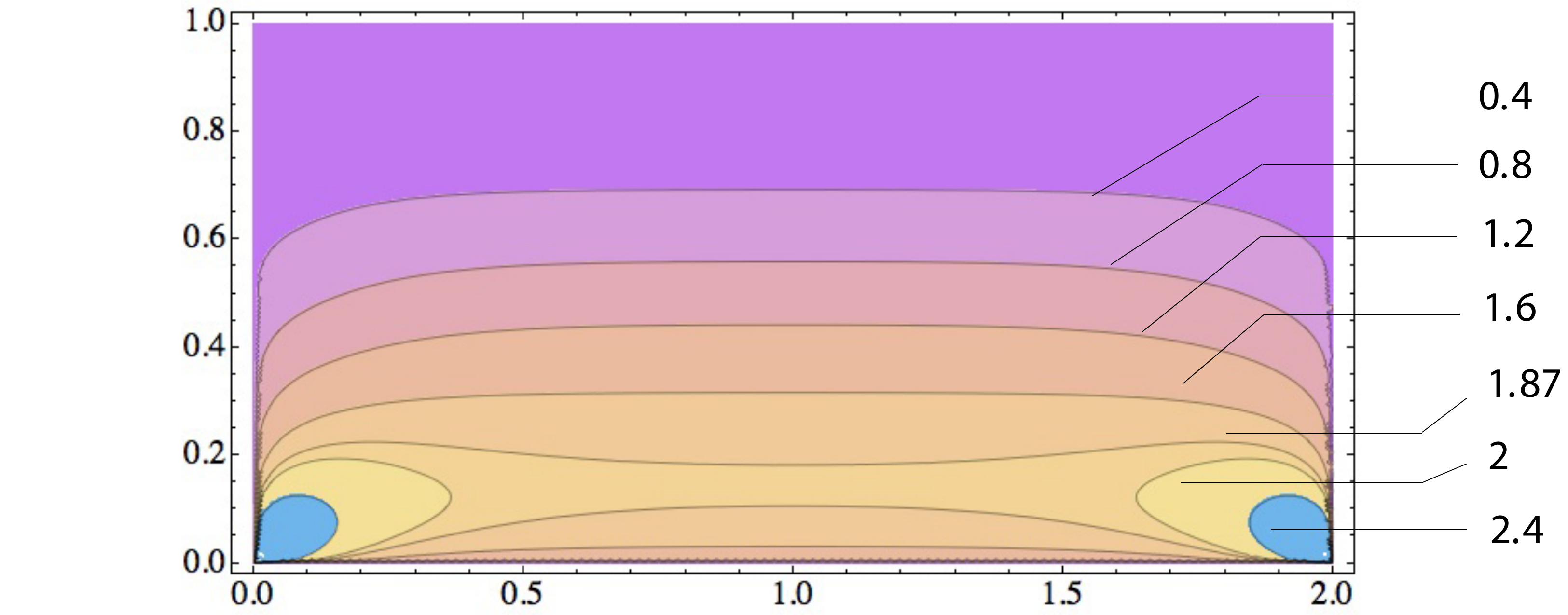}
\caption{Contour plot of Ising FK cluster $(\kappa=16/3)$ one-pinch-point density $\rho^\mathcal{R}_{(12:34|1)}$.}
\label{Rect1DensityPlots}
\end{figure}

The density of type-$(12:34)$ one-pinch points in the rectangle $\mathcal{R}$ with the left/right sides independently wired behaves as (figure \ref{Rect1DensityConfig})
\begin{multline}\label{1;12:34}\rho_{(12:34|1)}^\mathcal{R}(m;w,\bar{w})\underset{\delta\rightarrow0}{\sim}C_1^2\delta^{1-\kappa/8}\left[\frac{(1-m)^{6/\kappa-1}}{_2F_1(2-\frac{12}{\kappa},1-\frac{4}{\kappa};2-\frac{8}{\kappa}\,|\,1-m)}\right]\\
\times\left[\frac{\text{Im}[\,\text{sn}(w\,K'\,|\,m)^2]}{K'|\text{sn}(wK'\,|\,m)\text{cn}(wK'\,|\,m)\text{dn}(wK'\,|\,m)|}\right]^{\kappa/8-1}\left[\frac{2G_2+(n^2-2)G_4-nG_1-nG_3}{n(n^2-4)}\right]\left(m,z,\bar{z}\right).\end{multline}
The $G_i$ are given in (\ref{firstG}-\ref{lastG}), and $n$ is given in (\ref{fugacity}).  The density $\rho_{(12:34|2)}^\mathcal{R}$ with the left/right sides mutually wired is found by replacing the argument of the hypergeometric function with $1-m$ in (\ref{1;12:34}) and multiplying by $n$.  Figure \ref{Rect1DensityPlots} shows a contour plot of $\rho_{(12:34|1)}^\mathcal{R}$ for Ising FK boundary clusters ($\kappa=16/3$).

\begin{figure}[b]
 \centering
\includegraphics[scale=0.35]{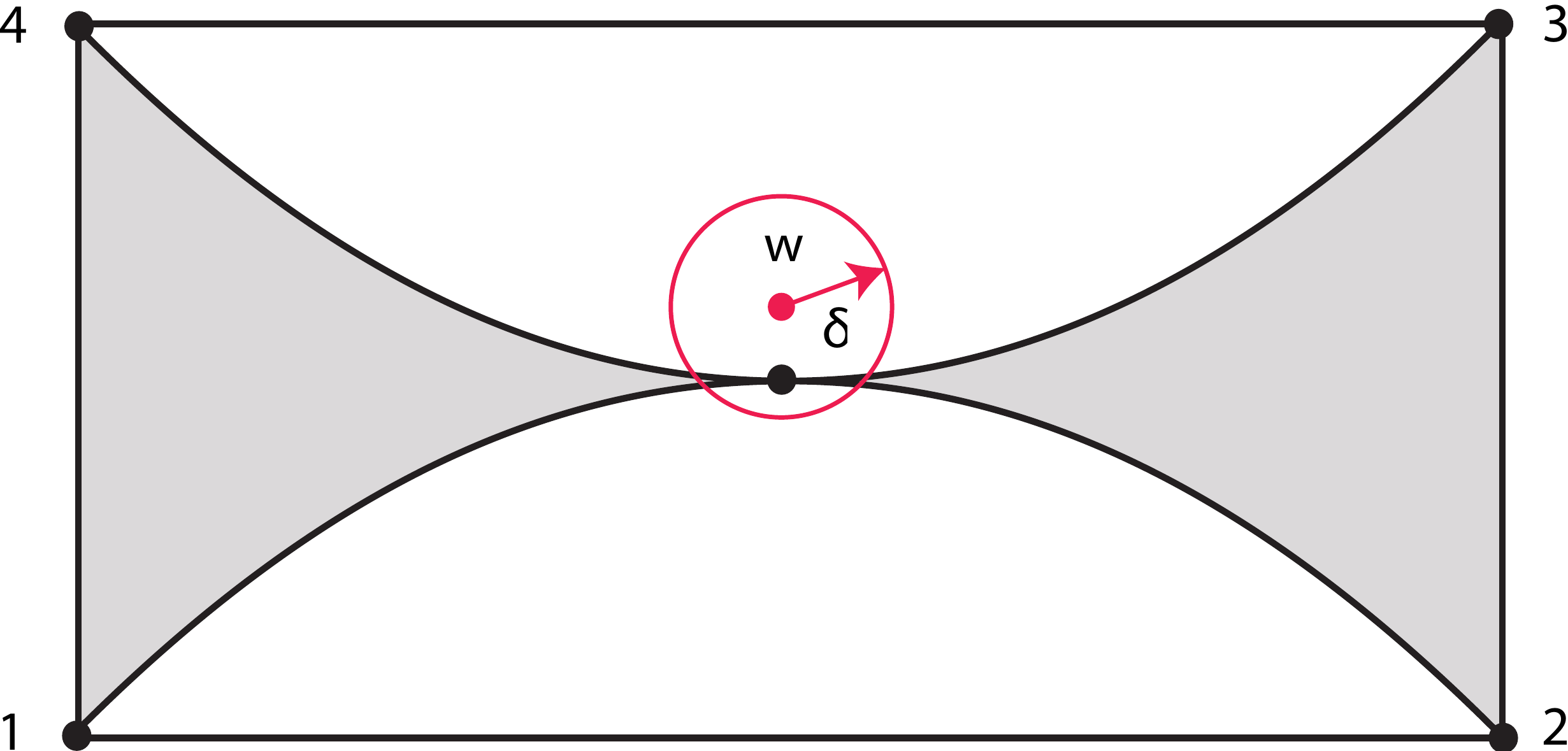}
\caption{Illustration of the two-pinch-point configuration in the rectangle.  Boundary clusters are shaded gray.}
\label{Rect2DensityConfig}
\end{figure}

\subsubsection{A two-pinch-point density for the rectangle}
\begin{figure}[h!]
\centering
\includegraphics[scale=0.35]{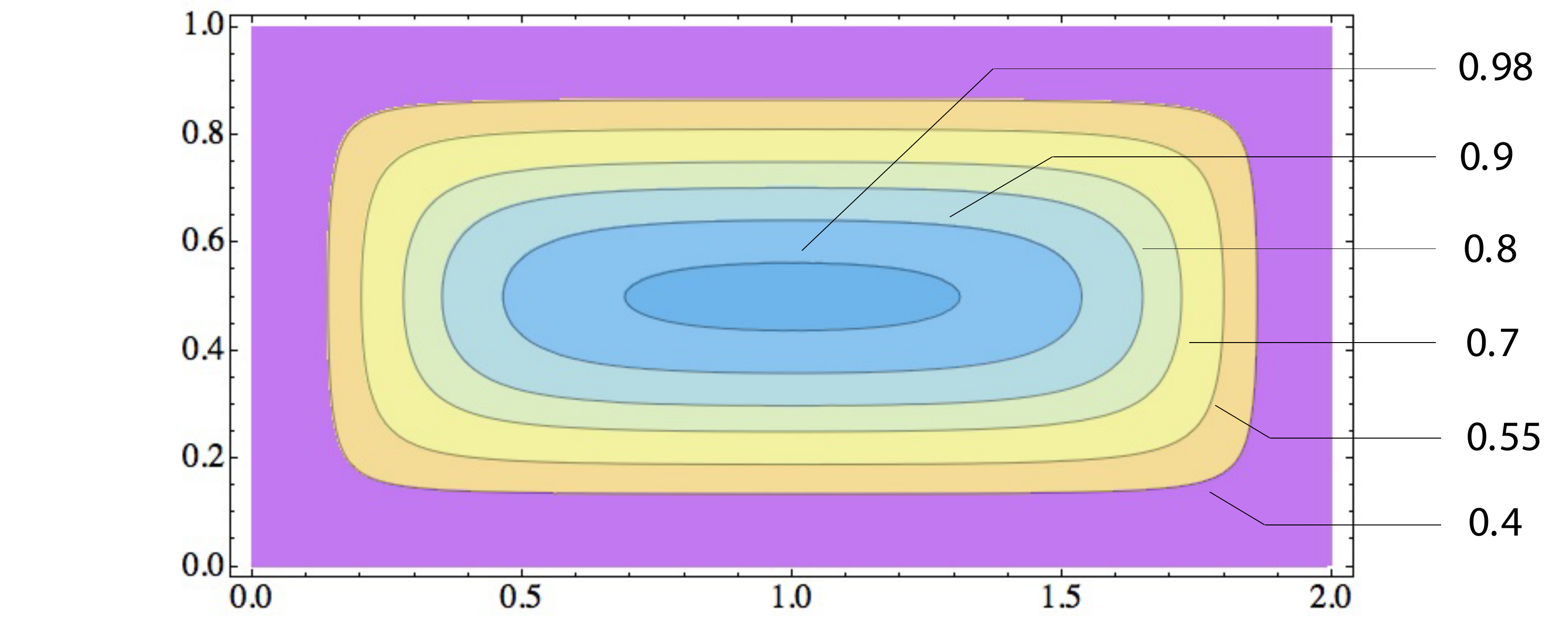}
\caption{Contour plot of Ising FK cluster $(\kappa=16/3)$ two-pinch-point density $\rho^\mathcal{R}_{(1234|1)}$.}
\label{Rect2DensityPlots}
\end{figure}

The density of two-pinch points in the rectangle $\mathcal{R}$ with the left/right sides independently wired behaves as (figure \ref{Rect2DensityConfig})
\begin{multline}\label{1;1234}\rho_{(1234|1)}^\mathcal{R}(m;w,\bar{w})\underset{\delta\rightarrow0}{\sim} C_2^2\delta^{6/\kappa-\kappa/8+1}\left[\frac{2^{24/\kappa-2}K'^{6/\kappa-\kappa/8+1}(n^{-1}+1)[m(1-m)]^{8/\kappa-1}}{\,_2F_1(2-\frac{12}{\kappa},1-\frac{4}{\kappa};2-\frac{8}{\kappa}\,|\,1-m)}\right]\\
\times\left[\frac{\text{Im}\left[\text{sn}(wK'\,|\,m)^2\right]}{|\text{sn}(wK'\,|\,m)\text{cn}(wK'\,|\,m)\text{dn}(wK'\,|\,m)|}\right]^{\kappa/8+18/\kappa-3}.\end{multline}
The density $\rho_{(12:34|2)}^\mathcal{R}$ with the left/right sides mutually wired is found by replacing the argument of the hypergeometric function with $1-m$ in (\ref{1;1234}).  Figure \ref{Rect2DensityPlots} shows a contour plot of $\rho_{(1234|1)}^\mathcal{R}$ for Ising FK boundary clusters ($\kappa=16/3$).

 \begin{figure}[b]
 \centering
\includegraphics[scale=0.35]{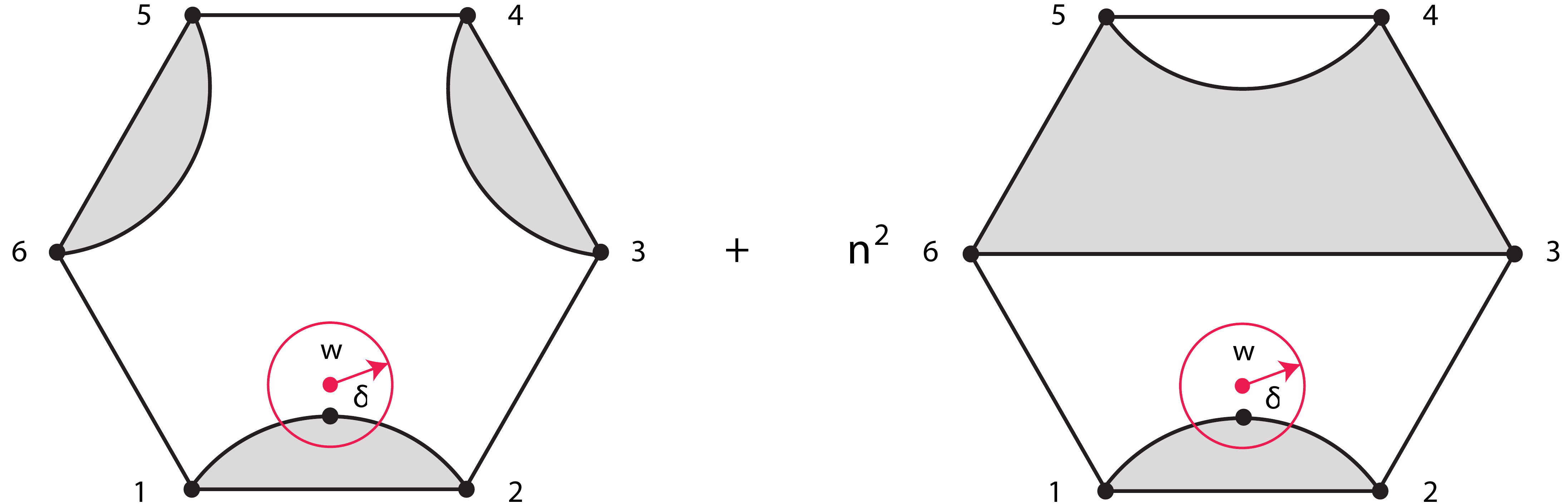}
\caption{Illustration of the type-$(12:34:56)$ one-pinch-point configuration plus $n^2$ times the $(12:36:45)$ one-pinch-point configuration in the hexagon.  Boundary clusters are shaded gray.}
\label{Hex1DensityConfig}
\end{figure}

\subsubsection{A combination of one-pinch-point densities for a hexagon}
Equation (\ref{1ppindwire}) is a combination of one-pinch-point densities in the hexagon $\mathcal{H}$ with the bottom and top-left/right sides independently wired (figure \ref{Hex1DensityConfig}).  We have
\begin{multline}\label{1pinchpointhexfinal}[\rho^\mathcal{H}_{(12:34:56|3)}+n^2\rho^\mathcal{H}_{(12:36:45|3)}](m_1,m_2,m_3;w,\bar{w})\underset{\delta\rightarrow0}{\sim}C_1^2\delta^{1-\kappa/8}n(4-n^2)^{-1/2}|z-\bar{z}|^{\kappa/8+8/\kappa-2}\\
\times|27z(m_1-z)(m_2-z)(m_3-z)(1-z)/8|^{1/3-\kappa/24}\frac{|I_1(m_1,m_2,m_3;z,\bar{z})|}{I_2(m_1,m_2,m_3)}.\end{multline}
Here, $I_1$ is the definite integral
\begin{multline}I_1(m_1,m_2,m_3;z,\bar{z}):=\sideset{}{_\Gamma}\int\sideset{}{_{m_3}^1}\int\mathcal{N}\left[\prod_{i=1,2}(u_i-z)^{8/\kappa-1}(u_i-\bar{z})^{8/\kappa-1}\right.\\
\left.\times(u_2-u_1)^{8/\kappa}\prod_{i=1,2}u_i^{-4/\kappa}(1-u_i)^{-4/\kappa}\prod_{j=1}^3(m_j-u_i)^{-4/\kappa}\right]\,du_1\,du_2\end{multline}
with the contour $\Gamma$ starting at $\bar{z}$, crossing the real axis through either $(m_2,m_3)$ or $(1,\infty)$, and ending at $z$.  One can show that $I_1$ is purely imaginary.  Also, $I_2$ is
\be\label{I2int}I_2(m_1,m_2,m_3):=\sideset{}{_0^{m_1}}\int\sideset{}{_{m_2}^{m_3}}\int\mathcal{N}\left[\prod_{i=1,2}u_i^{-4/\kappa}(1-u_i)^{-4/\kappa}\prod_{j=1,2,3}(m_j-u_i)^{-4/\kappa}\right](u_2-u_1)^{8/\kappa}\,du_2\,du_1.\ee

 \begin{figure}[h!]
 \centering
\includegraphics[scale=0.35]{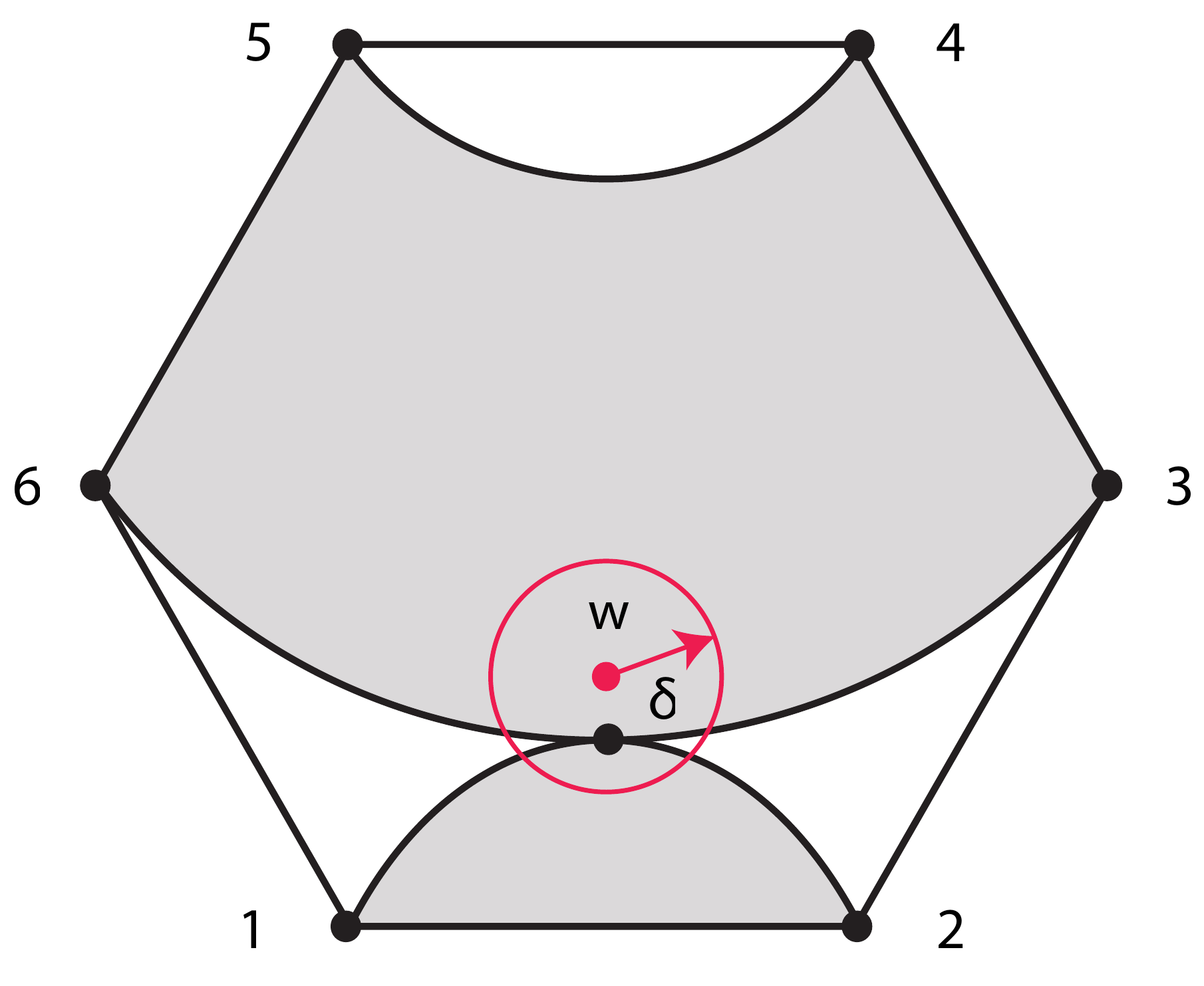}\includegraphics[scale=0.35]{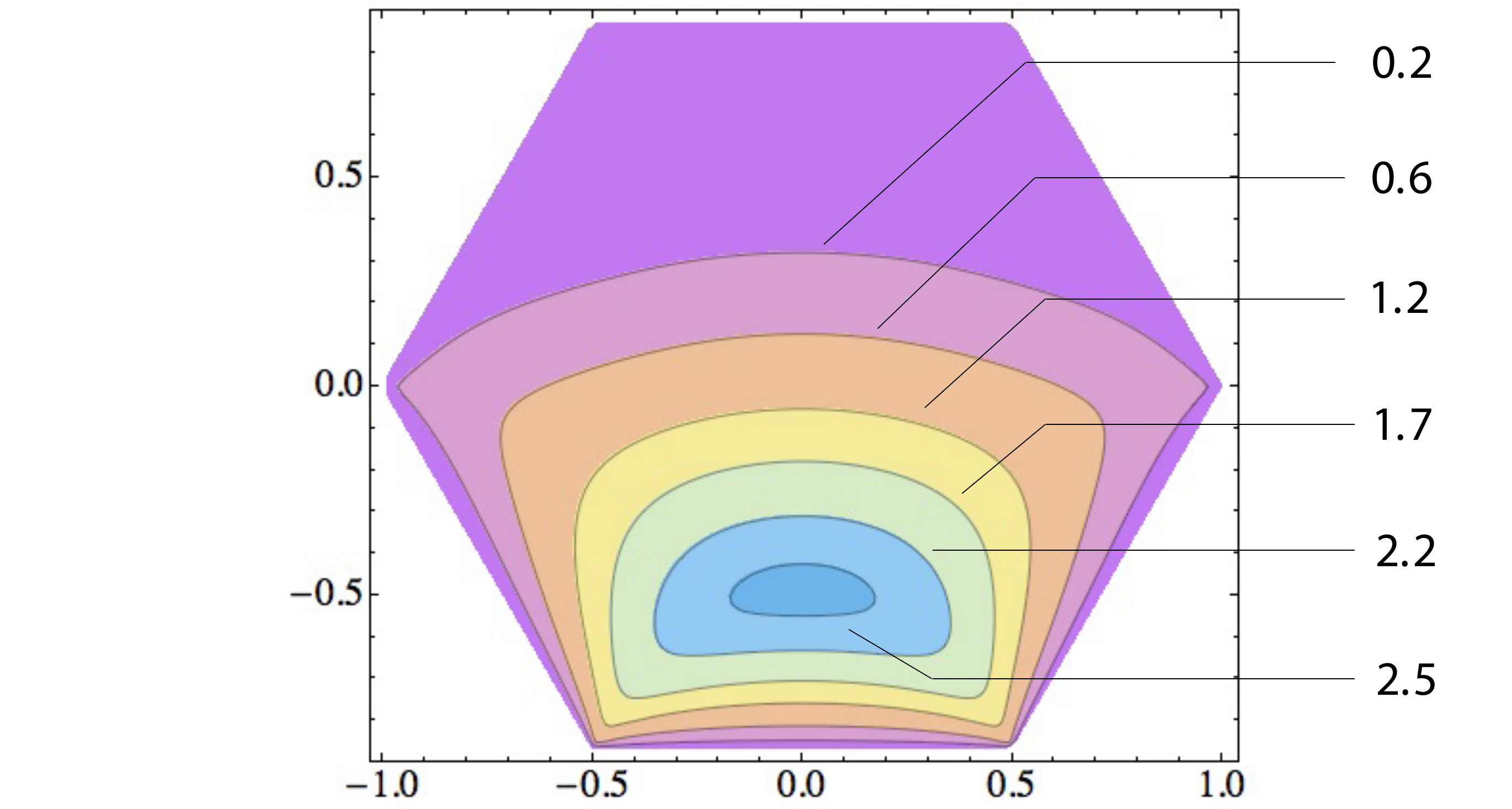}
\caption{Illustration of the type-$(6123:45)$ two-pinch-point configuration in the hexagon (left, boundary clusters are shaded gray), and a contour plot of the Ising FK cluster $(\kappa=16/3)$ two-pinch-point density $\rho^\mathcal{H}_{(123456|3)}$ (right).}
\label{Hex2DensityConfig}
\end{figure}

\subsubsection{A two-pinch-point density for a hexagon}
The density of two-pinch points in the hexagon $\mathcal{H}$ with the top and bottom-left/right sides independently wired behaves as (figure \ref{Hex2DensityConfig})
\begin{multline}\label{2ppindwire}\rho_{(6123:45|3)}^\mathcal{H}(m_1,m_2,m_3;w,\bar{w})\underset{\delta\rightarrow0}{\sim}C_2^2\delta^{6/\kappa-\kappa/8+1}(n^{-1}+1)\beta(-4/\kappa,-4/\kappa)^2|z-\bar{z}|^{\kappa/8-6/\kappa-1}I_2(m_1,m_2,m_3)^{-1}\\
\begin{aligned}&\times\frac{[m_1(m_3-m_2)]^{1-6/\kappa}|27z(m_1-z)(m_2-z)(m_3-z)(1-z)/8|^{1/3-\kappa/24+2/\kappa}}{[m_1m_2m_3(m_2-m_1)(m_3-m_1)(m_3-m_2)(1-m_1)(1-m_2)(1-m_3)]^{2/\kappa}}\\
&\times\left[\frac{(2-n^2)(H_6+H_4)+nH_1-2H_2+nH_3+n(n^2-3)H_5}{(n^2-4)(n^2-1)}\right](m_1,m_2,m_3;z,\bar{z}),\end{aligned}\end{multline}
where $I_2$ is given in (\ref{I2int}) and the $H_i$ are given in (\ref{Gi}, \ref{Ifirst}-\ref{Ilast}).  Figure \ref{Hex2DensityConfig} shows a contour plot of $\rho_{(6123:45|3)}^\mathcal{H}$ for Ising FK boundary clusters ($\kappa=16/3$).

\subsubsection{A three-pinch-point density for a hexagon}
\begin{figure}[h!]
 \centering
\includegraphics[scale=0.35]{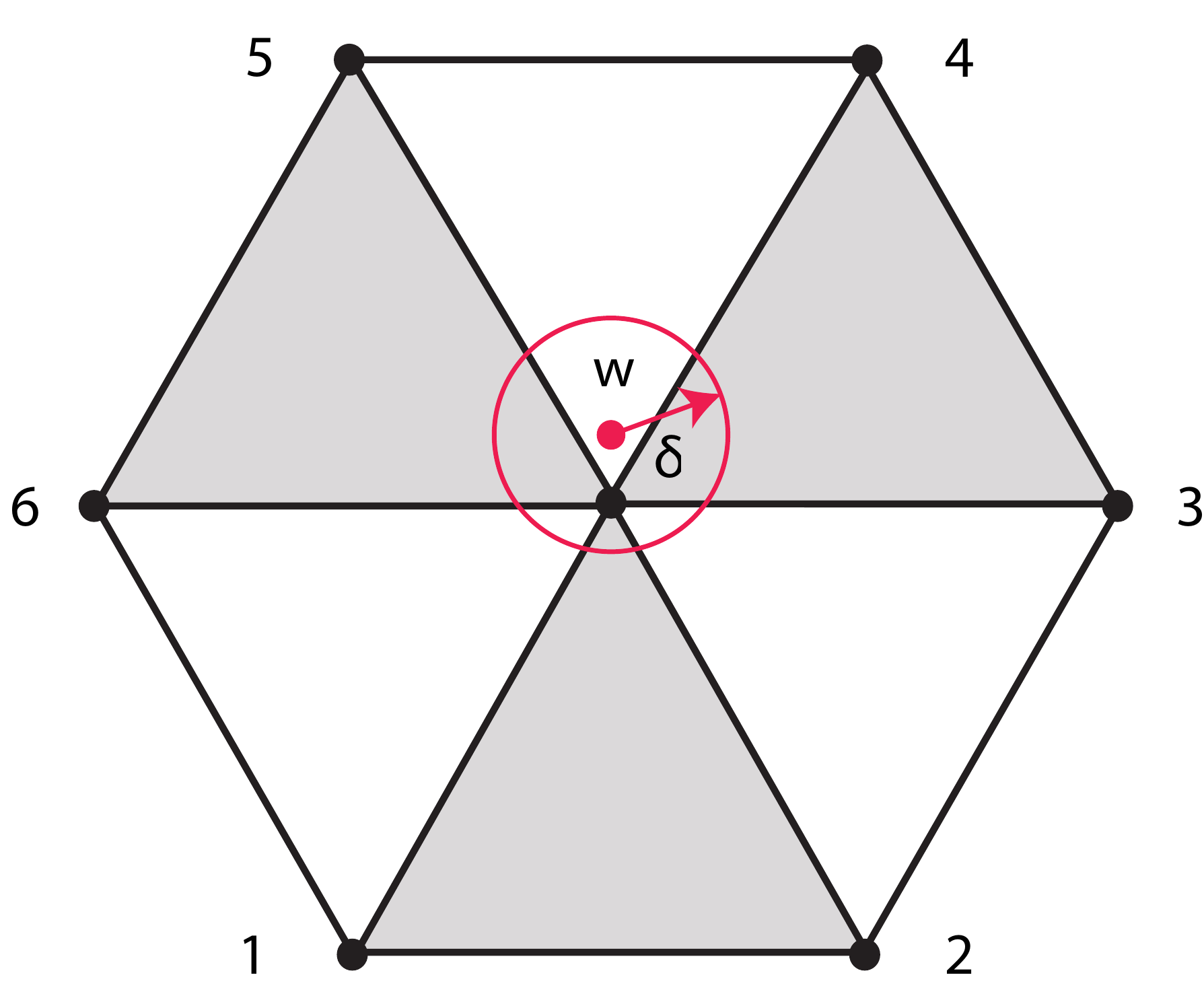}\includegraphics[scale=0.35]{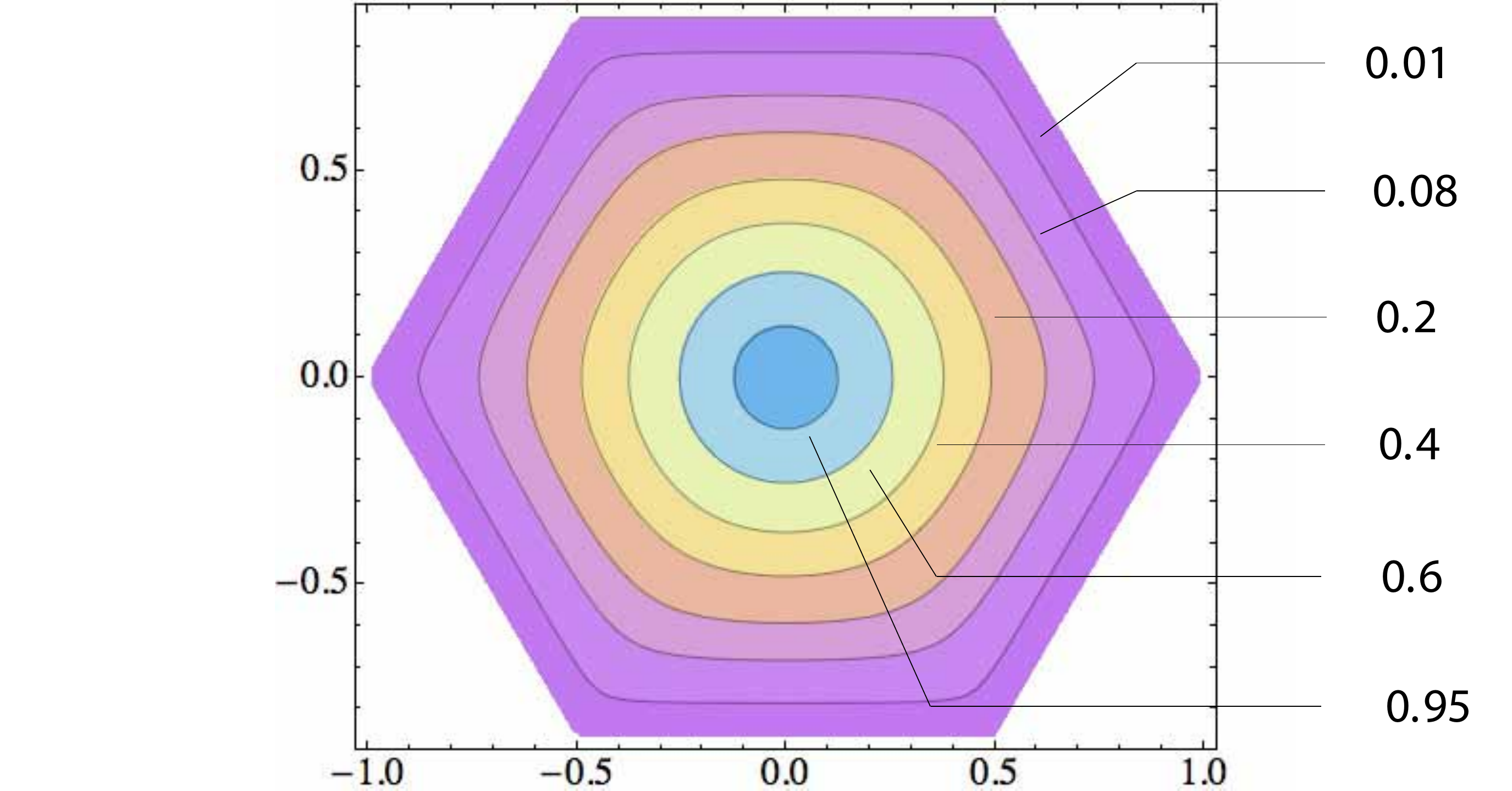}
\caption{Illustration of the three-pinch-point configuration in the hexagon (left, boundary clusters are shaded gray) and a contour plot of the Ising FK cluster $(\kappa=16/3)$ two-pinch-point density $\rho^\mathcal{H}_{(123456|3)}$ (right).}
\label{Hex3DensityConfig}
\end{figure}

The density of three-pinch points in the hexagon $\mathcal{H}$ with the top and bottom-left/right sides independently wired behaves as (figure \ref{Hex3DensityConfig})
\begin{multline}\label{threepppredict}\rho_{(6123:45|3)}^\mathcal{H}(m_1,m_2,m_3;w,\bar{w})\underset{\delta\rightarrow0}{\sim}C_3^2\delta^{16/\kappa-\kappa/8+1}(27/8)^{16/3\kappa-\kappa/24-1/3}(n^{-2}+3n^{-1}+1)\\
\begin{aligned}&\times|z(m_1-z)(m_2-z)(m_3-z)(1-z)|^{4/3-\kappa/24-32/3\kappa}\\
&\times|z-\bar{z}|^{\kappa/8+32/\kappa-4}\beta(-4/\kappa,-4/\kappa)^2I_2(m_1,m_2,m_3)^{-1},\end{aligned}\end{multline}
where $I_2$ is given in (\ref{I2int}).  Figure \ref{Hex3DensityConfig} shows a contour plot of $\rho_{(6123:45|3)}^\mathcal{H}$ for Ising FK boundary clusters ($\kappa=16/3$).

\section{Simulation results}\label{simresults}

In this section, we present simulation results that verify our predicted pinch-point densities $\rho^\mathcal{R}_{(12:34|1)}$ and $\rho^\mathcal{R}_{(1234|1)}$ (\ref{1;12:34}, \ref{1;1234}) for the rectangle and $\rho^\mathcal{H}_{(12:34:56|3)}+n^2\rho^\mathcal{H}_{(12:36:45|3)}$ and $\rho^\mathcal{H}_{(6123:45|3)}$ (\ref{1pinchpointhexfinal}, \ref{2ppindwire}) for the hexagon.  We simulated critical bond percolation on a 2000$\times$1000 square lattice (aspect ratio $R=2$) in a rectangle $\mathcal{R}$ and critical site percolation on a triangular lattice in a regular hexagon $\mathcal{H}$ inscribed in a 2000$\times$2000 rhombus.  (Actually, only half of the sites in $\mathcal{R}$ belonged to the physical lattice, and the other sites belonged to the medial lattice.  This was for the purpose of performing a hull-walk later.  This was also true when we sampled Ising FK clusters in $\mathcal{H}$, but it was not true when we sampled site percolation clusters in $\mathcal{H}$.  In the latter case, all of the sites in our simulation belonged to the physical lattice.)  Using the Swendsen-Wang (SW) algorithm \cite{sw}, we also sampled critical Ising FK clusters on the same lattices in $\mathcal{R}$ and $\mathcal{H}$ respectively.  We independently wired the left/right sides of $\mathcal{R}$, and we independently wired the bottom and top left/right sides of $\mathcal{H}$, leaving all other sides free in both situations.  Our simulation results show good agreement with our predictions.  We used Mathematica (version 8) to evaluate all definite integrals that appear in our analytic predictions via the ``NIntegrate" function, and we used the Schwarz-Christoffel toolbox \cite{dristref} for MATLAB to numerically perform the transformation from the upper half-plane to the interior of $\mathcal{H}$.

To approximate the continuum limit of these critical models, we used very large lattices in our simulations.  This suppressed the frequency of each $s$-pinch-point event at every lattice site, so we generated many samples in order to compensate this effect.  Overall, about sixteen months of computer time (with about a 2 GHz processor) were used to sample percolation clusters in $\mathcal{R}$ and $\mathcal{H}$ and Ising FK clusters in $\mathcal{R}$, and over thirty-two months of computer time were used to sample Ising FK clusters in $\mathcal{H}$, running simulations on many processors simultaneously.

As usual, we number the vertices of $\mathcal{R}$ and $\mathcal{H}$ counterclockwise, starting with the left vertex of the bottom side of either polygon as vertex one.  In this section, we also shift the hexagon so that its center coincides with the origin.

\subsection{The rectangle} 

To measure the density of percolation pinch points ($\kappa=6$), we simulated critical bond percolation on the square lattice in $\mathcal{R}$. One-pinch points are bonds inside $\mathcal{R}$ and on the perimeter of a boundary cluster, and two-pinch points are red bonds whose activation or deactivation respectively connects or disconnects the left and right boundary clusters.  Because pinch points occur only on boundary cluster perimeters, our simulations only sampled these perimeters via percolation hull-walks on the medial lattice \cite{gunno, grass, z}.  They did not generate entire percolation bond configurations.

 \begin{figure}[t]
 \centering
\includegraphics[scale=1.2]{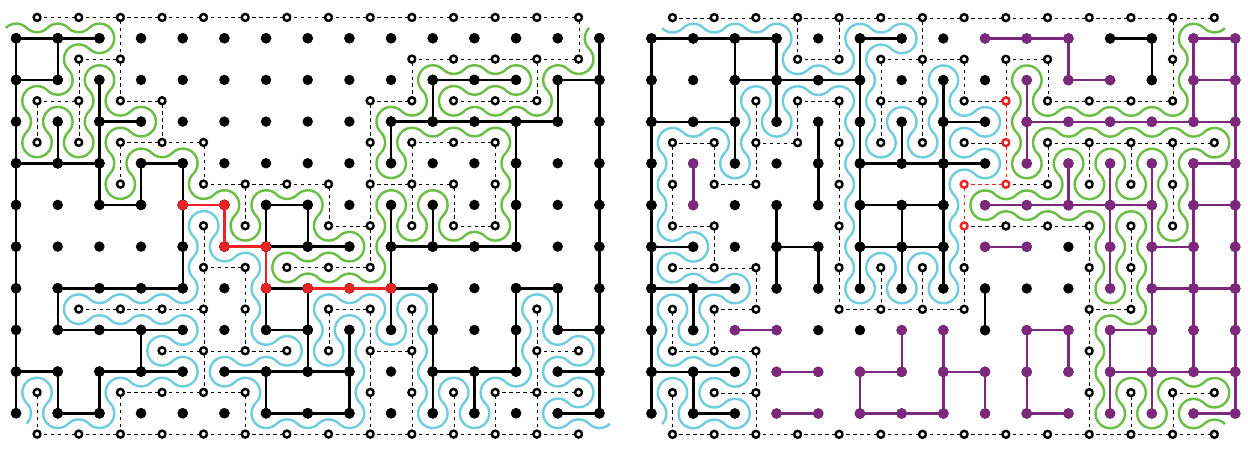}
\caption{Typical percolation (left) and Ising FK (right) cluster (black, purple, and red bonds) and hull-walk samples (blue and green) generated by our simulations in $\mathcal{R}$.  Two-pinch points are colored red.  We note that for percolation (left), our simulation only generates bonds that comprise the boundary clusters' perimeters.}
\label{HullWalks}
\end{figure}

\begin{figure}[b]
\centering
\includegraphics[scale=0.25]{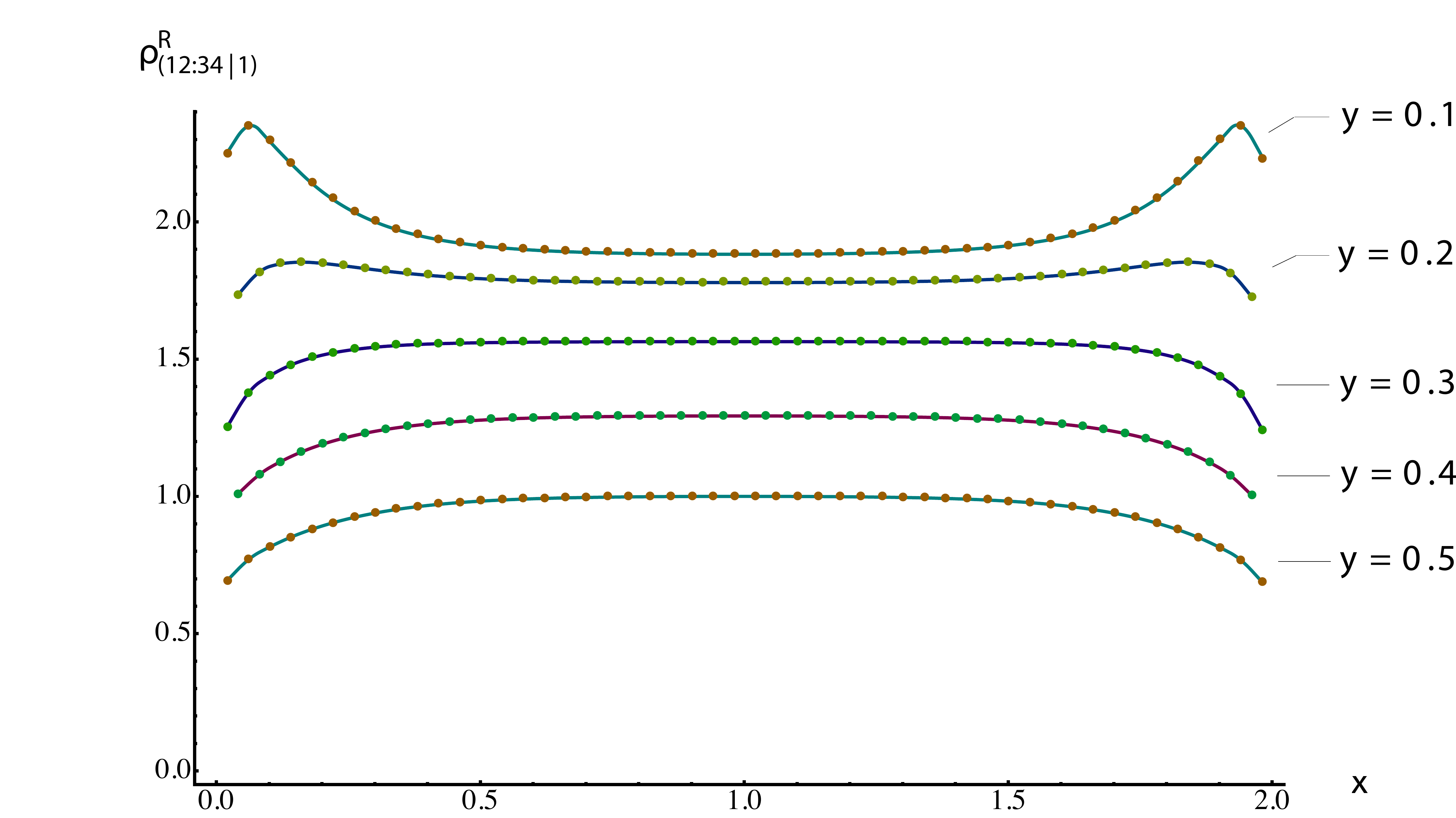}\vspace{.2cm}
\includegraphics[scale=0.25]{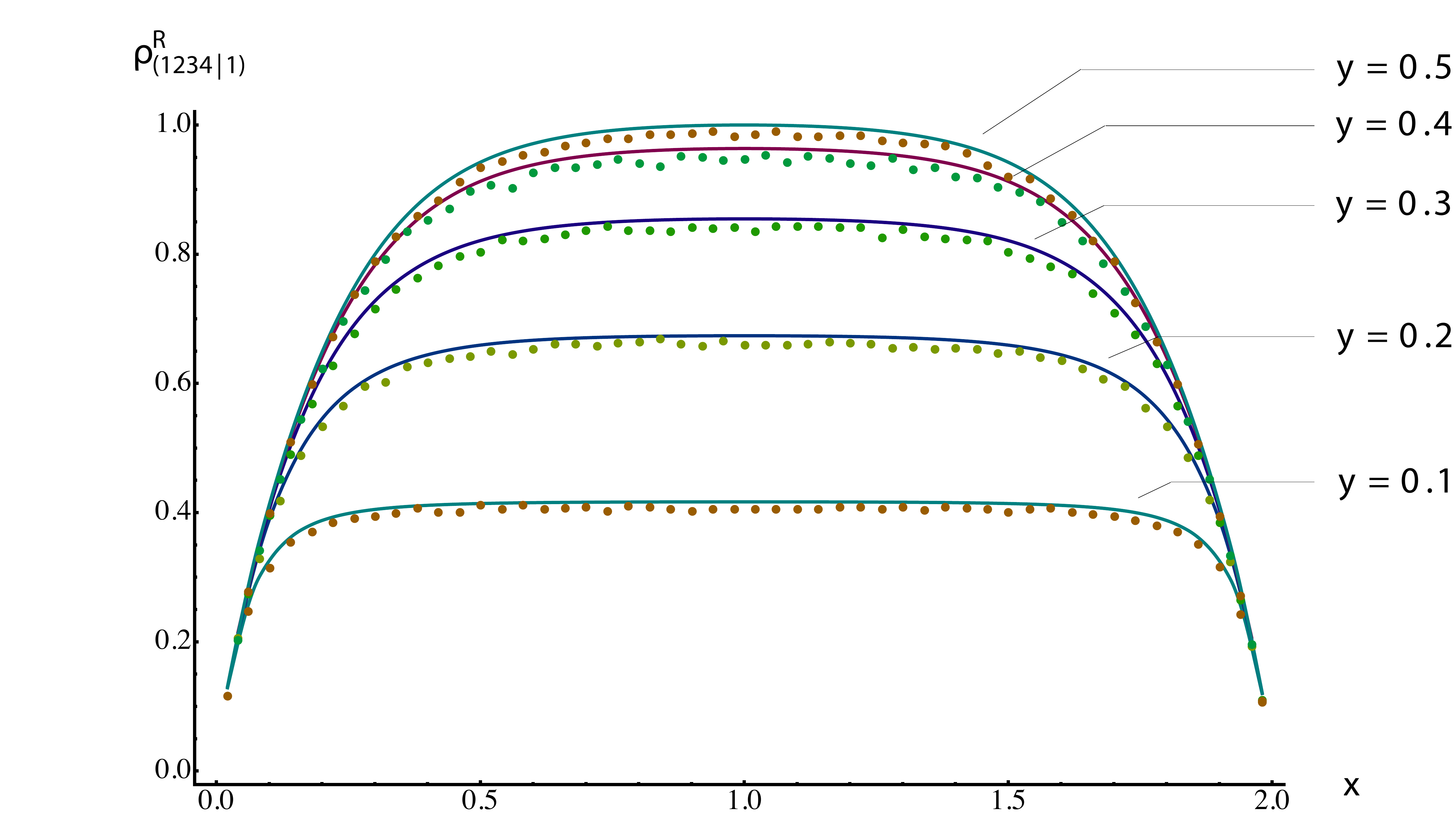}
\caption{Density of percolation ($\kappa=6$) one-pinch points $\rho^\mathcal{R}_{(12:34|1)}$ (top) and two-pinch points $\rho^\mathcal{R}_{(1234|1)}$ (bottom) versus $x$ in a 2000$\times$1000 rectangle rescaled to $2\times1$ for various values of $y$.  Both densities are normalized to equal one at the center $(1,0.5)$ of $\mathcal{R}$.}
\label{PercRectPinchPoint}
\end{figure}

We performed two hull-walks to generate the two boundary cluster perimeters in $\mathcal{R}$ (figure \ref{HullWalks}).  The first (resp.\,second) walk starts at the medial lattice site $a/2$ units below vertex one (resp.\,above vertex three), where $a$ is the lattice spacing.  One walk ends at the medial lattice site $a/2$ units below vertex two, and the other ends at $a/2$ units above vertex four.  Each step of the walk is located at the midpoint of a bond, and at each step we decide to activate or deactivate that bond with critical bond activation probability $p_c=1/2$.  If the bond is activated, then the walk turns right.  Otherwise it turns left.  Each site that is visited by either hull-walk is a one-pinch point, and each site that is visited by both hull-walks is a two-pinch point, or red bond.  (Actually, one-pinch-point events and two-pinch-point events at $z\in\mathcal{R}$ are mutually exclusive according to our definitions.  However, our method counts a two-pinch point as one-pinch points of both hull walks too.  This miscounting is insignificant though because the ratio of the two-pinch-point density to any one-pinch-point density vanishes as the system size increases, or equivalently, as the lattice spacing $a$ goes to zero, with power $a^{2(\Theta_2-\Theta_1)}=a^{6/\kappa}$.)  

When both of the hull-walks are finished, we note how they connect the vertices of $\mathcal{R}$ and bin one-pinch-point events accordingly.  If they connect vertex one with vertex two and vertex three with vertex four (resp.\,vertex one with vertex four and vertex two with vertex three) to create a horizontal (resp.\,vertical) crossing, then one-pinch-point events on the first and second hull-walks contribute to $\rho^\mathcal{R}_{(12:34|1)}$ and $\rho^\mathcal{R}_{(34:12|1)}$ (resp.\,$\rho^\mathcal{R}_{(41:23|1)}$ and $\rho^\mathcal{R}_{(23:41|1)}$) respectively.  Two-pinch-point events did not need to be sorted this way, although it is interesting to note that the red bonds are activated (resp.\,deactivated) in the event of a horizontal (resp.\,vertical) crossing.  After generating about $2\times10^8$ samples, we tallied the number of each kind of pinch-point event at each site in an array and divided the total by the corresponding array value at the center of $\mathcal{R}$ to eliminate the non-universal  constant and scaling factor appearing in (\ref{1;12:34}, \ref{1;1234}).  With this normalization scheme, our measured densities always equal one at the center of $\mathcal{R}$.

\begin{figure}[t]
\centering
\includegraphics[scale=0.25]{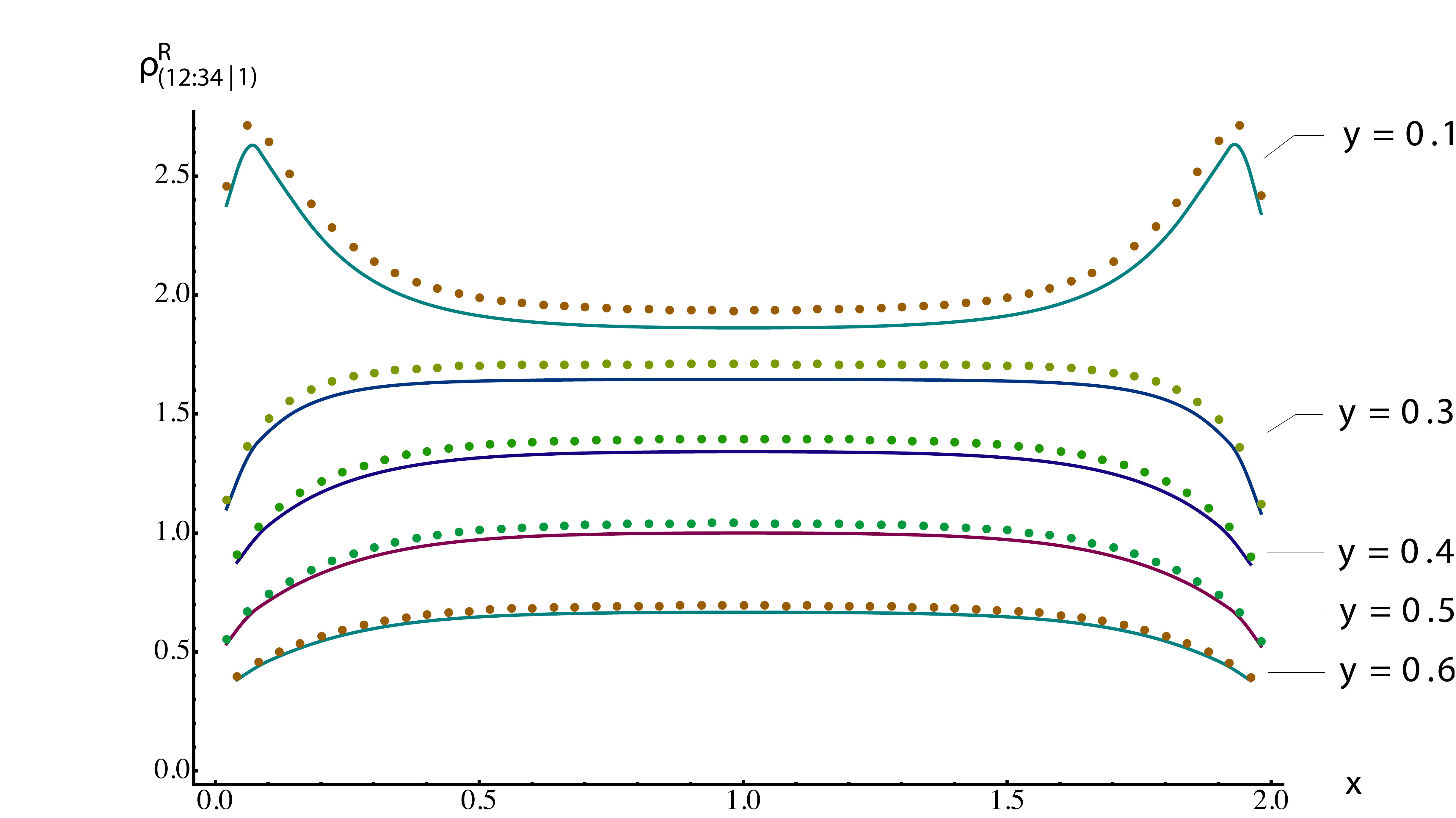}\vspace{.2cm}
\includegraphics[scale=0.25]{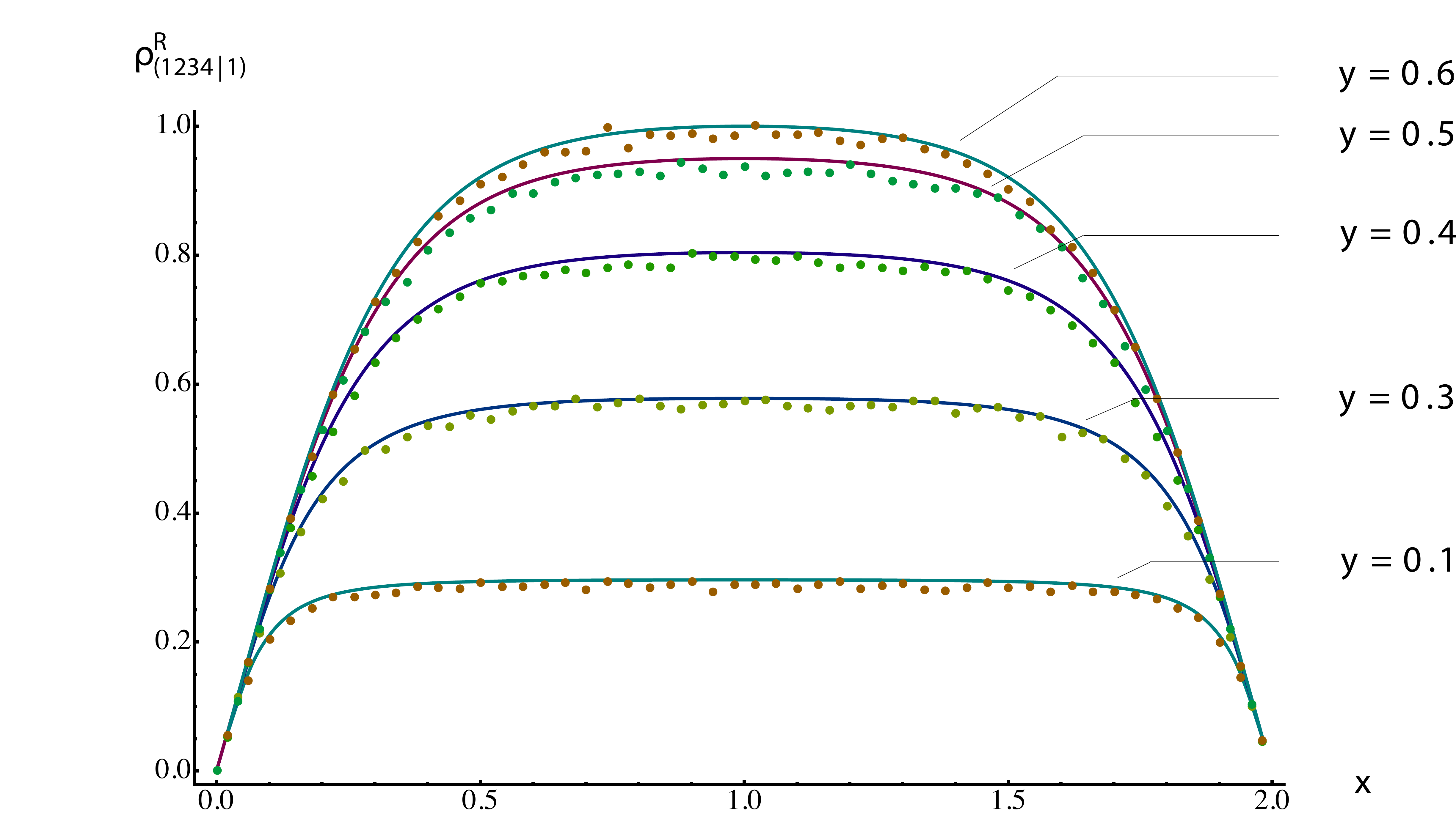}
\caption{Density of Ising FK ($\kappa=16/3$) one-pinch points $\rho^\mathcal{R}_{(12:34|1)}$ (top) and two-pinch points $\rho^\mathcal{R}_{(1234|1)}$ (bottom) versus $x$ in a 2000$\times$1000 rectangle rescaled to $2\times1$ for various values of $y$.  Both densities are normalized to equal one at the center $(1,0.5)$ of $\mathcal{R}$.}
\label{IsingRectPinchPoint}
\end{figure}

Our percolation pinch-point simulation results are compared with our theory predictions for $\rho^\mathcal{R}_{(12:34|1)}$ and $\rho^\mathcal{R}_{(1234|1)}$ (respectively (\ref{1;12:34}) and (\ref{1;1234}) with $\kappa=6$) in figure \ref{PercRectPinchPoint}.  In the figure, $\mathcal{R}$ has been centered at $(1,0.5)$ and rescaled to have length two and height one, and we have fixed $y$ to $0.1, 0.2, 0.3, 0.4,$ and $0.5$.  The top plot shows the density $\rho^\mathcal{R}_{(12:34|1)}$ of points touched by the hull-walk connecting vertices one and two.  The left/right peaks of the top curve in the figure show that these events are most likely to occur near vertices one and two, and this is expected because these vertices are connected by the hull-walk.  The bottom plot shows the density $\rho^\mathcal{R}_{(1234|1)}$ of two-pinch points, or red bonds.  Our results show that these events are most likely to occur near the center of $\mathcal{R}$, and this is expected because there is more room in the center of $\mathcal{R}$ for the hull-walks to collide.  Errors averaged over $x$ and the standard deviation from this average are shown in table \ref{RectError}.

To measure the density of Ising FK pinch points ($\kappa=16/3$), we generated about $4\times10^8$ samples of critical Ising FK clusters on the square lattice in $\mathcal{R}$ via the SW algorithm with critical bond activation probability $p_c=\sqrt{2}/(1+\sqrt{2})$.  In each sample, the left/right sides of $\mathcal{R}$ were independently wired.  That is, all FK bonds of either side were activated and necessarily the same color as the other bonds in its boundary cluster, but the bond color of the left boundary cluster was allowed to differ from that of the right.   The perimeters of the boundary clusters anchored to these wired sides were explored via the hull-walk used for percolation to detect the pinch points (without ever activating or deactivating FK bonds during the walk) (figure \ref{HullWalks}).  One-pinch-point events and two-pinch-point events were detected by the hull-walk in the same way as in bond percolation.  However, we note that the ``red bond definition" of a two-pinch point in percolation does not have a perfect analog with FK clusters.  If two boundary FK clusters of different colors are separated by a single unactivated bond, we cannot activate this bond and join the clusters, so such a bond is not ``red" according to the definition of a ``spin red bond" in \cite{apr}.  (However, because the $Q$ colors of the $Q$-state Potts model are distributed uniformly across the FK clusters, the fractal properties of the set of spin red bonds and the set of two-pinch-point bonds are the same.)

Our Ising pinch-point simulation results are compared with our theory predictions for $\rho^\mathcal{R}_{(12:34|1)}$ and $\rho^\mathcal{R}_{(1234|1)}$ (respectively (\ref{1;12:34}) and (\ref{1;1234}) with $\kappa=16/3$) in figure \ref{IsingRectPinchPoint}.  Explanations and notable features of these two plots are the same as those for figure \ref{PercRectPinchPoint}, except that $y$ is fixed to 0.1, 0.3, 0.4, 0.5, and 0.6.  In addition, we note that our measurements of the one-pinch-point density consistently exceeded our predictions by a small amount that increased as $y$ decreased.  We suspect that this is a finite-size effect that may be reduced by sampling from a larger system.  A similar effect was observed for the simulations in \cite{skfz}.

\begin{table}[h!]
\centering
\begin{tabular}{l|cccccc} 
Avg.\,error&$\hspace{0.15cm}y=0.1\hspace{0.15cm}$&$\hspace{0.15cm}y=0.2\hspace{0.15cm}$&$\hspace{0.15cm}y=0.3\hspace{0.15cm}$&$\hspace{0.15cm}y=0.4\hspace{0.15cm}$&$\hspace{0.15cm}y=0.5\hspace{0.15cm}$&$\hspace{0.15cm}y=0.6\hspace{0.15cm}$\\ 
\hline\hline
Perc.\,$s=1$ & $-0.004$ & 0.023 & $-0.003$ & 0.012 & $-0.002$ & --\\ 
Perc.\,$s=2$& 0.010 & 0.012 & 0.014 & 0.014 & 0.014 & --\\ 
Ising $s=1$& $-0.080$ & -- & $-0.062$ & $-0.048$ & $ -0.036$ & $-0.346$\\ 
Ising $s=2$ & 0.008 & -- & 0.009 & 0.012 & 0.014 & 0.046\\ 
\end{tabular}

\vspace{.5cm}

\begin{tabular}{l|cccccc} 
Std.\,dev.\,&$\hspace{0.15cm}y=0.1\hspace{0.15cm}$&$\hspace{0.15cm}y=0.2\hspace{0.15cm}$&$\hspace{0.15cm}y=0.3\hspace{0.15cm}$&$\hspace{0.15cm}y=0.4\hspace{0.15cm}$&$\hspace{0.15cm}y=0.5\hspace{0.15cm}$&$\hspace{0.15cm}y=0.6\hspace{0.15cm}$\\ 
\hline\hline
Perc.\,$s=1$ & 0.002 & 0.026 & 0.002 & 0.025 & 0.000 & --\\ 
Perc.\,$s=2$ & 0.003 & 0.003 & 0.003 & 0.005 & 0.004 & --\\ 
Ising $s=1$ & 0.008 & -- & 0.006 & 0.013 & 0.006 & 0.042\\ 
Ising $s=2$ & 0.004 & -- & 0.006 & 0.005 & 0.007 & 0.034\\ 
\end{tabular}
\caption{The error (theory minus simulation) averaged over $x$, and the standard deviation of the error from that average, of the data displayed in figures \ref{PercRectPinchPoint} and \ref{IsingRectPinchPoint}.}
\label{RectError}
\end{table}

\subsection{The hexagon}

\begin{figure}[t]
\centering
\includegraphics[scale=0.37]{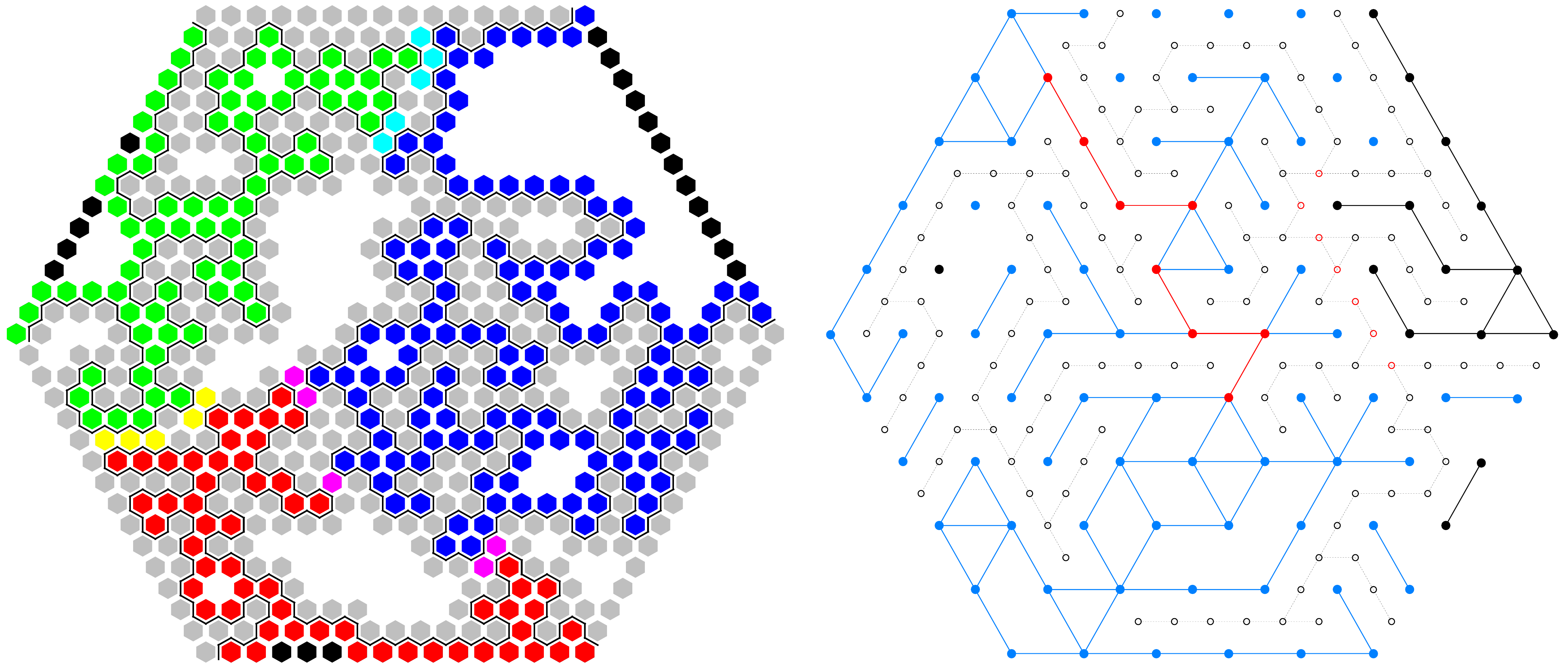}
\caption{Typical percolation (left) and Ising FK (right) cluster samples $\mathcal{H}$ generated by our simulations.  In the left illustration, black, red, blue, and green (resp.\,gray, yellow, pink, light blue, resp.\,white) sites are activated (resp.\,deactivated, resp.\,undecided), black paths are smart kinetic walks, red, blue, and green (resp.\,gray) sites form the inner (resp.\,outer) boundary arc of a boundary cluster, and yellow, pink, and light-blue sites are two-pinch points.  In the right illustration, red sites and centers of red bonds are two-pinch points.}
\label{HexPinchPoints}
\end{figure}

To measure the density of percolation pinch points, we generated about $6.56\times10^8$ samples of critical site percolation on the triangular lattice (site activation probability $p_c=1/2$) in $\mathcal{H}$.  Again, because pinch-point events occur on the perimeters of the three boundary clusters, our simulations sampled only these perimeters via three distinct site-percolation hull-walks on the triangular lattice (left picture in figure \ref{HexPinchPoints}).  The first, second, and third hull-walks respectively started at vertices one, three, and five and in the direction pointing into the adjacent free side of $\mathcal{H}$.  Each hull-walk necessarily ended at an even vertex, and no two hull-walks could end at the same vertex.  The finished hull-walk actually consists of two juxtaposed paths of neighboring activated and deactivated sites.  We will call the former (resp.\,latter) path the \emph{inner} (resp.\,\emph{outer) boundary arc} of the boundary cluster (figure \ref{HexPinchPoints}).  A path, called a \emph{smart kinetic walk}, passes between these juxtaposed paths with each step on the dual (honeycomb) lattice \cite{wt}.

Pinch-point events were counted in almost the same way as in the simulations for the rectangle.  Each point on the inner boundary arc is a one-pinch point, activated (resp.\,deactivated) sites shared between two inner or outer boundary arcs are two-pinch points, and activated (resp.\,deactivated) sites shared between three inner or outer boundary arcs are three-pinch points.   Again, although one-pinch-point, two-pinch-point, and three-pinch-point events are technically mutually exclusive, we counted three-pinch-point events as two-pinch-point events and two-pinch-point events as one-pinch-point events in our simulations, knowing that this over-counting has a negligible effect on our results.

\begin{figure}[t]
\centering
\includegraphics[scale=0.25]{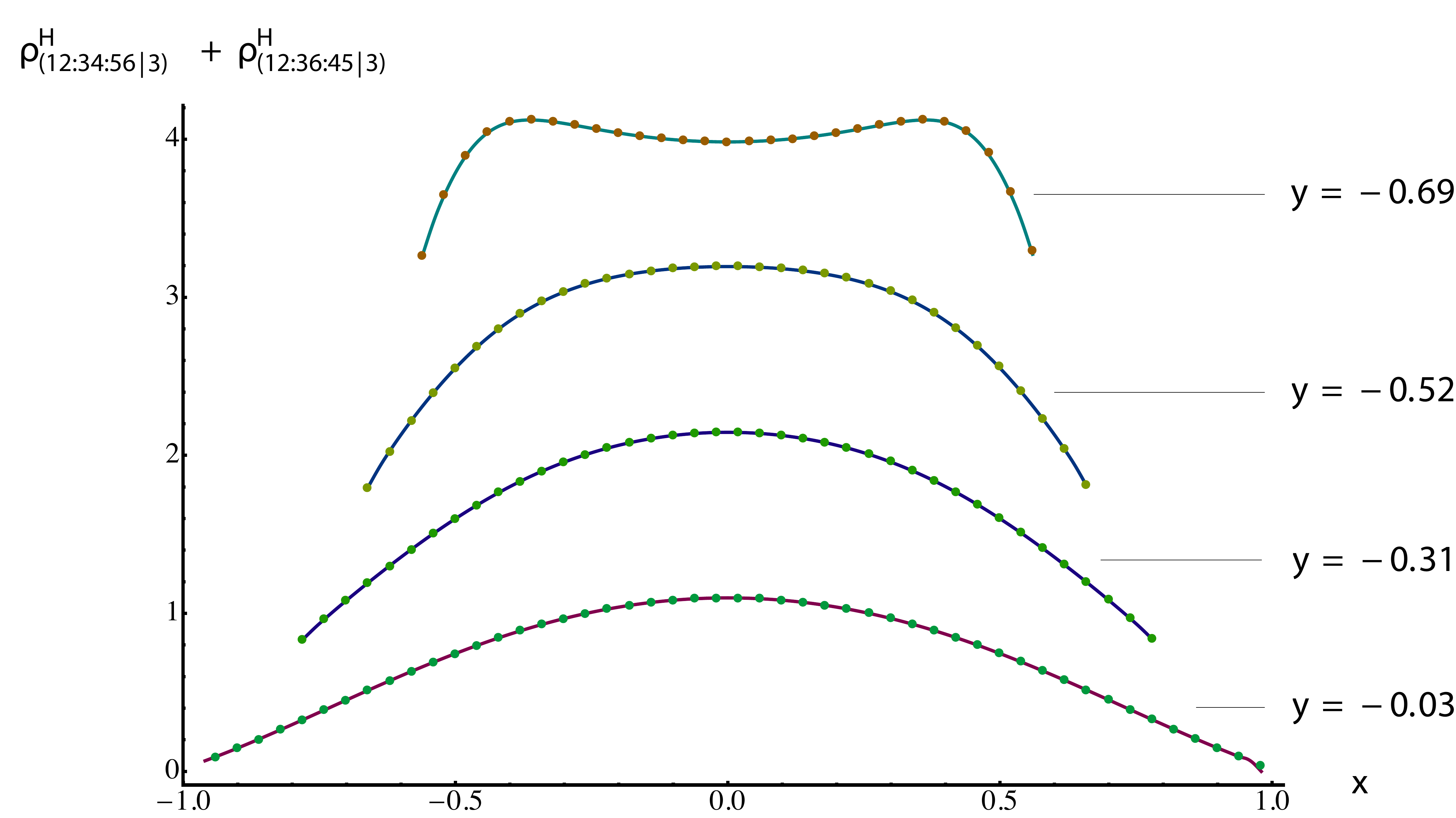}\vspace{.2cm}
\includegraphics[scale=0.25]{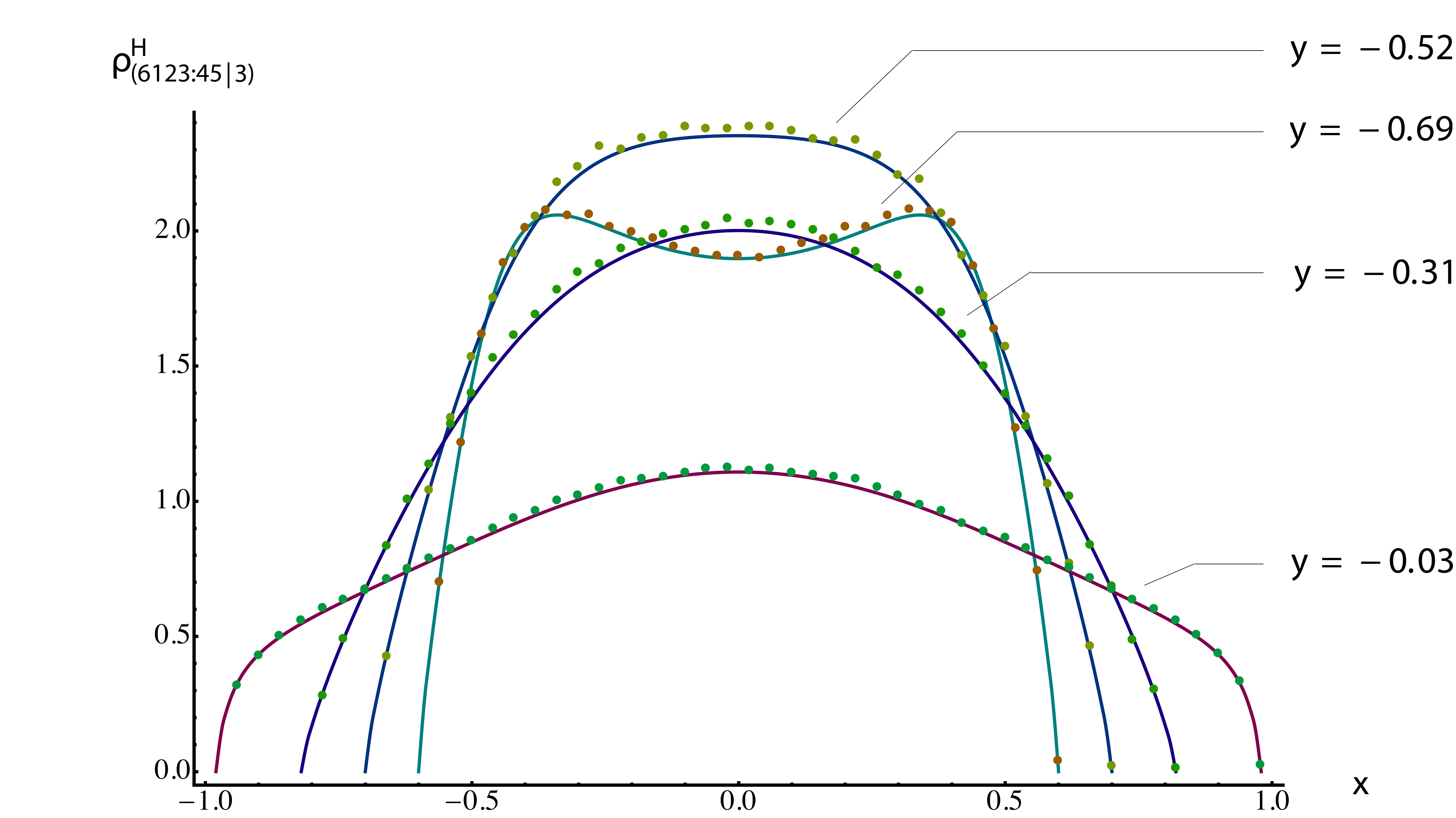}
\caption{Density of percolation ($\kappa=6$) one-pinch points $\rho^\mathcal{H}_{(12:34:56|3)}+\rho^\mathcal{H}_{(12:36:45|3)}$ (top) and two-pinch points $\rho^\mathcal{H}_{(6123:45|3)}$ (bottom) versus $x$ in a regular hexagon inscribed in 2000$\times$2000 rhombus and adjusted to have side-length one and center at the origin.  Both densities are normalized to one at the center of $\mathcal{H}$.}
\label{PercHexPinchPoint}
\end{figure}

Our percolation pinch-point simulation results are compared with our theory predictions for $\rho^\mathcal{H}_{(12:34:56|3)}+\rho^\mathcal{H}_{(12:36:45|3)}$ and $\rho^\mathcal{H}_{(6123:45|3)}$ (respectively (\ref{1pinchpointhexfinal}) and (\ref{2ppindwire}) with $\kappa=6$) in figure \ref{PercHexPinchPoint}.  Three-pinch-point events were so rare that we could not generate enough samples to verify our three-pinch-point density prediction (\ref{threepppredict}).  In the figure, $\mathcal{H}$ has been centered at the origin, rescaled to have side-length one, and oriented as in figure \ref{HexPinchPoints}, and we have fixed $y$ to $ -0.69, -0.52, -0.31,$ and $-0.03$.  The top plot shows the density $\rho^\mathcal{H}_{(12:34:56|3)}+\rho^\mathcal{H}_{(12:36:45|3)}$ of one-pinch points touched by the hull-walk connecting vertices one and two.  (This density includes all such events without regard to how the other two walks connect vertices three through six.)  As expected, this density is greatest near vertices one and two which are connected by the hull-walk.  The bottom plot shows the density $\rho^\mathcal{H}_{(6123:45|3)}$ of two-pinch points, or sites touched by a hull-walk connecting vertices one and two and another hull walk connecting vertices three and six.  This density grows at first as we move from the bottom side of $\mathcal{H}$ towards its center, as expected, and then it diminishes as the center is approached.  This diminishing is also expected since $\rho^\mathcal{H}_{(6123:45|3)}$ is greatest below the center of the rectangle conformally equivalent to $\mathcal{H}$ with vertices one, two, three, and six, and the image of this point in $\mathcal{H}$ is below the origin and on the $y$-axis.

To measure the density of Ising FK pinch points, we generated $5.68\times10^8$ samples of critical Ising FK clusters on the triangular lattice in $\mathcal{H}$ via the SW algorithm.  The critical probability of FK bond activation on the triangular lattice is $p_c=(\sqrt{3}-1)/\sqrt{3}\approx0.42265$ \cite{kj}.   Because the triangular lattice is not self-dual, a hull-walk similar to that of $\mathcal{R}$ that steps on both regular and dual sites equally often is not so easy to implement.  For this reason, we chose our hull-walk to step along the bonds in $\mathcal{H}$ forming the perimeter of a boundary FK cluster (i.e., the inner boundary arcs).  We also tracked the activated dual bonds that surrounded the boundary clusters (i.e., the outer boundary arcs).  The $s$-pinch-point events were identified with intersections between $s$ distinct inner or outer boundary arcs in the same way as with percolation (figure \ref{HexPinchPoints}).

\begin{figure}[t]
\centering
\includegraphics[scale=0.25]{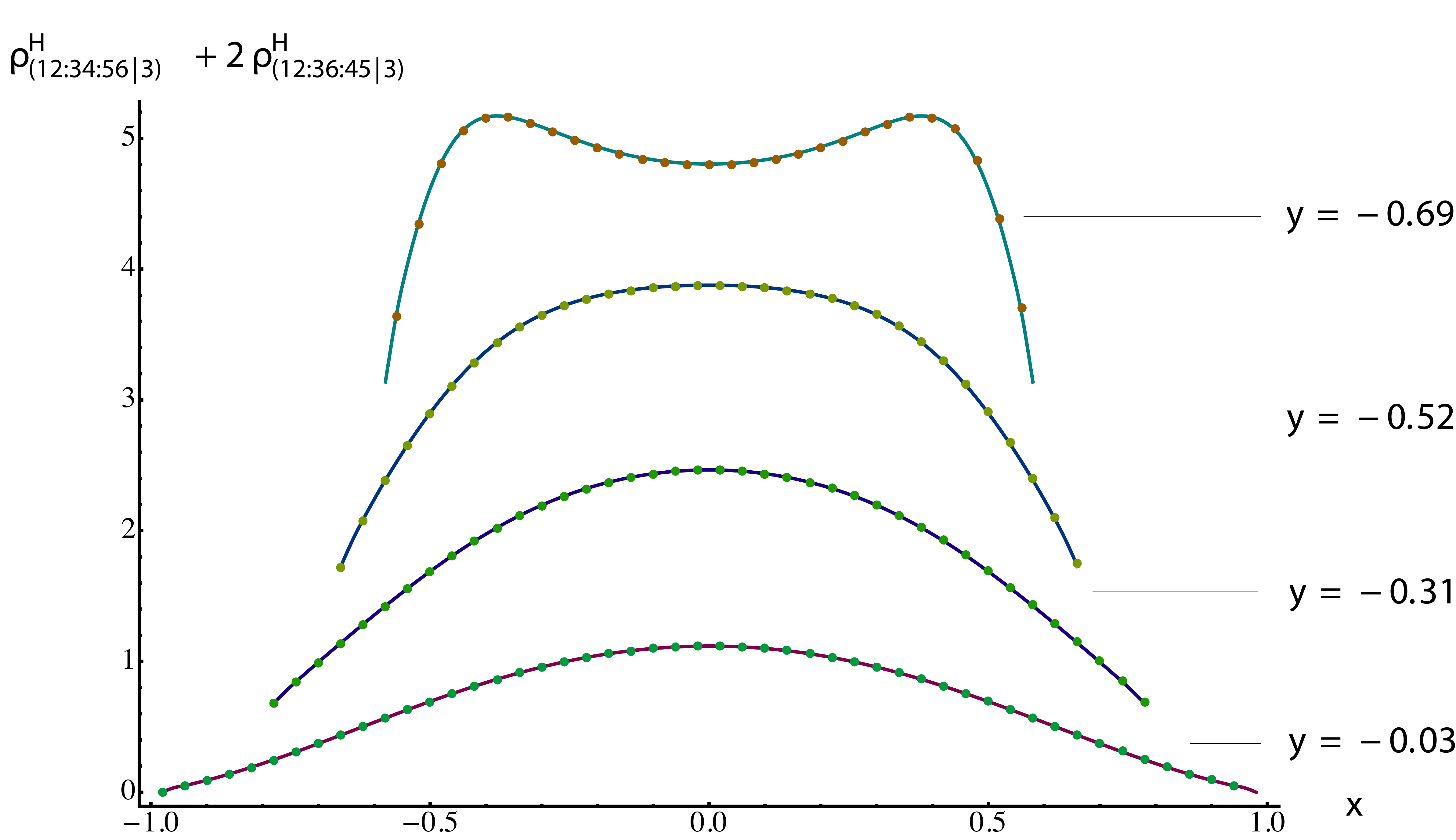}\vspace{.2cm}
\includegraphics[scale=0.25]{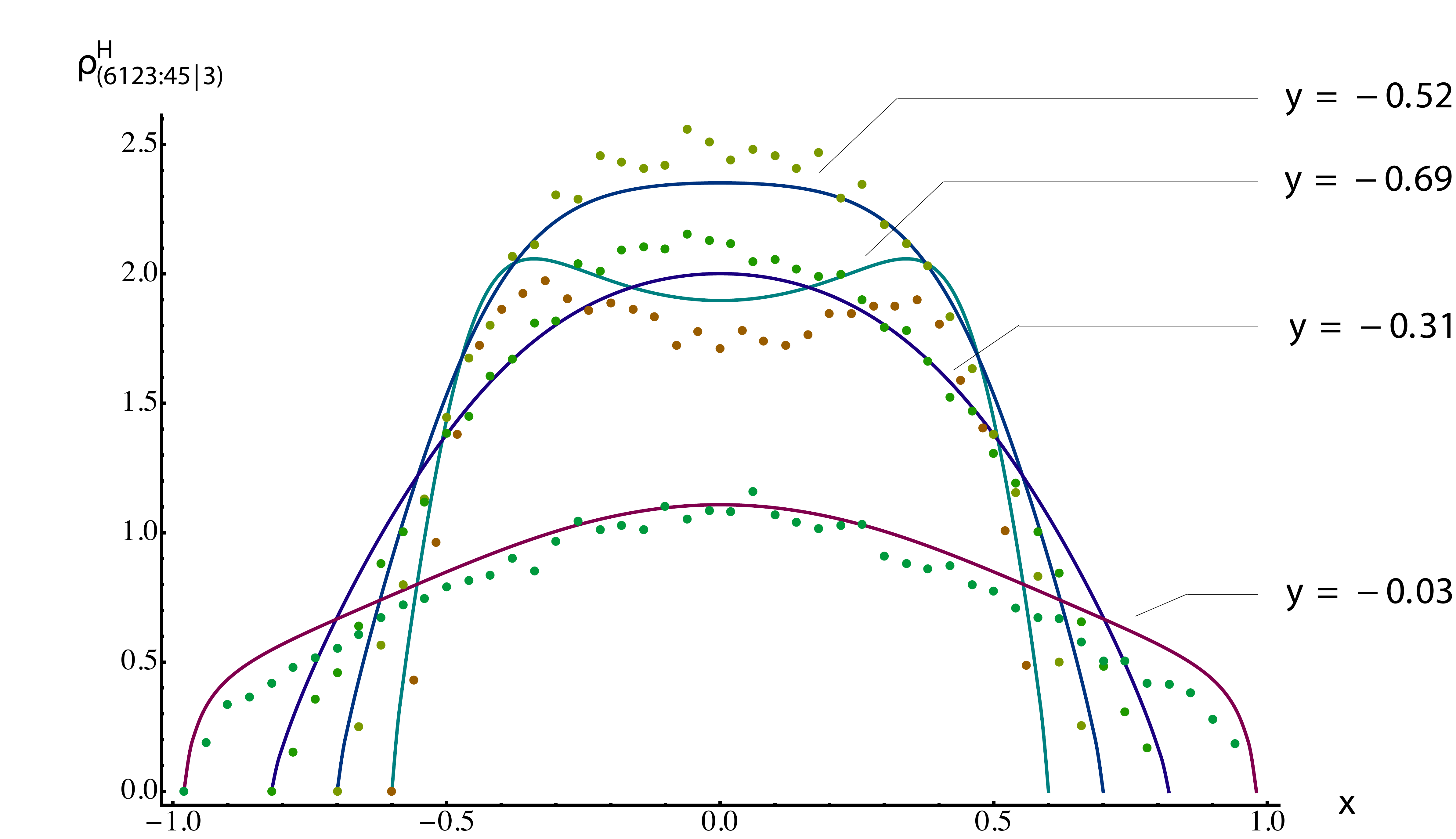}
\caption{Density of Ising FK ($\kappa=16/3$) one-pinch points $\rho^\mathcal{H}_{(12:34:56|3)}+2\rho^\mathcal{H}_{(12:36:45|3)}$ (top) and two-pinch points $\rho^\mathcal{H}_{(6123:45|3)}$ (bottom) versus $x$ in a regular hexagon inscribed in 2000$\times$2000 rhombus and adjusted to have side-length one and center at the origin.  Both densities are normalized to one at the center of $\mathcal{H}$.}
\label{IsingHexPinchPoint}
\end{figure}

Our Ising FK cluster pinch-point simulation results are compared with our theory predictions for $\rho^\mathcal{H}_{(12:34:56|3)}+2\rho^\mathcal{H}_{(12:36:45|3)}$ and $\rho^\mathcal{H}_{(6123:45|3)}$ (respectively (\ref{1pinchpointhexfinal}) and (\ref{2ppindwire}) with $\kappa=16/3$) in figure \ref{IsingHexPinchPoint}.  Explanations and noticeable features of these two plots are the same as those for figure \ref{PercHexPinchPoint}.  The top plot shows the density $\rho^\mathcal{H}_{(12:34:56|3)}+2\rho^\mathcal{H}_{(12:36:45|3)}$ of one-pinch points touched by the hull-walk connecting vertices one and two.  The factor of two arises from the factor of $n^2$ in (\ref{1pinchpointhexfinal}) and because $n=\sqrt{2}$ when $\kappa=16/3$.  This factor may seem unnatural, but omitting it considerably worsens the agreement of prediction and simulation.  Thus, this agreement gives a non-trivial verification of our prediction for the relative coefficients between the two densities in (\ref{1pinchpointhexfinal}).  We note that this one-pinch-point density deviates from our prediction by much less than that in the rectangle.  This seems peculiar since the system size in the rectangle and in the hexagon are about the same.  One might suspect that the larger coordination number of the triangular lattice over that of the rectangular lattice causes this effect, but this suspicion is contradicted by the large errors in the hexagon two-pinch-point case.  We therefore do not propose a possible explanation for this observation.  The  deviations of our two-pinch-point simulation results from our prediction (\ref{2ppindwire}) with $\kappa=16/3$ are larger than those for percolation.  This may be understood as follows.  The two-pinch-point density scales as $a^{2\Theta_2}$ as the lattice spacing $a$ goes to zero \cite{cardydomb}.  For percolation $(\kappa=6)$, the power is $5/4$, but for Ising FK clusters ($\kappa=16/3$), the power increases to $35/24$.  Thus on a very large lattice, the frequency of Ising FK two-pinch-point events is suppressed even more than that of the percolation two-pinch-point events.  Indeed, this is what have observed.  We expect that more samples would lessen the deviation.  However, such simulations would require very considerable computer resources.

\section{Summary}
We summarize the main results of this article.  The half-plane weight $\Pi_\Lambda$ of the type-$\Lambda$ $s$-pinch-point configuration is given by (\ref{1HRectweight}, \ref{2ppcovform}) for the cases $N=2$ and $s=1$ and $2$ and is given by (\ref{2configs}, \ref{2pphexcov}, \ref{3ppcovform}) for the cases $N=3$ and $s=1,2,$ and $3$ also for various pinch-point events $\Lambda$.  After specifying a particular ffbc event  $\varsigma$, we construct the universal partition function $\Upsilon_{(\Lambda|\varsigma)}$ that sums exclusively over the event $\Lambda\cap\varsigma$, and we transform it into a universal partition function with the rectangle $\mathcal{R}$ or the hexagon $\mathcal{H}$ for its domain.  The results are (\ref{2pprect}) and  (\ref{2pphex}) respectively.  Then to within non-universal  factors, the type-$\Lambda$ pinch-point density $\rho^\mathcal{P}_{(\Lambda|\varsigma)}$ is found by dividing the transformed universal partition function by the universal partition function $\Upsilon_\varsigma^\mathcal{P}$ summing exclusively over the ffbc event $\varsigma$.  This prediction agrees well with measurements by simulation that sampled one-pinch-point and two-pinch-point events between critical percolation and Ising FK boundary clusters in rectangles and hexagons with every other side independently wired.

\begin{table}[h!]
\centering
\begin{tabular}{l|cccccc} 
Avg.\,error\,& \hspace{.15cm}$y=-0.69$\hspace{.15cm} & \hspace{.15cm}$y=-0.52$\hspace{.15cm} & \hspace{.15cm}$y=-0.31$\hspace{.15cm} & \hspace{.15cm}$y=-0.03$\hspace{.15cm}  \\
\hline\hline
Perc.\,$s=1$ & $-0.005$ & $-0.004$ & $-0.003$ & $-0.002$ \\ 
Perc.\,$s=2$ & $-0.017$ & $-0.024$ & $-0.024$ & $-0.013$ \\ 
Ising $s=1$ & 0.006 & 0.001 & $-0.000$ &  $-0.000$ \\ 
Ising $s=2$ & 0.140 & $-0.002$ & 0.005 & 0.068 \\ 
\end{tabular}

\vspace{.5cm}

\begin{tabular}{l|cccc} 
Std.\,dev.\,& \hspace{.15cm}$y=-0.69$\hspace{.15cm} & \hspace{.15cm}$y=-0.52$\hspace{.15cm} & \hspace{.15cm}$y=-0.31$\hspace{.15cm} & \hspace{.15cm}$y=-0.03$\hspace{.15cm}  \\
\hline\hline
Perc.\,$s=1$ & 0.004 & 0.004 & 0.003 & 0.002 \\ 
Perc.\,$s=2$ & 0.011 & 0.015 & 0.013 & 0.006 \\ 
Ising $s=1$ & 0.006 & 0.006  & 0.004 & 0.002  \\ 
Ising $s=2$ & 0.052 & 0.114 & 0.107 & 0.045 \\ 
\end{tabular}
\caption{The error (theory minus simulation) averaged over $x$, and the standard deviation of the error from that average, of the data displayed in figures \ref{PercHexPinchPoint} and \ref{IsingHexPinchPoint}.}
\label{HexError}
\end{table}

\section{Acknowledgements}

The authors thank J.\,J.\,H.\,Simmons for insightful conversations.

This work was supported by National Science Foundation Grants Nos.\,PHY-0855335 (SMF), DMR-0536927 (PK and SMF), and DMS-0553487 (RMZ).

\appendix{}

\section{A solution space of the system (\ref{nullstatebulkbdy}-\ref{sct})}\label{appendix}

We demonstrate that the Coulomb gas solution (\ref{chiralcorr}) solves the null-state PDEs (\ref{nullstatebulkbdy}) and the Ward identities (\ref{transl}-\ref{sct}).  This calculation is modeled after a similar calculation presented in \cite{dub}.  Throughout this section, we use the dense phase $(\kappa>4)$ conventions for our Kac charge and screening charge notations (\ref{kaccharge}), we modify the notation of (\ref{chiralcorr}) by letting $x_{2N+1}=z,x_{2N+2}=\bar{z},$ and $x_{2N+2+j}:=u_j$, and we assume that $s<N$.  When $s=N$, (\ref{chiralcorr}) reduces to (\ref{s=N}), and because this formula is purely algebraic, it is easy to directly verify that it solves the system.  As usual, $x_1<x_2<\ldots<x_{2N-1}<x_{2N}.$  

To begin, we consider the function
\be\label{Phi}\Phi(\,x_1,\ldots,x_{2N+2+M})=\prod_{i<j}^{2N+2+M}(x_j-x_i)^{2\alpha_i\alpha_j}.\ee
Our strategy is to choose the $\alpha_i$ (i.e., the charges of the chiral operators) such that for $1\leq i\leq2N$,
\begin{multline}\label{act}\bigg[\frac{\kappa}{4}\partial_i^2+\sum_{j\neq i}^{2N}\left(\frac{\partial_j}{x_j-x_i}-\frac{\theta_1}{(x_j-x_i)^2}\right)+\frac{\partial_{2N+1}}{x_{2N+1}-x_i}\\
+\frac{\partial_{2N+2}}{x_{2N+2}-x_i}-\frac{\Theta_s}{(x_{2N+1}-x_i)^2}-\frac{\Theta_s}{(x_{2N+2}-x_i)^2}\bigg]\Phi=\sum_{k=2N+3}^{2N+2+M}\partial_k(\,\,\ldots\,\,),\end{multline}
where we recognize the differential operator of the $i$-th null-state PDE (\ref{nullstatebulkbdy}) on the left side of (\ref{act}), and then integrate $x_{2N+3},\ldots,x_{2N+2+M}$ on both sides of (\ref{act}) around nonintersecting closed contours $\Gamma_1,\ldots,\Gamma_M$ such as Pochhammer contours entwining pairs of the points $x_1,\ldots,x_{2N+2}$ (figure \ref{PochhammerContour}).  Integration on the right side gives zero.  On the left side, the integrand is a smooth function of $x_1,\ldots,x_{2N+M+2}$ because the contours do not intersect.  Thus, we may commute each differentiation with each integration to find that the $M$-fold integral $\oint\Phi$ solves the system (\ref{nullstatebulkbdy}). 

With some algebra, we find that for any choice of $\{h_j\},\{\alpha_j\}$, $M\in\mathbb{Z}^+$, and $1\leq i\leq2N$,
\begin{multline}\label{algebra}\left[\frac{\kappa}{4}\partial_i^2+\sum_{j\neq i}^{2N+2+M}\left(\frac{\partial_j}{x_j-x_i}-\frac{h_j}{(x_j-x_i)^2}\right)\right]\Phi\\
=\left[\sum_{\substack{j,k\neq i\\ j\neq k}}^{2N+2+M}\frac{\alpha_j\alpha_k(\kappa\alpha_i^2-1)}{(x_j-x_i)(x_k-x_i)}+\sum_{j\neq i}^{2N+2+M}\frac{\alpha_i\alpha_j(\kappa\alpha_i\alpha_j-\kappa/2+2)-h_j}{(x_j-x_i)^2}\right]\Phi.\end{multline}
If we set $h_j=\theta_1$ for $1\leq j\leq2N$, $h_j=\Theta_s$ for $j=2N+1,2N+2$, and $h_j=1$ for $j>2N+2$ (the conformal weight of the $V_\pm$ chiral operators that generate screening operators), then for $1\leq i\leq2N$, we can write (\ref{algebra}) as
\begin{multline}\label{totalderiv+}\left[\frac{\kappa}{4}\partial_i^2+\sum_{j\neq i}^{2N}\left(\frac{\partial_j}{x_j-x_i}-\frac{\theta_1}{(x_j-x_i)^2}\right)+\,\frac{\partial_{2N+1}}{x_{2N+1}-x_i}+\frac{\partial_{2N+2}}{x_{2N+2}-x_i}-\frac{\Theta_s}{(x_{2N+1}-x_i)^2}-\frac{\Theta_s}{(x_{2N+2}-x_i)^2}\right]\Phi\\
=\left[\sum_{\substack{j,k\neq i\\ j\neq k}}^{2N+2+M}\frac{\alpha_j\alpha_k(\kappa\alpha_i^2-1)}{(x_j-x_i)(x_k-x_i)}+\sum_{j\neq i}^{2N+2+M}\frac{\alpha_i\alpha_j(\kappa\alpha_i\alpha_j-\kappa/2+2)-h_j}{(x_j-x_i)^2}\right]\Phi+\sum_{k=2N+3}^{2N+2+M} \partial_k\left(-\frac{\Phi}{x_k-x_i}\right).\end{multline}
Next, we choose a particular $i\in\{1,\ldots,2N\}$.  If we set
\be\label{branches}\alpha_i=\alpha_{1,2}^-=1/\sqrt{\kappa},\quad\alpha_j=\alpha_0\pm\sqrt{\alpha_0+h_j},\quad j\neq i,\ee
then the term in brackets on the right side of (\ref{totalderiv+}) vanishes, giving the desired form (\ref{act}).  In order to satisfy (\ref{branches}) for all $1\leq i\leq2N$, we need
\begin{align}\label{charges}&\alpha_j=\alpha_{1,2}^-,&&1\leq j\leq2N,\\
\label{middle}&\alpha_j=\alpha_{0,s}^\pm,&&2N+1\leq j\leq2N+2,\\
\label{screen}&\alpha_j=\alpha_\pm,&&2N+3\leq j\leq2N+2+M.\end{align}
Explicit formulas for (\ref{charges}-\ref{screen}) are given by (\ref{kaccharge}).  If we make this choice, then $\oint\Phi$ solves the $2N$ null-state equations.

To finish, we choose $M$ and the signs of the square roots in (\ref{middle}, \ref{screen}) so that $F:=\oint\Phi$ solves the conformal Ward identities (\ref{transl}-\ref{sct}) too.  This happens if the sum of the powers in $\Phi$ involving the screening charge $\alpha_j$ equals negative two for each $j\geq2N+3$.  We now motivate this claim.  The Ward identities dictate that $F$ must have the conformally covariant form (\ref{covform}), which is equivalent to demanding that the function
\bea\label{Psi}G(x_1,\ldots,x_{2N};x_{2N+1},x_{2N+2})&:=&(x_{2N+2}-x_{2N+1})^{2\Theta_s}\prod_{j=1,\,\,\text{odd}}^{2N-1}(x_{j+1}-x_j)^{2\theta_1}F(x_1,\ldots,x_{2N+2})\\
&=&(x_{2N+2}-x_{2N+1})^{[16s^2-(\kappa-4)^2]/8\kappa}\prod_{j=1,\,\,\text{odd}}^{2N-1}(x_{j+1}-x_j)^{6/\kappa-1}\\
&\times&\oint\ldots\oint\Phi(x_1,\ldots,x_{2N+2+M})\,dx_{2N+3}\ldots\,dx_{2N+2+M}\hspace{1cm}\eea
only depends on a set of $2N-1$ independent cross-ratios that can be formed from $x_1,\ldots,x_{2N+2}$.  In particular, we choose
\be\label{f}\eta_j=f(x_j),\hspace{.5cm}j=2,\ldots,2N-2,2N+1,2N+2,\hspace{.5cm}\text{where}\hspace{.5cm}f(x):=\frac{(x-x_1)(x_{2N}-x_{2N-1})}{(x_{2N-1}-x_1)(x_{2N}-x)},\ee
and we note that $f(x_1)=0<\eta_2<\eta_3<\ldots<\eta_{2N-2}<f(x_{2N-1})=1<f(x_{2N})=\infty$.  Then this condition amounts to requiring that $G$ satisfy
\be\label{PsiRequirement}G(x_1,x_2,x_3,\ldots,x_{2N-2},x_{2N-1},x_{2N};x_{2N+1},x_{2N+2})=G(0,\eta_2,\eta_3,\ldots,\eta_{2N-2},1,\infty;\eta_{2N+1},\eta_{2N+2}).\ee
Because the right side must ultimately be finite (and this requirement fixes $M$ once the signs in (\ref{middle}, \ref{screen}) are chosen), we momentarily ignore all of its infinite factors.  Thus, the $j$-th integral on the left side of (\ref{PsiRequirement}) has the form
\be\label{secondint}\int\prod_{k=1}^{2N+2}(x_k-x_j)^{\beta_k}\prod_{\substack{l=2N+3 \\ l\neq j}}^{2N+2+M}(x_l-x_j)^{\beta_l}\,dx_j\ee 
while the $j$-th integral on the right side of (\ref{PsiRequirement}) has the form (using $\eta_j$ instead of $x_j$ for our symbol of integration)
\be\label{firstint}\int \eta_j^{\beta_1}(1-\eta_j)^{\beta_{2N-1}}\prod_{\substack{k=2 \\ k\neq 2N-1,2N}}^{2N+2}(\eta_k-\eta_j)^{\beta_k}\prod_{\substack{l=2N+3 \\ l\neq j}}^{2N+2+M}(\eta_l-\eta_j)^{\beta_l}\,d\eta_j.\ee
(We note that the integrand of (\ref{secondint}) contains an extra factor that was ignored in (\ref{firstint}) when $x_{2N}$ was set to infinity.)  The simplest way to achieve the equality in (\ref{PsiRequirement}) is for these two integrals to be the same up to algebraic prefactors.  Upon introducing the change of variables 
\be\eta_j=f(x_j),\hspace{.5cm}j\neq 1,2N-1,2N,\ee
with $f$ defined in (\ref{f}), the integral (\ref{firstint}) transforms into 
\be\label{transint}\mathcal{P}(x_1,\ldots,x_{2N+2})\int\prod_{\substack{k=1\\ k\neq 2N}}^{2N+2}\left(\frac{x_k-x_j}{x_{2N}-x_j}\right)^{\beta_k}\prod_{\substack{l=2N+3 \\ l\neq j}}^{2N+2+M}\left(\frac{x_l-x_j}{x_{2N}-x_j}\right)^{\beta_l}\frac{dx_j}{(x_{2N}-x_j)^2},\ee
where $\mathcal{P}(x_1,\ldots,x_{2N+2})$ is an algebraic prefactor whose explicit formula is presently unimportant.  The main point for now is that in order to match the integral in (\ref{transint}) with that in (\ref{secondint}), we must have
\be \beta_{2N}=-\beta_1-\ldots-\beta_{j-1}-\beta_{j+1}-\ldots-\beta_{2N-1}-\beta_{2N+1}-\ldots-\beta_{2N+M+2}-2.\ee
Thus the sum of the powers in the integrand of the $j$-th integral (\ref{secondint}) appearing in $\oint\Phi$ (when that integral is considered to be evaluated first) should equal negative two.  This must be true for all $j\in\{2N+3,\ldots,2N+M\}$.  We note that this requirement is satisfied if all of the $\alpha_j$ in (\ref{Phi}) sum to $2\alpha_0$, with $2\alpha_0=\alpha_++\alpha_-$ the background charge; this is the Coulomb gas neutrality condition \cite{df}:
\be\label{neut}-2=\sum_{i\neq j}\beta_i=2\alpha_j\sum_{i\neq j}\alpha_i=2\alpha_j\left(\sum_i\alpha_i-\alpha_j\right)=2\alpha_\pm\left(\sum_i\alpha_i-\alpha_\pm\right)\hspace{1cm}\Longleftrightarrow\hspace{1cm}\sum_i\alpha_i=2\alpha_0.\ee
(In (\ref{neut}), we have used the identities $\alpha_++\alpha_-=2\alpha_0$ and $\alpha_+\alpha_-=-1$.)  Thus, we should choose $M$ and the signs of the square roots in (\ref{middle}, \ref{screen}) so that the total charge $\sum_j\alpha_j$ equals $2\alpha_0$ and $\oint\Phi$ therefore solves the Ward identities.  We suppose that $m_+$ (resp.\,$m_-$) of the screening charges equal $\alpha_+$ (resp.\,$\alpha_-$) with $m_++m_-=M$ necessarily.  Then the total charge is
\bea 2\alpha_0&=&\sum_j\alpha_j=2N\alpha_{1,2}^-+\alpha_{0,s}^\pm+\alpha_{0,s}^\pm+m_+\alpha_++m_-\alpha_-\\
&=&-N\alpha_-+\alpha_++(1\pm s)\alpha_-/2+(1\pm s)\alpha_-/2+m_+\alpha_++m_-\alpha_-\\
&=&\label{threecases}\begin{cases}(m_++1)\alpha_++(-N+s+m_-+1)\alpha_-&++\\
(m_++1)\alpha_++(-N-s+m_-+1)\alpha_-&--\\
(m_++1)\alpha_++(-N+m_-+1)\alpha_-&+-\end{cases}.\eea
The $++$ (resp.\,$--$) case corresponds to using the $+$ (resp.\,$-$) sign for both $\alpha_{0,s}$, and the $+-$ case corresponds to using the $+$ sign for one of the $\alpha_{0,s}$ and the $-$ sign for the other.  In order for these expressions to equal $2\alpha_0$, we need the coefficients of $\alpha_+$ and $\alpha_-$ in (\ref{threecases}) to equal one.  Therefore, $m_+=0$ and $m_-=M$, so all $M$ screening charges use the $-$ sign.  The number of screening charges is thus given by
\be\label{M}M=\begin{cases}N-s&++\\ N+s&--\\ N&+-\end{cases}.\ee
The $++$ case is used exclusively in this article since it apparently contains all of the pinch-point densities when $N=1,2,$ and $3$.  We anticipate that the $++$ case contains all pinch-point densities for all $N\in\mathbb{Z}^+$.  Moreover, we see that these solutions span a subspace of a larger solution space containing all cases in (\ref{M}) (among possibly more solutions).  

To explicitly show that $\oint\Phi$ satisfies the conformal Ward identities (\ref{transl}-\ref{sct}), we can check that relation (\ref{PsiRequirement}) is satisfied by introducing the substitution $\eta_j=f(x_j)$, with $j=2,\ldots,2N-2,2N+1,\ldots,2N+2+M$ and with $f$ defined in (\ref{f}), on the right side and performing some straightforward algebra.

\end{document}